\documentclass[final,11pt]{article}
\pdfoutput=1
\usepackage{MRnotes}

\newcommand{\RNAdS}[1]{Reissner-Nordstr\"om-AdS$_{#1}$}

\newcommand{\harm}[1]{\operatorname{Har}\left(#1\right)}

\newcommand{\WeylR}{\tensor[^{\mathcal{W}}]{R}{}}
\newcommand{\ann}{\mathscr{M}}   
\newcommand{\sen}[1]{\varphi_{_{#1}}} 

\newcommand{\bwt}{\mathfrak{w}}   
\newcommand{\bqt}{\mathfrak{q}}    
\newcommand{\bk}{\vb{k}}  
\newcommand{\bx}{\vb{x}}   

\newcommand{\ctor}{\zeta}  
\newcommand{\br}{\xi} 
\newcommand{\ri}{\varrho}   
\newcommand{\Dz}{\mathbb{D}}  
\newcommand{\Dann}[1]{\mathfrak{D}_{#1}}  

\newcommand{\ScS}{\mathbb{S}} 
\newcommand{\ScST}{\mathbb{S}^{\scriptscriptstyle{\text{T}}} } 

\newcommand{\HH}{\Psi}
\newcommand{\HHT}{\Psi_{_\text{T}}}
\newcommand{\FX}{\mathbb{X}}

\newcommand{\FY}{\mathbb{Y}}
\newcommand{\FYE}{\mathbb{Y}_{_\text{E}}}
\newcommand{\FYO}{\mathbb{Y}_{_\text{O}}}
\newcommand{\FYB}{\mathbb{Y}_{_\text{B}}}
\newcommand{\FYS}{\mathbb{Y}_{_\text{S}}}
\newcommand{\gOrb}{\Upsilon}

\newcommand{\HS}{\Psi_{_\text{S}}}
\newcommand{\PHE}{\Phi_{_\text{E}}}
\newcommand{\PHO}{\Phi_{_\text{O}}}
\newcommand{\PHW}{\Phi_{_\text{W}}}

\newcommand{\PHB}{\Phi_{_\text{B}}}

\newcommand{\EEq}{\mathbb{E}}
\newcommand{\EEqT}{\mathbb{E}_{_\text{T}}}

\newcommand{\EEqB}{\mathbb{E}_{_\text{B}}}

\newcommand{\TcftD}{T^{\text{\tiny{CFT}}}}
\newcommand{\TcftU}{T_{\text{\tiny{CFT}}}}

\newcommand{\Lk}{\Lambda_k}
\newcommand{\MW}{\Theta}
\newcommand{\MZ}{\mathscr{Z}}
\newcommand{\Zmt}{\widetilde{\mathscr{Z}}}
\newcommand{\PiZ}{\Pi_{_\MZ}}
\newcommand{\Tone}{\mathscr{T}}

\newcommand{\OpZ}{\breve{\mathcal{O}}_{_\MZ}}
\newcommand{\PoZ}{\breve{\mathcal{Z}}}
\newcommand{\EOp}{\breve{\mathcal{E}} }
\newcommand{\JoZ}{\breve{\bm{\zeta}}}

\newcommand{\Gatt}{\Gamma_s}
\newcommand{\Ost}{\widetilde{\Gamma}_s(\omega,\bk)}
\newcommand{\kaps}{\mathfrak{p}_s}
\newcommand{\KS}{K_s}

\newcommand{\In}{\text{\tiny{in}}}     
\newcommand{\Rev}{\text{\tiny{rev}}}  

\newcommand{\nB}{n_{_B}}   
\newcommand{\Gin}[1]{G_{_{#1}}^\In}     
\newcommand{\Kin}[1]{K_{_{#1}}^\In}      
\newcommand{\Grev}[1]{G_{_#1}^\Rev}   
\newcommand{\Krev}[1]{K_{_{#1}}^\Rev}    

\newcommand{\Mser}[2]{\varphi_{_{#1}}^{#2}}  
\newcommand{\Mserh}[2]{\hat{\varphi}_{_{#1}}^{#2}} 
\newcommand{\Dfn}[2]{\Delta_{_{#1}}^{#2}} 
\newcommand{\Dfnh}[2]{\hat{\Delta}_{_{#1}}^{#2}} 

\newcommand{\skR}{\text{\tiny R}}
\newcommand{\skL}{\text{\tiny L}}

\title{The timbre of Hawking gravitons: an effective description of energy transport from holography}
\author[a]{Temple He,}
\author[b]{R. Loganayagam,}
\author[a]{Mukund Rangamani,}
\author[b]{Akhil Sivakumar,}
\author[a]{Julio Virrueta}

\affiliation[a]{
	Center for Quantum Mathematics and Physics (QMAP)\\
	Department of Physics \& Astronomy, University of California, Davis, CA 95616 USA}
\affiliation[b]{
	International Centre for Theoretical Sciences (ICTS-TIFR), \\ 
	Tata Institute of Fundamental Research, Shivakote, Hesaraghatta, Bangalore 560089, India.}

\emailAdd{tmhe@ucdavis.edu}
\emailAdd{nayagam@icts.res.in}
\emailAdd{mukund@physics.ucdavis.edu}
\emailAdd{akhil.sivakumar@icts.res.in}
\emailAdd{jvirrueta@ucdavis.edu}

\abstract{
Planar black holes in AdS, which are holographically dual to compressible relativistic fluids,  have a long-lived phonon mode that captures the physics of attenuated sound propagation and transports energy in the plasma. We describe the open effective field theory of this fluctuating phonon degree of freedom. The dynamics of the phonon is encoded in a single scalar field whose gravitational coupling has non-trivial spatial momentum dependence. This description fits neatly into the paradigm of classifying gravitational modes by their Markovianity index, depending on whether they are long-lived. The sound scalar is a non-Markovian field with index $3-d$ for a $d$-dimensional fluid. We reproduce (and extend) the  dispersion relation of the holographic sound mode to quartic order in derivatives, constructing in the process the effective field theory governing its attenuated dynamics and associated stochastic fluctuations. We also remark on the presence of additional spatially homogeneous zero modes in the gravitational problem, which remain disconnected from the phonon Goldstone mode.
}

\begin{document}
\maketitle


\section{Introduction}
\label{sec:intro}

Perturbations of asymptotically  AdS black holes relate to the thermal stochastic dynamics of the dual boundary conformal field theory plasma \cite{Horowitz:1999jd,Policastro:2001yc,Policastro:2002se,Policastro:2002tn}. The classical decay of perturbations via quasinormal modes is accompanied by the thermal and quantum fluctuations, manifested in the form of Hawking quanta. The physics of such dissipative systems should be encoded in a real-time open effective field theory (EFT). Such a theory can be constructed as the dynamics induced onto a probe system coupled to the thermal environment after integrating out the plasma degrees of freedom.  Recently, such open EFTs were constructed from holography: for instance, \cite{Jana:2020vyx} describes the non-linear dynamics of quasinormal modes associated with probes that have short lifetimes of order the thermal scale, and \cite{Ghosh:2020lel,He:2021jna}  (cf., also \cite{Glorioso:2018mmw,deBoer:2018qqm,Bu:2020jfo}) describe the physics of diffusion which originates from the fact that the conserved currents of plasma have long-lived modes. Other related work includes \cite{Chakrabarty:2019aeu,Loganayagam:2020eue,Loganayagam:2020iol,Chakrabarty:2020ohe,Bu:2021clf}. Progress in this direction was spurred by \cite{Glorioso:2018mmw} who gave a simple geometric proposal for computing real-time, Schwinger-Keldysh ordered observables in holography (cf.,  \cite{Son:2002sd,Herzog:2002pc,Skenderis:2008dh,vanRees:2009rw} for earlier work in this direction). 

In the previous works \cite{Ghosh:2020lel,He:2021jna} the focus was on constructing an open effective field theory of diffusive modes focusing on  momentum diffusion. This occurs through shear modes, which carry transverse momentum and shear the fluid elements  (hence their name) without compressing the plasma as they propagate.  Hence, the shear  hydrodynamic mode is present in compressible as well as incompressible fluids alike. The shearing causes  transverse viscous drag, resulting in the diffusion of the transverse momentum. These shear modes are long-lived and diffuse slowly through the medium. 
 
On the other hand, compressible fluids  have an additional degree of freedom: the sound mode. Sound travels by carrying a longitudinal momentum, applying pressure on the fluid elements, which in turn results in a wave of compressions and rarefactions. As the fluid gets compressed, there is a local change in pressure and energy density, unlike in the case of shear modes (where the local pressure and energy density remain unperturbed). We remind the reader that the relativistic fluids are always compressible since incompressibility requires an instantaneous transmission of pressure which is forbidden within special relativity. Thus, relativistic fluids  always have sound modes. The physics of shear and  sound are qualitatively different. While shear modes are diffusive and obey parabolic PDEs, sound modes are oscillatory and obey hyperbolic PDEs.  The difference owes to the fact that a fluid at rest already has an energy density and pressure. A perturbation over this background results in the  sound mode.\footnote{
	We will  only discuss perturbations which do not change the flat spacetime energy density. A static homogeneous change of temperature or energy density is IR divergent  and not included in our analysis. \label{fn:noEchange}} 
	In contrast, a fluid at rest has no momentum density, so the transverse momentum diffuses,  resulting in the shear mode.

We have described until now the sound and the shear mode as being governed by second order PDEs. In a realistic system, higher derivative corrections appear and  the effects  of the thermal fluctuations  need to be considered, resulting in a higher derivative stochastic PDE. In the case of  the sound mode, higher derivative corrections describe the sound relaxation whereas fluctuations describe the  noise background. More precisely, the sound mode (and the shear mode) should be thought in terms of an \emph{open effective theory} incorporating both fluctuation and dissipation.  Such a theory is obtained by integrating out the fast modes in the plasma. This is more easily said than done: one first needs to systematically separate out the fast modes from the slow modes. In weakly coupled theories, we need to further deal with the fact that sound relaxes over non-perturbatively long time scales. We will sidestep these issues by considering a strongly coupled CFT plasma and study it using holography. Sound and shear modes in holography appear as linearized metric perturbations over planar \SAdS{d+1} black holes \cite{Policastro:2002se,Policastro:2002tn}. Non-linearly, the plasma dynamics is captured by the fluid/gravity correspondence \cite{Bhattacharyya:2008jc,Hubeny:2011hd} (which as constructed to date however only captures the dissipative but not the fluctuation effects).

The holographic set-up using gravitational dynamics to model the dynamics of a strongly coupled plasma has some important features highlighted in \cite{Ghosh:2020lel}. Firstly, it makes clear that the dynamics of short-lived and long-lived modes, dubbed Markovian and non-Markovian, respectively,  are qualitatively different. At the linearized level, each such mode is described by a scalar field propagating in the AdS black hole background. These fields were called designer scalars as they are non-minimally coupled to the  gravitational background, with the coupling  modeled as a (auxiliary) background dilaton. This dilatonic coupling modulates the interaction the field has in the radial direction. Heuristically, the short-lived modes are repelled from the boundary of the spacetime, while the long-lived modes are floppy and have a large wavefunction support near the boundary, which prevents them from decaying away rapidly.  Technically, the Markovian nature is captured by the asymptotic behaviour of the dilaton and can be encoded in a single number, the \emph{Markovianity index}, $\ann$. 

The Markovian nature of a bulk field is characterized by the boundary conditions imposed on it in order to compute correlation functions via the GKPW dictionary. 
\begin{itemize}[wide,left=0pt]
\item Markovian fields are quantized with Dirichlet boundary conditions and have index $\ann >-1$.
\item Non-Markovian fields are quantized with Neumann boundary conditions and have index $\ann <1$. 
\end{itemize}  
This definition mildly updates the definition given in \cite{Ghosh:2020lel}. The main distinction is that fields in the window $\ann \in (-1,1)$ can either be Markovian or non-Markovian, depending on the boundary conditions imposed. This window is similar to the usual discussion of relevant operators close to the unitarity bound in AdS/CFT \cite{Klebanov:1999tb}, when near-boundary fall-offs are slow enough  that one can switch to general multi-trace boundary conditions \cite{Witten:2001ua}. 

Secondly, the linearized metric  perturbations which contain both short and long-lived modes can be decoupled by working with diffeomorphism/gauge   invariant degrees of freedom. This provides  a clean separation of fast and slow modes, allowing one to integrate out the former. It  also makes clear what the natural gauge choices are for analyzing perturbations in the real-time gravitational Schwinger-Keldysh (grSK) geometry of \cite{Glorioso:2018mmw}. For instance, use of standard gauge fixing in the AdS black hole background leads to solutions which have spurious singularities at the horizon, necessitating artificial boundary conditions in the interior of the spacetime, in tension with the rules of the AdS/CFT correspondence (this is an issue for instance in the solutions discussed in \cite{Glorioso:2018mmw} and \cite{Bu:2020jfo}). 
In contrast, the gauge invariant variables chosen in  \cite{Ghosh:2020lel,He:2021jna} leads to a smooth solution on the grSK geometry with appropriate boundary conditions on the two asymptotic boundaries (corresponding to the bra and ket pieces of the Schwinger-Keldysh evolution). 

Let us understand these statements by examining the dynamics of the plasma stress tensor. Given a direction of propagation we can identify polarizations labeled by the little group in the transverse spatial geometry.  The stress tensor operator has traceless spin-$2$ polarizations which are short-lived, and hence are Markovian with index $\ann = d-1$. The transverse vector spin-$1$ polarizations are non-Markovian with index $\ann =1-d$ and correspond to the shear modes of the plasma. This leaves the single longitudinal mode which is the focus of the present paper -- it relates to the sound mode resulting from energy transport. These statements translate directly in the dual gravity picture, for the selfsame decomposition can be applied to the linearized gravitons  \cite{Kodama:2003jz,Kovtun:2005ev}. 

In \cite{Ghosh:2020lel} and \cite{He:2021jna} we derived the effective dynamics from the dual gravitational perspective for the spin-$2$ transverse traceless tensor and spin-$1$ transverse vector modes.  The non-Markovian shear mode was captured by a designer scalar with index $1-d$. This scalar, which is weakly coupled near the AdS boundary due to the dilatonic modulation, should be quantized with Neumann boundary conditions for purposes of computing the generating function of dual stress tensor correlators. This is not an ad-hoc choice, but rather one enforced by the underlying Einstein-Hilbert dynamics which, when distilled through the field redefinitions necessary to arrive at the decoupled designer scalar dynamics, ensures that the variational principle arises with suitable boundary terms for this choice. 

Since the shear mode is long-lived, constructing a local effective field theory requires that we treat it in a Wilsonian fashion. Rather than compute the generating function of Schwinger-Keldysh correlators one therefore chooses to compute a Wilsonian influence functional (WIF) parameterized by the long-lived boundary modulus field --  the momentum flux operator for shear modes. In practice, one can obtain this WIF by Legendre transforming the generating function of correlators. Happily, the Legendre transformation is  simple and can be obtained by quantizing the designer non-Markovian field with (renormalized) Dirichlet boundary conditions. 

Based on the description of sound propagation in relativistic plasma, one expects that the longitudinal modes of the gravitons are similarly captured by an effective designer non-Markovian scalar, which is dual to the boundary energy flux operator. While this observation is morally correct, the technical details are significantly more involved. For one, the field dual to the energy flux operator by itself (which we denote as $\MW$ below) does not have a nice dual geometric description. Rather, a linear combination of this field and the conformal mode of gravitational perturbations has simple autonomous non-Markovian dynamics. Even more curiously, this designer field, denoted $\MZ$, has its coupling to the background geometry modulated non-trivially as a function of momentum.

While the technical reason for these statements can be traced through the derivation, we do not understand the physical reason for why this should happen. On the contrary, the field $\MW$, while exhibiting no such pathologies, does not have simple autonomous dynamics. All of this is in stark contrast with the physics of diffusive modes; even for charged plasmas where the shear mode mixes with the Markovian charge current mode, the dynamical system allows for clear decoupling and relatively simple dynamics for the resulting non-Markovian degrees of freedom \cite{He:2021jna}. 

We will primarily focus on the spatially inhomogeneous modes of the plasma. Physically, we imagine cutting-off spatial momenta as $k \geq k_{_\text{IR}}$ and examining the dynamics of propagating sound modes above this  cut-off. This is sensible for the plasma on $\mathbb{R}^{d-1,1}$ to mitigate IR effects.\footnote{
	The Goldstone mode for sound in Minkowski spacetime has soft modes which may be tamed by considering the plasma on a large compact spatial sphere, i.e., using global AdS to provide a regulator. While this is an interesting problem to study, working with a momentum cut-off will suffice to extract the physics of fluctuating phonons.} 
This perspective will be important for us, since the dynamical system we analyze exhibits a somewhat discontinuous behaviour as a function of momentum. Spatially homogeneous modes (zero spatial momentum) have a qualitatively different behaviour. For one, their dynamics appears to be Markovian, and further there are zero modes that do not merge into $\MZ$. One such is the mode which corresponds to a homogeneous static heating  of the plasma, which as explained in \cref{fn:noEchange}  is unphysical (it changes the background solution). Since these modes do not directly affect the dynamics of sound, we will mostly not discuss them in the main text. Nevertheless, for the sake of completeness, we include an analysis of the homogeneous solution space. We demonstrate that it can be understood as the space of large diffeomorphisms, and moreover demarcate the part which is captured  by our designer field  in \cref{sec:zeromodes} (see \cite{deOliveira:2018jhc} for  an earlier analysis).

Part of the complication in the sound mode sector has to do with the fact that there are many degrees of freedom  in the dual gravitational description. One has seven metric functions which can be combined into diffeomorphism invariant combinations. Four of these can be eliminated a-priori, three by gauge fixing and one by using an algebraic constraint arising from the dynamical equations of motion (latter for $k\neq 0$). The remaining three fields satisfy three linearly independent equations of motion,  of which two can be solved by introducing the field $\MW$ and enabling elimination of two functions. The final step is to show that this field combines with the conformal mode to produce the designer field $\MZ$ for spatially inhomogeneous modes. Somewhat amazingly, the off-shell Einstein-Hilbert action simplifies considerably when we parameterize the linearized gravitons using $\MZ$. 

The complexities are all pushed to pure boundary terms (a consequence of the large degree of redundancy in the classical system). These also simplify significantly when we consider asymptotically locally AdS boundary conditions, allowing one to show that for purposes of computing correlation functions, $\MZ$ should be quantized with Neumann boundary conditions.  Earlier analyses of scalar sector quasinormal modes in AdS black hole geometries  have not carefully analyzed the variational principle, leading to some inaccurate statements in the literature. 

The reduction to the single scalar field $\MZ$ was first ascertained by \cite{Kodama:2003jz} at the level of equations of motion. Using their results and examining the asymptotics in global \SAdS{4}, \cite{Michalogiorgakis:2006jc} argued that one should impose a Robin boundary condition for the global analog of our field $\MZ$.  During their study of planar
 black hole quasinormal modes, \cite{Morgan:2009pn} argued that the field $\MZ$ should have Robin boundary conditions in $d=3,4$ but should have Dirichlet boundary conditions in $d\geq5$.\footnote{
	A  different combination of the linearized metric components with decoupled dynamics was constructed in \cite{Kovtun:2005ev} in radial gauge. This combination should be quantized with Dirichlet boundary conditions as noted by the authors. We have not made direct contact with their variables, but note that the relation to the field $\MZ$ can be recovered from   Eq. (4.7) of  \cite{Morgan:2009pn}.}  
Since we follow the field redefinitions and examine the variational principle, we have a clear prescription to obtain an unambiguous answer: $\MZ$ should be quantized with Neumann boundary conditions to obtain boundary  stress tensor correlation functions.\footnote{
	This is true even in low dimensions $d=3,4$ where the index for $\MZ$ lies in the window  $\ann_{\MZ} \in (-1,1)$ mentioned earlier, allowing for both sets of boundary conditions. So while general boundary conditions are technically allowed, they don't compute stress tensor correlation functions.} 
One way to see the issue is to note that the field $\MZ$, and not its conjugate momentum, gets renormalized by higher order counterterms. The renormalization of $\MZ$ starts with the counterterm at quartic order in boundary gradients (the leading cosmological constant and boundary Einstein-Hilbert counterterms do not renormalize $\MZ$). Usual AdS/CFT dictionary with Dirichlet boundary conditions renormalizes the field momentum and not the field, which is held frozen as the source.

Once we have isolated the dynamics in terms of a single designer scalar degree of freedom, the rest of the analysis follows along the lines described in \cite{Ghosh:2020lel}. We first obtain the ingoing boundary to bulk (inverse) Green's function by solving the dynamical equation of motion, order by order in boundary gradient expansion. We discover in this process another surprise:  the solution for $\MZ$  can be written in terms of the solution to the minimally coupled scalar wave equation (equivalently the tensor mode solution) with some simple replacement rules up to cubic order in gradients. At quartic order there are new functions which reflect the change in the nature of the dynamics. 

Armed with the boundary to bulk (inverse) ingoing Green's function we can construct the full boundary to bulk solution on the grSK geometry, parameterized by the expectation value of the corresponding boundary plasma operator. This information then suffices to obtain the WIF at the Gaussian order by evaluating the on-shell action (which gives the saddle point semiclassical answer). This expression can be equivalently written in terms of the energy flux operator by a suitable field redefinition. 

This is the  primary result of the paper: we reproduce in the process the expected correlation function of the energy-momentum tensor isolating the locus of the sound pole and obtaining thus the sound dispersion relation of the holographic plasma. Our results are consistent with the earlier computations of \cite{Baier:2007ix}  and \cite{Diles:2019uft}, who obtained the non-linear sound dispersion for $\mathcal{N} =4$ SYM ($d=4$) and ABJM plasma ($d=3)$, respectively, and generalize them to arbitrary dimensions.

The outline of the paper is as follows. We begin with a quick summary of the grSK geometry which forms the arena of the computation in \cref{sec:background}. In \cref{sec:dynamics} we argue that the dynamics of Einstein's equations can be distilled into one designer field. We primarily present the final result of the analysis and give a bird's-eye view of the arguments leading up to capturing the dynamics into a single field. We describe in \cref{sec:ZSK}  the solutions of the designer field and its on-shell action on the grSK geometry. We then use this data in \cref{sec:results} to write down the open effective action for the  energy flux operator, completing thereby the task initiated in \cite{Ghosh:2020lel} for the neutral plasma. 

To avoid cluttering the text with technical details, we have tried to summarize the essential points as much as possible. Readers interested in understanding the steps leading to our statements in detail are invited to consult the appendices. \cref{sec:dynamicsderive} gives a detailed argument of how to assemble the gauge invariant data and use the dynamics implied by Einstein's equations to deduce the field $\MZ$ for spatially inhomogeneous modes. While the final result is not original, with a close relative of $\MZ$ having already been motivated in \cite{Kodama:2003jz}, we have tried to make more transparent the origins of the designer field.  On the other hand, \cref{sec:actionderive} contains a careful examination of the Einstein-Hilbert dynamics at the level of the off-shell action, which has hitherto not appeared in the literature. For completeness, we give two presentations, one in terms of metric functions after suitable gauge fixing to make connection with the dynamical equations, and another in terms of the designer field $\MZ$. The latter is important for our analysis, since we wish to establish the boundary conditions for $\MZ$ -- we prove that it satisfies Neumann boundary conditions asymptotically  for  computing the  generating function of correlations. In \cref{sec:bdyobs} we give the expressions for the boundary observables which we use in \cref{sec:results}. Finally, \cref{sec:gradexpfns} compiles details of the solution we obtain in gradient expansion. In particular, we give the near-boundary asymptotics of the functions which enter the  computation of the boundary observables.

For completeness, we also include \cref{sec:zeromodes} where we characterize the dynamics of the spatially homogeneous modes and relate them to the large diffeomorphisms of the background geometry. We demonstrate that there are additional zero modes in the problem, which do not  smoothly connect to the solution space parameterized by $\MZ$.

\section{Background}
\label{sec:background}

We are interested in analyzing the real-time dynamics of energy current and the associated propagation of sound in a neutral plasma. Holographically, a neutral conformal plasma is dual to the \SAdS{d+1} geometry, which is a solution of the Einstein-Hilbert action (with a negative cosmological constant). In ingoing Eddington-Finkelstein coordinates the background metric reads
\begin{equation}\label{eq:efads}
ds_{(0)}^2 = g_{AB}^{(0)} \,dx^A\, dx^B = -r^2\, f(br) \, dv^2 + 2\, dv\, dr + r^2\, d\vb{x}^2 \,.
\end{equation}	

\begin{figure}[h!]
\begin{center}
\begin{tikzpicture}[scale=0.6]
\draw[thick,color=rust,fill=rust] (-5,0) circle (0.45ex);
\draw[thick,color=black,fill=black] (5,1) circle (0.45ex);
\draw[thick,color=black,fill=black] (5,-1) circle (0.45ex);
\draw[very thick,snake it, color=orange] (-5,0) node [below] {$\scriptstyle{r_h}$} -- (5,0) node [right] {$\scriptstyle{r_c}$};
\draw[thick,color=black, ->-] (5,1)  node [right] {$\scriptstyle{r_c+i\varepsilon}$} -- (0,1) node [above] {$\scriptstyle{\Re(\ctor) =0}$} -- (-4,1);
\draw[thick,color=black,->-] (-4,-1) -- (0,-1) node [below] {$\scriptstyle{\Re(\ctor) =1}$} -- (5,-1) node [right] {$\scriptstyle{r_c-i\varepsilon}$};
\draw[thick,color=black,->-] (-4,1) arc (45:315:1.414);
\draw[thin, color=black,  ->] (9,-0.5) -- (9,0.5) node [above] {$\scriptstyle{\Im(r)}$};
\draw[thin, color=black,  ->] (9,-0.5) -- (10,-0.5) node [right] {$\scriptstyle{\Re(r)}$};  
\end{tikzpicture}
\caption{ The complex $r$ plane with the locations of the two boundaries and the horizon marked. The grSK contour is a codimension-1 surface in this plane (drawn at fixed $v$). As indicated the direction of the contour is counter-clockwise and it encircles the branch point at the horizon.}
\label{fig:mockt}
\end{center}
\end{figure}
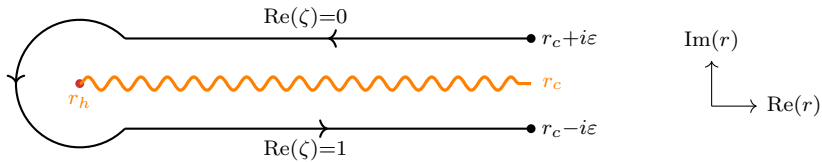

For analyzing  real-time dynamics in the $d$ dimensional field theory we want to use the grSK geometry. In the conventions introduced in  \cite{Jana:2020vyx,Ghosh:2020lel} the metric reads 
\begin{equation}\label{eq:sadsct}
ds_{(0)}^2 = -r^2\, f(br) \, dv^2 +  i\, \beta\, r^2 \, f(br)\,  dv\, d\ctor + r^2\, d\vb{x}^2 \,, \qquad f(\br) = 1 - \frac{1}{\br^d} \,.
\end{equation}	
Here $\ctor$ is the \emph{mock tortoise coordinate}, a real-time contour in the bulk complexified radial coordinate \cref{fig:mockt} and is defined  by 
\begin{equation}\label{eq:ctordef}
\frac{dr}{d\ctor} = \frac{i\,\beta}{2} \, r^2\, f(br) \,, \qquad \beta = \frac{4\pi b}{d} \equiv \frac{4\pi}{d \,r_h}\,,
\end{equation}	
subject to the following boundary conditions at the cut-off surface $r=r_c$
\begin{equation}\label{eq:ctorbc}
\ctor(r_c+i\,\varepsilon) = 0 \,, \qquad \ctor(r_c-i\,\varepsilon) = 1\,.
\end{equation}	
For further details we refer the reader to the aforementioned papers.\footnote{
	As in those references, uppercase Latin indices are used for the bulk AdS spacetime, with lowercase Greek indices reserved for the timelike boundary. Spatial directions along the boundary are further indexed by lowercase mid-alphabet Latin indices. Furthermore, $g_{AB}$ is the bulk metric, $\gamma_{\mu\nu}$ the induced boundary metric, and $n^A$ the unit normal to the boundary. \label{fn:conventions}} 

When we analyze the dynamics of gravitational fluctuations, it will be convenient to work with suitably dressed radial derivative operators that make Schwinger-Keldysh time reversal properties manifest.  Time-reversal is a $\mathbb{Z}_2$ involution acting as $v\mapsto i\beta\ctor-v$ and thus a natural basis  tangent and cotangent spaces are $\{\Dz_+,\partial_v, \partial_i\}$ and $\{\frac{dr}{r^2 f} , dv - \frac{dr}{r^2 f}, dx^i\}$, respectively.  We defined following \cite{Ghosh:2020lel} the operators $\Dz_\pm$
\begin{equation}\label{eq:Dz}
\Dz_\pm = r^2f \, \pdv{}{r} \pm \, \pdv{}{v} \,, \qquad \Dz_\pm = r^2f \, \pdv{}{r} \mp \, i \, \omega \,, 
\end{equation}	
in the time and frequency domain, respectively.

In presenting our results we find it helpful to scale out dimensions using the horizon scale set by $b^{-1}$ and work with dimensionless parameters 
\begin{equation}\label{eq:dimless}
\bwt = b\,\omega \,, \qquad \bqt = b\,k \,, \qquad \br = br \,.
\end{equation}  
%

\section{Dynamics and the designer sound field}
\label{sec:dynamics}

As discussed in \cref{sec:intro}, to understand energy transport and the sound modes in the plasma, it suffices to focus on metric perturbations involving  scalar plane waves and their derivatives. These scalar polarized gravitons will be the only set of modes we analyze here; the discussion of tensor and vector polarizations was previously described in \cite{Ghosh:2020lel,He:2021jna} for neutral and charged plasmas, respectively. For this decomposition we  pick a direction for the spatial momentum $\bk$ and define harmonics on $\mathbb{R}^{d-1,1}$ to be the $SO(d-2)$ harmonics of the corresponding little group. 

Concretely, we consider metric perturbations of the form
\begin{equation}\label{eq:hscalarpert}
\begin{split}
ds_{(1)}^2 
&= 
	\left(h_{AB} \,dx^A dx^B\right)^{_\text{Scal}}  \\
&= 
        \int_k\,  \Bigg\{\left(2\,\HS  \, ds_{(0)}^2 	+ 
        \HH_{vv}\, dv^2+2\,\HH_{vr} \, dv dr+\HH_{rr} \,dr^2\right)\ScS \\
& \qquad \qquad 
    -\left[ 2\,r\, (\HH_{vx} \,dv+\HH_{rx} \,dr)\, \ScS_i\, dx^i- 
 2\,r^2 \,\HHT\, \ScST_{ij}\ dx^i dx^j \right] \Bigg\} .
\end{split}
\end{equation}  
Here $\ScS=e^{i\bk\cdot\bx-i\omega v}$ is the scalar plane wave on $\mathbb{R}^{d-1,1}$ and $\ScS_i$ and $\ScST_{ij}$ are derived harmonics, defined as 
\begin{equation}\label{eq:SiSij}
\ScS_i = \frac{1}{k}\, \partial_i\, \ScS \,, \qquad \ScST_{ij} = \frac{1}{k^2}\left(\partial_i\, \partial_j - \frac{\delta_{ij}}{d-1}\, \partial^2 \right)\, \ScS\,.
\end{equation}	
$\ScST_{ij}$ is traceless but not transverse, $\partial_i\ScST_{ij} =\frac{d-2}{d-1}\, \partial^2\,\partial_j$. It is thus distinguished from the derived harmonic $\ScS_{ij}$ defined in \cite{Ghosh:2020lel}, which is neither transverse nor traceless.  We have also defined a short-hand for the measure on the Fourier domain
\begin{equation}
\begin{split}
\int_k \equiv \int\frac{d\omega}{2\pi}\int\frac{d^{d-1}\bk}{(2\pi)^{d-1}} 
\end{split}
\end{equation}
to keep expressions compact. 

There are seven metric components in the perturbation above. A-priori we expect that these seven functions obey seven coupled radial ODEs arising from the linearized Einstein equations. Owing to  diffeomorphism invariance not all of these dynamical equations are independent. One must first identify the pure gauge modes from the physical perturbations and focus on their dynamics. This problem has been analyzed in detail in \cite{Kodama:2003jz}, whose discussion we can adapt for our purposes.  Following their analysis we will refer to the two-dimensional $\{v,r\}$ spacetime as the \emph{orbit space}. The physical sound mode ends up being described as a non-Markovian scalar field in this orbit space.

We now explain how to efficiently distill the dynamics into a single gauge invariant degree of freedom inspired by the previous analysis of gauge dynamics in 
\cite{Ghosh:2020lel,He:2021jna}.\footnote{
	We will exclusively describe the dynamics of modes carrying non-vanishing spatial momentum in the text, relegating the analysis of spatially homogeneous modes to \cref{sec:zeromodes}. } 
The key point is that there are seven metric functions while the classical phase space, parameterized by gauge invariant data, has only one physical degree of freedom in the scalar sector of gravitational perturbations.  To arrive at this result we make use of the following observations which are   elaborated upon in \cref{sec:dynamicsderive}:
\begin{itemize}[wide,left=0pt]
\item There are seven diffeomorphism invariant combinations of the metric perturbations $\HH_{AB}$ which can be organized as an orbit space traceless tensor, an orbit space vector, and two orbit space scalars. One first deduces that the orbit space vector ($\HH_{vx}$ and $\HH_{rx}$) and one scalar ($\HHT$) can be gauge fixed to vanish, leaving four functions, which are essentially $\HS, \HH_{vv}, \HH_{vr}, \HH_{rr}$. 
\item Time-reversal involution is an orbit space diffeomorphism: $\HS, \HH_{vv}, \HH_{vr}+\frac{1}{2}\,r^2f\, \HH_{rr}$ are time-reversal even, while $\HH_{vv} + r^2f\, \HH_{vr}$ is time-reversal odd. This information is useful to constrain the structure of the equations of motion and the action.
\item Among the equations of motion we find an algebraic constraint, which allows elimination of $\HH_{rr}$ for non-zero spatial momentum. Two other equations are the momentum constraint equation (which is the  boundary energy-momentum tensor  conservation) and a first order radial equation. These two equations can be used to solve for $\HH_{vv}$ and $\HH_{vr}$ in terms of a function $\MW$, which is related to  the boundary stress tensor component $(\TcftU)\indices{_v^i}$.
\end{itemize}

One therefore finds that the metric can be parameterized by two functions:  an overall Weyl rescaling of the background geometry and a function which encodes the physical data of the boundary stress tensor. We will label these fields as $\PHW$ and $\MW$, respectively, and judiciously define them with suitable factors of $r$ to simplify the dynamical equations;
\begin{equation}\label{eq:PHWMWdef}
\HS \equiv \frac{1}{2r^{d-2}}\, \PHW \,, \qquad \HH_{vv} + r^2f\, \HH_{vr} \equiv  -\frac{i\omega}{r^{d-3}}\, \MW\,.
\end{equation}	

The metric parameterized by these fields, subject to the gauge fixing where all the metric components involving derived scalar spherical harmonics are set to zero, is said to be in the \emph{Debye gauge}. Our scalar perturbations are then captured by\footnote{
	We have written the metric directly in position space as the fields $\PHW$ and $\MW$ are simply Fourier transformed with the scalar harmonic $\ScS$.} 
\begin{equation}\label{eq:MZscalarpert}
\begin{split}
ds_{(1)}^2 
&=
   \frac{\PHW}{r^{d-2}} \,ds_{(0)}^2 +2\, \frac{ f}{r^{d-5}}\,\dv{\MW}{r} \,\frac{dr}{r^2f} \left(\frac{dr}{r^2f}-dv\right) 
     +\frac{\Dz_+\MW}{r^{d-3} } dv^2   	-  (d-1)\, \frac{f\,\PHW}{r^{d-4}}\, \left(\frac{dr}{r^2f} \right)^2.
\end{split}
\end{equation}
We have written the metric in the basis of the cotangent space that is  adapted to the time-reversal involution of the background.

The last step involves analyzing the remaining dynamical equations of motion and discerning that they can be solved if one further introduces a field $\MZ$ to parameterize $\PHW$ and $\MW$ as 
\begin{equation}\label{eq:XiHSZ}
\MW=\frac{r}{\Lk}\, \left(\Dz_+ -\frac{1}{2}\, r^2\, f'\right) \MZ \,, \qquad 
\PHW = \frac{1}{\Lk}\left(r\, \Dz_++ \frac{k^2}{d-1} \right)\, \MZ\,.
\end{equation}  
The function $\Lambda_k$ is curious. It is a non-trivial function of spatial momentum (indicated by the subscript). It will turn out to be a designer dilaton for the field $\MZ$ and is given by 
\begin{equation}\label{eq:Lambak}
\Lk(r) \equiv k^2 + \frac{d-1}{2}\, r^3\, f' = k^2 + \frac{d(d-1)}{2\,b^d\, r^{d-2}}\,.
\end{equation}  
Because of the momentum dependence, the field $\MZ$ should be seen as residing in the orbit space and not in the entire background geometry.\footnote{
	The origin of $\Lk$ is analogous to the Ohmic function $h(r)$ which appears in the analysis of vector perturbations of the \RNAdS{d+1} black hole  \cite{He:2021jna}.  The modulation there was due to the background charge whereas here it directly relates to the momentum carried by the perturbation. }
Note that there  is a linear relation between $\MW$ and $\PHW$ from \eqref{eq:XiHSZ}
\begin{equation}\label{eq:linearXSZmain}
\MW = \PHW- \frac{1}{(d-1)} \, \MZ \,.
\end{equation}  

Our main claim is the following: linearized Einstein equations for scalar perturbations of the \SAdS{d+1} geometry are satisfied provided the field  $\MZ$ obeys a second order linear differential equation
\begin{equation}\label{eq:ZMeqn}
\begin{split}
r^{d-3}\,\Lk(r)^2 \, \Dz_+
        \left( \frac{1}{r^{d-3}\,\Lk(r)^2} \, \Dz_+  \MZ \right)    
            + \left(\omega^2 - k^2 f  \left[1-\frac{d\,(d-2)}{b^d\, r^{d-2}\, \Lk(r)}\right]\right) \MZ  &=0 \,.\\
\end{split}
\end{equation}  
This equation is the `master field equation' obtained in \cite{Kodama:2003jz}, which we have obtained in a somewhat different parameterization.\footnote{ 
	The authors of \cite{Kodama:2003jz} prefer to write the equations in Schwarzschild coordinates and express it as  a Schr\"odinger equation in orbit space for a field $\mathcal{S}_\text{KI}$. To bring our equation to their form, one first transforms from our ingoing coordinates to Schwarzschild coordinates and implements a field redefinition: $\MZ = r^\frac{d-5}{2}\, \Lambda_k\, \mathcal{S}_\text{KI}$. }  

We have written the effective dynamics in a manifest time-reversal invariant form. This implies that for analyzing  the solution on the grSK geometry  \eqref{eq:sadsct} we only need to solve \eqref{eq:ZMeqn} in a single copy background \eqref{eq:efads} with ingoing boundary conditions. One can then use the time-reversal properties to construct the outgoing solution and thence the full grSK solution with boundary conditions specified on the two asymptotic boundaries at $r\to \infty \pm i0$ as described in \cite{Jana:2020vyx}. Arriving here was a key step in the analysis: had one worked with other gauge fixing methods traditionally employed in AdS/CFT such as radial gauge, while one would have been able to solve the ingoing problem, the corresponding outgoing solution would be singular. The issue is similar to the problems encountered with gauge fields and momentum diffusion analyzed in \cite{Ghosh:2020lel,He:2021jna}.

Not only does one end up with a single second order differential equation to solve in the scalar sector of gravity, but there is also a remarkable simplification of the Einstein-Hilbert action. Plugging in the parameterization, we find after a series of algebraic simplifications the dynamics of $\MZ$ to be governed by a simple dynamical system\footnote{
	The effective central charge is defined as $c_\text{eff} = \frac{\lads^{d-1}}{16\pi G_N}$.} 
\begin{equation}\label{eq:Zaction}
\begin{split}
\frac{1}{c_\text{eff}} \, S[\MZ] 
&= 
    - \int_k\, \frac{d}{8}\,\nu_s \, k^4\int dr\,  \sqrt{-g}\; \, 
    e^{\chi_s}\,     \left[
        \frac{1}{r^2f} \, \Dz_+ \MZ^\dag\, \Dz_+ \MZ + V_{\MZ}(r) \MZ^\dag\, \MZ \, \right]+ S_{_\text{bdy}}[\MZ] \,, \\
 V_{\MZ}(r) 
 &=
     -\frac{\omega^2}{r^2f} + \frac{k^2}{r^2} \left(1-\frac{(d-2)\, r^3\,f'}{\Lk}\right) .
\end{split}
\end{equation}
The dilaton $\chi_s$ which modulates the gravitational interaction is 
\begin{equation}\label{eq:dilchi}
e^{\chi_s} \equiv  \frac{1}{r^{2(d-2)} \, \Lk^2}  \,,
\end{equation}  
while the normalization is fixed by a parameter $\nu_s$, which in turn is 
\begin{equation}\label{eq:nusdef}
\nu_s \equiv \frac{2\, (d-2)}{d\,(d-1)}\,.
\end{equation}  
This is a remarkable simplification given the complexities inherent in the scalar perturbation; our field redefinitions in \eqref{eq:XiHSZ} imply that the metric in Debye gauge is a function of 
$\{\MZ, \Dz_+\MZ,\Dz_+^2\MZ\}$, so the truncation of two derivative dynamics is indeed a welcome surprise.  As one might expect, much of the complication is hidden in the boundary term $S_{_\text{bdy}}[\MZ]$ in \eqref{eq:Zaction}. It captures all the contributions from the Gibbons-Hawking term, additional boundary terms encountered while writing the action in terms of $\MZ$, and counterterms. In \cref{sec:actionderive}, we explain how to obtain the action and the variational principle for the field $\MZ$ starting from the Einstein-Hilbert dynamics.

This action \eqref{eq:Zaction} is of the general form of a non-Markovian designer scalar introduced in \cite{Ghosh:2020lel}, albeit with some additional novelties. Firstly, the dilaton $\chi_s$ modulates the gravitational interaction non-trivially as a function of spatial momentum $\bk$. The Markovianity index,\footnote{
	The Markovianity index was defined in \cite{Ghosh:2020lel} with a minimally coupled massless scalar having index $d-1$.} 
which depends on the asymptotic behaviour of the dilaton, is determined to be 
\begin{equation}\label{eq:dilasym}
\begin{aligned}
    \lim_{r\to \infty} \, e^{\chi_s} 
    =
    \frac{1}{k^4\, r^{2(d-2)}}  \;\; \Longrightarrow \;\; \ann_{k\geq k_{_\text{IR}}} = 3-d \,.
\end{aligned}
\end{equation}  
We have made explicit our spatial momentum cut-off for clarity. Thus, we see that for non-zero momentum the field $\MZ$ is a  non-Markovian designer  scalar with index $\ann = 3-d$.  Note that while the field $\MZ$ has a non-trivial potential which we will need to take into account, the potential does not, as in the analysis of \cite{He:2021jna}, modify the Markovianity properties. The latter is purely governed by the radial kinetic operator. 

While our distillation of the dynamics into the designer field $\MZ$ is only valid for spatially inhomogeneous modes, for purposes of finding the solution, we can examine the behaviour of the dilaton and the wave equation \eqref{eq:ZMeqn}  at zero spatial momentum. On this locus Markovianity index changes to $\ann =d-1$, which suggests that the zero mode sector comprises of short-lived modes (sound of course doesn't propagate without momentum). The full dynamics of $k=0$ modes is a bit more involved, though a part of the solution space is indeed captured by $\MZ$ with an effective Markovian dynamics as \eqref{eq:ZMeqn} suggests. 

From a pragmatic standpoint, this switch between Markovian and non-Markovian behaviour in \cref{sec:MZgradexp} will prove very useful when we solve \eqref{eq:ZMeqn} in a gradient expansion. Thanks to this observation, we will be able to write down the solution directly in terms of known solutions \cite{Ghosh:2020lel} of the Markovian wave equation with $\ann = d-1$ at low orders in the gradient expansion. As mentioned earlier, an analysis of zero modes can be found in \cref{sec:zeromodes}, where we describe how $\MZ(r,\omega,\vb{0})$ connects onto the set of large diffeomorphisms.

\section{Sounding out the grSK geometry}
\label{sec:ZSK}

For the rest of the discussion we will focus on the dynamics of the gravitational system encoded in $\MZ$ and recover the physics of sound propagation with attenuation.  We begin with the solution  the equation \eqref{eq:ZMeqn} with ingoing boundary conditions. Subsequently, using the time-reversal involution of the grSK geometry we will construct the full linearized solution parameterized by the expectation value of a boundary stress tensor component at the L and R boundaries.

\subsection{Solving the designer equation}
\label{sec:Zsolving}

To better understand the nature of the designer scalar $\MZ(r,\omega,\bk)$, we first analyze the asymptotics of the wave equation \eqref{eq:ZMeqn}.  As described above, while the 
derivation of the $\MZ$ equation is valid for $\bk\neq0$, we can consider $k \geq 0$, since part of the homogeneous mode solutions merge into $\MZ$. 

Let us first focus on zero frequency solutions, but examine both the zero momentum and non-zero momentum behaviour separately, owing to \eqref{eq:dilasym}.  At zero spatial momentum, we have the asymptotic behaviour determined by
\begin{equation}\label{eq:Z0sol}
\begin{split}
\MZ(r,0,\vb{0}) = \tilde{c}_a + \tilde{c}_m \,  \frac{1}{r^d} \,,
\end{split}
\end{equation}
where the constant mode $c_a$ is analytic, but the monodromy mode $c_m$ typically has a logarithmic branch cut emanating from the horizon. On the other hand at  non-zero spatial momentum one finds the expected non-Markovian behaviour:
\begin{equation}\label{eq:Zksol}
\MZ(r,0,\bk) = c_a + c_m\, r^{d-4} \,.
\end{equation}  
Per se, this is not unexpected given the general analysis of Markovian and non-Markovian degrees of freedom as described in \cite{Ghosh:2020lel}. This is an explicit realization of the observation above that naively there is a change in character as a function of momentum. 

Let us pause here to comment on two special cases: $d=3$ and $d=4$. The situation in $d=4$ is marginal;  the monodromy mode behaves logarithmically. When we present solutions we will be careful to normalize them as $\frac{r^{d-4}}{d-4}$ so the limit $d\to 4$ can be easily taken by replacing this function by $\log r$. On the other hand, in $d=3$, it appears from the fall-offs  that the field $\MZ$ is Markovian since the index vanishes. This is misleading since, as noted in \cref{sec:intro}, the field $\MZ$ is actually non-Markovian for all $d\geq 3$. This is not manifest from just analyzing fall-offs, for we also need to take into consideration the variational principle and the boundary conditions for $\MZ$, which we  discuss in \cref{sec:ZvarP}. We demonstrate there that $\MZ$ has Neumann boundary conditions imposed on it for purposes of computing the generating function of correlators  and hence it is non-Markovian for all $d \geq 3$.

Returning to the dynamical problem we can solve \eqref{eq:ZMeqn} by disentangling the physical non-Markovian behaviour from the auxiliary Markovian one. We first rewrite the differential equation for $\MZ$ in terms of the designer scalar wave operator \cite{Ghosh:2020lel,He:2021jna}:
\begin{equation}\label{eq:Dmark}
\Dann\ann =     r^{-\ann}\, \Dz_+ \left(r^\ann \, \Dz_+ \right)     + \left(\omega^2 - k^2 f \right) \,.
\end{equation}  
When expanded in powers of momentum we find a remarkably simple form for \eqref{eq:ZMeqn}: 
\begin{equation}\label{eq:ZMeqnMark}
\begin{split}
\Dann{d-1}\, \MZ - k^2\nu_s\, f  \left(1-\frac{\nu_s\,k^2}{(d-2)\, r^2\, (1-f)}  + \order{k^4}\right)\left(\frac{2}{r (1-f)} \, \Dz_+ -d\right)\MZ  = 0\,,
 \end{split}
\end{equation}
where the parameter $\nu_s$ was defined earlier in \eqref{eq:nusdef}.

Since the operator $\Dann{d-1}$ annihilates a Markovian scalar of index $d-1$, which we write as the field $\sen{d-1}$ in the notation of \cite{Ghosh:2020lel}, one can subtract out this piece from $\MZ$ and write a general solution as 
\begin{equation}\label{eq:Zsplit}
\MZ(r,\omega,\bk) = \sen{d-1}(r,\omega,\bk) - \nu_s\, \bqt^2 \, \Zmt(r,\omega,\bk)\,.
\end{equation}  

We are almost done: $\Zmt$ satisfies an inhomogeneous linear differential equation of the form $\Dann{d-1} \Zmt = \text{source}$. The source is simply determined in terms of a Markovian field  $\sen{d-1}$ in the background \SAdS{d+1} geometry, which has already been solved for in  \cite{Ghosh:2020lel,He:2021jna}. That analysis has already inverted the operator $\Dann{d-1}$, so one simply needs to add in a particular solution to determine the full behaviour. 

Not only do we have access to the solution space, but the nature of the change in asymptotics is now transparent.   At zero momentum the second term in \eqref{eq:ZMeqnMark} vanishes and  simply has $\Dann{d-1}$ annihilating $\MZ(r,\omega,\mathbf{0})$. At non-zero momentum however, the source provided by the Markovian solution is scaled up because of the  $\frac{2}{r(1-f)}\, \Dz_+$ term. This accounts for the change in behaviour. We will parameterize the solutions in a particularly convenient manner so that the  non-Markovian behaviour in fact only sets in at quartic order.\footnote{ 
	In fact, this behaviour is quite similar conceptually to that observed for transverse vector polarizations of photons and gravitons in \cite{He:2021jna}. There it was found that one had a Markovian mode  which mixed with a non-Markovian mode and picked up additional divergent terms. In that discussion there were two independent degrees of freedom, which were decoupled to isolate the long-lived and short-lived modes. Here there is only a single mode whose character changes owing to the underlying gauge invariance.} 
This will turn out to be another artifact of the field $\MZ$; translating back to the metric function $\PHW$, or even $\MW$, we will see a change at quadratic order in momenta.

We now present the ingoing solution for the field $\MZ$ and use it to determine the full solution on the grSK geometry. Our analysis will be accurate to quartic order in the boundary gradient expansion.

\subsection{Ingoing solution in gradient expansion}
\label{sec:MZgradexp}

Working with the dimensionless variables \eqref{eq:dimless}, we obtain the ingoing inverse Green's function for the field $\MZ$, normalized to satisfy  $\lim_{\br\to \infty} \Gin{\MZ}(\br,\omega,\mathbf{0})  = 1$:  
\begin{equation}\label{eq:Zsolution}
\begin{split}
\Gin{\MZ}(\br,\omega,\bk) 
&= 
     e^{-i \,\bwt \, F(\br)}\bigg\{
        1- \bwt^2\, H_\omega(\br)-\kaps^2  \, H_k(\br)  + i  \, \bwt\,  \kaps^2 \, I_k(\br)+i  \, \bwt^3 I_\omega(\br)\\
&
    \qquad \qquad +\; 
     \kaps^4 \, J_k(\br)+ \bwt^2 \, \kaps^2 \, J_{\omega k}(\br) + \bwt^4 \, J_\omega (\br) \\
& \qquad\qquad
    -\frac{\bqt^2}{d(d-1)}\left[4\,(d-2)^2\bwt^2\,  J_k(\br)+2\,\frac{\KS(\omega,\bk)}{d-2} \, V_k(\br) \right]\  +\cdots \bigg\} \,.
\end{split}
\end{equation}
We have written the solution using the notation employed in \cite{Ghosh:2020lel} since the functions appearing below are essentially those already encountered there.\footnote{
	In \cite{Ghosh:2020lel} we had only obtained the solution to cubic order in gradients. Solutions accurate to the quartic order were obtained in \cite{He:2021jna} both for the neutral and charged black holes. In the latter reference the gradient expansion used a slightly different grouping of terms in the solution; we describe this form for completeness in \cref{sec:exppargrad}. } 

We have introduced two new functions of frequencies and momenta in parameterizing \eqref{eq:Zsolution}. The first, $\KS(\omega,\bk)$, which we will later confirm to be the sound dispersion function, is defined as
\begin{equation}\label{eq:KinS}
\begin{split}
\KS (\omega,\bk)
&\equiv 
    -\bwt^2+\frac{\bqt^2}{d-1} +    \nu_s\, \bqt^2\, \Gatt (\omega,\bk) \,, \\
\Gatt (\omega,\bk)
&=
    -i\, \bwt - \bwt^2 \left[(d-2)\, H_k(1)-\frac{1}{d-2}\right]+\frac{d-3}{(d-1)\,(d-2)}\, \bqt^2+\cdots \,.
\end{split}
\end{equation}
Up to quadratic order $\KS$ captures the propagation of sound while $\Gatt$ encodes its attenuation. The second parameter, $\kaps$, may be viewed as a `deformed momentum' parameter arising from the spatial modulation of the dilaton and is
\begin{equation}\label{eq:kappas}
\kaps^2 \equiv \bqt^2 \left(-\frac{d-3}{d-1} + 2\, \nu_s\, \Gatt\right) .
\end{equation}  
We have judiciously combined terms from the solution for $\sen{d-1}$ and $\Zmt$ to write the result in this compact form (introducing the  parameter $\kaps$ in the process).\footnote{
	One way to observe that $\kaps^2 $ starts off as $-\frac{d-3}{d-1}\, \bqt^2$ is to note that setting $\Lk = \frac{d-1}{2}\, r^3f'$ in \eqref{eq:ZMeqn} reduces it to
	\[
		\frac{1}{r^{d-1}}\, \Dz_+\left(r^{d-1}\, \Dz_+ \MZ\right) + \left(\omega^2 + \frac{d-3}{d-1}\,k^2 \right)\, \MZ =0\,,
	\]
		which has the form of a Markovian wave operator with analytically continued momenta.
		\label{fn:Zkapsversion}
}

There is one important subtlety which the reader should be aware of in the way we have presented the solution for $\MZ$. The solution to \eqref{eq:ZMeqn} up to the quartic order in gradients does not determine $\Gatt$ owing to the explicit factor of $\bqt^2$  multiplying $V_k(\br)$ in \eqref{eq:Zsolution}. This would be unfortunate since $\Gatt$ will turn out to be the sound attenuation function. However, if we compute the functions $\PHW$ and $\MW$ from $\MZ$ (or any other metric function) then this overall factor of $\bqt^2$ disappears and we obtain $\Gatt$ accurate to quadratic order by examining the coefficient of the non-normalizable mode in the solution. A physical way to say this is to note that the  boundary source is the conjugate momentum with a factor of $k^2$ stripped off, see \eqref{eq:JZdef}.  This behaviour is manifest in the ingoing Green's function of the field $\PHW$ which is given in \eqref{eq:PHWasym} (the information in $\MW$ is similar, but not independent owing to the linear relation \eqref{eq:linearXSZmain}). To obtain $\Gatt$ directly from $\MZ$  as written above, we would have to compute the solution accurately to sextic order, another peculiarity of the fact that it is related to radial derivatives of the metric functions.

In fact, we would like to conjecture that one can actually get $\Gatt$ to quartic order in gradients. From the leading divergent mode in $\PHW$ (or $\MW$), which scales as $r^{d-2}$, we get an expression for $\KS$ to quartic order. However, we can also examine the constant mode, scaling as $r^{0}$, near the boundary. It has a coefficient which is a non-trivial function of $\omega$ and $\bk$. By judiciously parameterizing the solution for $\PHW$, one finds that this function starts off as $\Gatt$ and gets corrected at  cubic and quartic corrections (the same quantity appears as the  constant mode of $\Gin{\MZ}$). If one uses a parameterization of the solution in terms of $\Gatt$, simply noting that it starts  at linear order, then the on-shell action is finite up to sextic order. Based on this observation and the nature of the explicit solution, we predict an expression for $\Gatt$ up to the fourth order in boundary gradients, which we record for completeness in \eqref{eq:Gatt4} (where for clarity it is denoted as $\Ost$).  

All of the functions that appear above, except  for $V_k(\br)$, are defined using the ingoing solution for massless Klein-Gordon scalar $\sen{d-1}$ in the  \SAdS{d+1} background. More precisely,
\begin{equation}\label{eq:minscalsol}
\begin{split}
\Gin{d-1}(\br,\omega,\bk)
&\equiv  
    e^{-i\, \bwt\,  F(\br)}  \, \bigg\{
        1-\, \bwt^2 \,  H_\omega(\br)- \bqt^2\, H_k(\br) + i  \, \bwt \, \bqt^2 \, I_k(\br)+i \, \bwt^3\,  I_\omega(\br) \\
&\qquad \qquad \qquad 
        +\bqt^4 \, J_k(\br)+\bwt^2 \, \bqt^2\,  J_{\omega k}(\br) +\bwt^4 \, J_\omega (\br)+\cdots\bigg\}  
\end{split}
\end{equation}
solves  $\Dann{d-1}  \sen{d-1} =0 $ with ingoing boundary conditions,  
$\lim_{r\to\infty}\, \sen{d-1} =1$, and thus is the ingoing bulk to boundary Green function for the massless scalar described in \cite{Ghosh:2020lel}. The asymptotic boundary condition along with the regularity at the horizon (ingoing boundary condition) uniquely fixes the eight functions $\{F,H_\omega,H_k,I_\omega,I_k,J_\omega,J_{\omega k},J_k\}$. The asymptotic boundary condition implies that they all vanish as $r\to\infty$.  
The conversion from the massless scalar to the field $\MZ$ involves replacing factors $\bqt^2$ with $\kaps^2$ in various places up to the quartic order: this in particular means that $\Gin{\MZ}$ remains Markovian up to that order (for reasons explained in \cref{fn:Zkapsversion}).

The key function that controls non-Markovianity is $V_k(\xi)$, which is defined by the integral
\begin{equation}\label{eq:Vksol}
\begin{split}
V_k(\xi)\equiv - \frac{\xi^{d-4}}{d-4}+\int_\xi^\infty\frac{y^{d-4}-1}{y(y^d-1)}\ .
\end{split}
\end{equation}
For $d>4$ this function grows as $r\to\infty$, unlike the other functions. We have extracted the leading divergence as $r^{d-4}$, which is precisely what one expects for a non-Markovian field of index $\ann =3-d$ as described around  \eqref{eq:Zksol}.

Since we are effectively inverting the Markovian operator $\Dann{d-1}$ to find the functions above, they can all be given formal integral representations, assuming that the sources are regular on the horizon and do not grow too fast at infinity. We can write in general following \cite{Ghosh:2020lel}  
\begin{equation}\label{eq:gensol}
\mathfrak{F}(\br) = \int_\br^\infty \, \frac{dy}{y\,(y^d-1)} \, \int_1^y \frac{dy'}{y'\,(y'^d-1)}\, \widehat{\mathfrak{J}}(y') \,,
\end{equation}  
with $\widehat{\mathfrak{F}}(\br) = \mathfrak{F}(\br)  - \mathfrak{F}(1)$ defined to measure function values relative to that on the horizon.  We tabulate the data for the functions appearing in the gradient expansion originating from the minimally coupled scalar in \cref{tab:gradfns}. 
\begin{table}
\centering
\begin{tabular}{ ||c|c||c|c|| }
\hline\hline
\shadeB{Function} & \shadeB{Source} & \shadeR{Function} & \shadeR{Source} \\
\hline\hline
$F$ & 
    $(d-1)\, \br^{d-1}\,  (\br^d-1)$
    &  
    $I_\omega$ 
        & $\frac{2}{d-2}\, (\br^{d-2}-1)-2\, (d-2)\, \widehat{H}_k(\br)$
 \\ \hline  
$H_\omega$ & 
    $1-\br^{2(d-1)}$
        & 
        $J_{\omega k}$ 
            & $\left(1-\br^{2(d-1)}\right)H_k(\br) -4\,\widehat{H}_k(\br)$ 
\\
 & &  & $+\br^{d-2} \, (\br^d-1)\, H_\omega(\br)$ 
 \\ \hline 
 $H_k$  & 
    $\br^{d-2} (\br^d-1)$ 
        & $J_\omega$ 
            & $\left(1-\br^{2(d-1)}\right) \, H_\omega(\br) -4\,\widehat{H}_\omega(\br) $ 
\\ \hline 
$I_k$ & 
    $-\frac{2}{d-2}\, (\br^{d-2}-1)$  
    & 
        $J_k$ 
            & $\br^{d-2} \, (\br^d-1)\, H_k(\br)$  \\
\hline  \hline 
\end{tabular}
\caption{The sources for the functions parameterizing the ingoing solution for $\MZ$ in (\ref{eq:Zsolution}) which enter their integral representation (\ref{eq:gensol}).}
\label{tab:gradfns}
\end{table}
Further details of the solution, including asymptotic expansions and expressions for the gravitational data from our solution for $\MZ$, are compiled in \cref{sec:gradexpfns}.

\subsection{The grSK solution for the designer field}
\label{sec:grsKZ}

The ingoing Green's function for $\MZ$ suffices for us to determine the full solution on the grSK geometry thanks to the time-reversal invariance of \eqref{eq:ZMeqn}. We want to impose suitable boundary conditions at the two boundaries of the grSK geometry at $r\to \infty \pm i0$. It was argued in \cite{Ghosh:2020lel} that a non-Markovian fields should be quantized with Neumann boundary conditions if we wish to compute their correlation functions. Equivalently, the asymptotic field value does not correspond to the boundary source, but rather gives the dual boundary operator (akin to the alternate quantization of low lying operators). This was cleanly formulated for probe non-Markovian fields and  established to be the case for diffusive modes in the aforementioned reference and \cite{He:2021jna}. 

The question we should first ask is what the Einstein-Hilbert dynamics with its usual Dirichlet boundary conditions for computing correlation functions of dual energy-momentum tensor correlation functions implies for the field $\MZ$. It turns out that while the dynamics of $\MZ$ is pretty simple in the bulk, as evidenced from \eqref{eq:Zaction}, it has a pretty involved set of boundary terms owing to the redefinitions in \eqref{eq:EOSMZ}. The boundary terms are a general quadratic form in $\MZ$, $\Dz_+\MZ$, and $\Dz_+^2\MZ$, implying that generically we need to fix a combination of these three quantities for stationarity of the action. However, one can do better: armed with the asymptotic fall-offs of the field, one learns that the leading set of boundary terms in a near-boundary expansion are simpler, and one indeed finds that $\MZ$ ought to be quantized with Neumann boundary conditions to obtain the dual stress tensor correlators.  The detailed argument analyzing the variational principle for the designer field $\MZ$ is given in \cref{sec:ZvarP}.

While the field $\MZ$ is quantized with Neumann boundary conditions  in order to compute the generating function of boundary correlators, the resulting correlators have a sound pole, reflecting the non-Markovian nature of the field. For this reason, it was proposed in \cite{Ghosh:2020lel} that one should, for purposes of computing a real-time Wilsonian influence functional,  parameterize the solution for $\MZ$ in terms of the normalizable mode, which for a non-Markovian field  corresponds to the conjugate momentum. 

With this understanding we will now parameterize the general solution on the grSK geometry in terms of  the sound modulus, which is the expectation value of the dual boundary operator $\OpZ$ (thinking of $\MZ$ probe field in the fixed \SAdS{d+1} background). We will denote the modulus associated with the auxiliary field $\MZ$ as $\PoZ$ and write
\begin{equation}\label{eq:ZOpDef}
\expval{(\OpZ)_\skL} = \PoZ_\skL \,, \qquad \expval{(\OpZ)_\skR} = \PoZ_\skR \,, 
\end{equation}  
with 
\begin{equation}\label{eq:PoZdef}
\PoZ_{\skL/\skR} = \lim_{r\to \infty \pm i0} \left[ \MZ  + \text{counterterms}\right] .
\end{equation}  

In terms this boundary modulus field $\PoZ$ we can write the full grSK solution in the average-difference basis as
\begin{equation}\label{eq:Zsksol}
\MZ^\text{SK}(\ctor, \omega, \bk) = \Gin{\MZ} \, \PoZ_a +\left[ \left(\nB+\frac{1}{2}\right) \, \Gin{\MZ } -\nB\, e^{\beta\omega(1-\ctor)} \, \Grev{\MZ} \right] \PoZ_d \,,
\end{equation}  
with $\Grev{\MZ}(\ctor,\omega,\bk)= \Gin{\MZ}(\ctor,-\omega,\bk)$ being the time-reversed propagator. It is useful to note that $\MZ$ has mass dimension $d-2$ and hence $\PoZ_{a,d}$ are likewise boundary fields with this dimension.  We will use this information in the next section to present the effective action for the phonon modes in the relativistic plasma.

\section{Effective dynamics of sound and energy transport}
\label{sec:results}

With the grSK solution for the designer sound mode, we can evaluate the on-shell action parameterized by the fields $\PoZ_{a,d}$, which then gives us the open effective field theory of sound propagation in the holographic plasma at the Gaussian order in amplitudes. We first outline the details of this effective action and then turn to the boundary stress tensor, which we will express in terms of $\PoZ$ and background polarization terms involving the sources $\JoZ$.

\subsection{The sound Wilsonian influence functional }
\label{sec:Zwif}

Since the background \SAdS{d+1} geometry has a non-vanishing free energy, we expect to see two contributions to the on-shell action. One is a piece from the background, which based on fluid/gravity intuition should correspond to the ideal fluid free energy. In addition, there will be the true dynamical data corresponding to the Wilsonian influence functional (WIF) of the field $\MZ$. These two contributions are cleanly separated in the shear and tensor sectors of a neutral fluid \cite{Ghosh:2020lel} because, in those sectors, we do not have a propagating mode. 

A relativistic ideal fluid has a propagating phonon mode, so the split in this sound sector is not a-priori manifest. We will therefore not identify the ideal fluid contribution at this stage, but simply separate the on-shell action into the WIF and contact terms. Subsequently, working out the stress tensor will enable us to  understand which pieces should be regarded as part of the ideal fluid contribution. With this preamble, let us write the full action as a sum of two pieces, viz., 
\begin{equation}\label{eq:SoSZsplit}
\begin{split}
S[\MZ] &=
	 S_\text{contact}[\MZ] +  S_\text{WIF}[\MZ] \,. 
 \end{split}
\end{equation}  
We now summarize the final result for the Schwinger-Keldysh effective action obtained by computing the on-shell action for $\MZ$  with the gradient expansion solution given in \cref{sec:MZgradexp}. The reader can find details of the evaluation in  \cref{sec:Zosbdy}. 

Let us start with the contribution to the WIF, which is ascertained to be
\begin{equation}\label{eq:SZwif}
\frac{1}{c_\text{eff}}\, S_\text{WIF}[\MZ] 
= 
	-\int_k\, k^2\, \left(\PoZ_d^\dag\, \Kin{\MZ}\,  \left[\PoZ_a + \left(\nB+\frac{1}{2}\right) \PoZ_d\right] + \text{cc} \right) ,
\end{equation}  
where 
\begin{equation}\label{eq:Zretcorr}
\Kin{\MZ}(\omega,\bk) = \frac{b^{d-2}}{2\,d\,(d-1)^2}\, \KS(\omega,\bk) \,.
\end{equation}  
From this expression we solve for the boundary source of the field $\MZ$ in terms of the moduli field $\PoZ$ and obtain 
\begin{equation}\label{eq:JZsksol}
\begin{split}
\JoZ_a 
&= 
    \Kin{\MZ} \, \PoZ_a + \left(\nB+\frac{1}{2}\right) \left[\Kin{\MZ} -\Krev{\MZ}\right]\PoZ_d \,,
 \\
\JoZ_d &= 
    \Krev{\MZ}\, \PoZ_d \,.
\end{split}
\end{equation}

One can give an alternate expression for the source directly from the conjugate momentum of the field $\MZ$,  after stripping off a factor of $k^2$, viz.,
\begin{equation}\label{eq:JZdef}
\JoZ_{\skL/\skR} = - \lim_{r\to \infty \pm i0} \, \frac{\PiZ}{k^2}   \,.
\end{equation}	
It can be checked that our identification of the source in \eqref{eq:JZdef} agrees with the expectation from the WIF \eqref{eq:JZsksol} (which we expect on general grounds from \cite{Ghosh:2020lel}) and is verified in \cref{sec:actionderive}. Isolating a factor of $k^2$ in the WIF results in stripping off a similar factor from the conjugate momentum \eqref{eq:JZdef}.\footnote
	{In the vector sector, the passage to Debye gauge already factors out a piece proportional to $k^2$,  even in the off-shell action for the designer fields \cite{Ghosh:2020lel,He:2021jna}. This does not happen in the off-shell action \eqref{eq:Zaction} for $\MZ$ due to the momentum dependent dilaton, but is reinstated in the on-shell action. }
The rescaling by a factor of $k^2$  is only allowed since we are focusing on spatially inhomogeneous modes and  are  implicitly working with $k> k_{_\text{IR}}$.

The source for the designer field can be given several equivalent expressions in terms of the metric functions. Using the asymptotics of the solutions obtained in \cref{sec:ZWasym} (see in particular \eqref{eq:PHWasym} and \eqref{eq:PHEasym}) one can show
\begin{equation}\label{eq:JZmultiple}
\JoZ_{\skL/\skR}
	= \frac{1}{4\,(d-1)}\, \lim_{r\to \infty \pm i0} \, \frac{\Dz_+\MW}{r^{d-1}}
	= \frac{d\, \nu_s}{8}\, \lim_{r\to \infty \pm i0} \, \frac{\PHW}{r^{d-2}} \,.
\end{equation}	
These source terms are basically capturing the deformation of the boundary metric order by order in the gradient expansion. Indeed, upon examining the induced metric $\gamma_{\mu\nu}$ on the boundary
\begin{equation}\label{eq:gammaLR}
\begin{split}
(\gamma_{\skL/\skR})_{\mu\nu}
&=
	 \lim_{r\to\infty \pm i0} \left[ \frac{\Dz_+\MW}{r^{d-1} }\, dv^2   +\left(1+ \frac{\PHW}{r^{d-2}}\right) \, \eta_{\mu\nu}\, dx^\mu\, dx^\nu \right]  \\
&=
	-\left(1-\frac{4\,(d-1)\, (d-3)}{d-2} \, \JoZ_{\skL/\skR}\right) dv^2 + \left(1+\frac{4\,(d-1)}{d-2} \, \JoZ_{\skL/\skR}\right) d\vb{x}^2\,,
\end{split}
\end{equation}	
we see that spatial and temporal components of the perturbed boundary metric can be viewed as sources for $\MZ$. Since we have only one physical degree of freedom in the longitudinal sector, we do not have independent metric perturbations, but rather see that the red-shift factor captured by the temporal term is related up to a dimension dependent constant to Weyl rescaling of the background.

With the identification of the sources we can  now present the contact term part of  the action,  which is a functional of these sources. Ignoring the background free energy term for simplicity, we find  the following result at linear and quadratic order in amplitudes:
\begin{equation}\label{eq:ScontactZ}
\frac{1}{ c_\text{eff}} \, S_\text{contact}[\MZ] 
	= \frac{2\,(d-1)^2}{b^d}\, 
		\int_k  \left[
			\JoZ_\skR - \JoZ_\skL  
			+ \frac{(d-1)\, (d-6) }{(d-2) }\, \left(\JoZ^\dag_\skR\, \JoZ_\skR -\JoZ^\dag_\skL\, \JoZ_\skL \right)
	\right] .
\end{equation}	
In \cref{sec:ideal} we will argue that this contact term can be understood as arising from the on-shell action of an ideal fluid propagating on \eqref{eq:gammaLR}. This includes the somewhat counter-intuitive numerical factor in the quadratic term, which vanishes in $d=6$. 

We will explain the individual contribution to the action, in particular the split between ideal and non-ideal parts, after we  discuss the stress tensor. With that understanding we will be able to cleanly identify the ideal fluid contribution. Along the way, we will also argue that the non-dissipative part of the hydrodynamic action, the Class L terms in the terminology of \cite{Haehl:2015pja}, can be extracted from the WIF. For the scalar sector we will for example see the curvature coupling of the fluid at quadratic order in gradients.  These statements will be elaborated in \cref{sec:ideal}.

Let us take stock of some physical implications from the grSK solution for $\MZ$, and the results for the Wilsonian influence functional. We see that the inverse Green's function computed from the WIF is proportional to $\KS(\omega,\bk)$. In other words, the retarded Green's function of $\OpZ$ has a pole at the vanishing locus of $\KS$. This function also appears as the coefficient of the divergent (non-normalizable) mode in $\MZ$. As described in \cite{Ghosh:2020lel}, non-Markovian fields have a completely normalizable solution on a codimension-1 locus in the boundary Fourier domain. For $\MZ$ this is the vanishing locus of $\KS$; it will end up defining the dispersion function for sound. This also  implies that $\Gatt$ defined in \eqref{eq:KinS} is the rate of attenuation of sound speed due to viscosity. We will elaborate on this further below when we discuss the physical sound degree of freedom and compute boundary energy-momentum tensor correlators.

\subsection{The boundary stress tensor}
\label{sec:stensor}

With the Wilsonian influence functional parameterized by the boundary value of the field 
$\MZ$ at hand, we can now turn to the physical boundary observables, which are the scalar polarizations of the boundary energy-momentum tensor density.  The stress tensor has both  a background contact piece and a contribution given in terms of the fluctuating field $\OpZ$. In the scalar sector, the presence of a non-trivial boundary metric \eqref{eq:gammaLR} means that the result we quote for the contact terms depends on the index positions and whether or not we work with stress tensor densities. We will work with  tensor densities, quoting the mixed components for the stress tensor operator, but compute the correlator for the operator with both indices raised.\footnote{
	In the AdS/CFT context, it is natural, as from any effective action, to extract the boundary stress tensor density, since it only requires variation with respect to the boundary metric and no removal of metric determinants (recall that the stress tensor operator is $T^{\mu\nu} = \frac{2}{\sqrt{-\gamma}} \, \fdv{S}{\gamma_{\mu\nu}}$). \label{fn:tensordensity} }
For the sake of notational simplicity, we define
\begin{equation}\label{eq:Tfull}
\begin{split}
	 \widehat{T}^{\mu\nu}   \equiv \frac{1}{c_\text{eff}} \, \TcftU^{\mu\nu} 
\,.            
\end{split}
\end{equation}	
Here $\TcftU^{\mu\nu}$ is the counterterm corrected Brown-York stress tensor density given in \eqref{eq:TBYcft}.

We will take the viewpoint that the standard rules of the extrapolate dictionary in holography (which can be extended to grSK geometry \cite{Jana:2020vyx}) as applied to non-Markovian operators dictates a canonical split between operator (or vev) and source contributions.  As explained in \cref{sec:BYT}, this can be seen by examining the Brown-York stress tensor (corrected by the counterterms) directly in terms of the metric fields $\PHW$ and $\MW$. This tensor must satisfy conservation and be traceless (up to  the conformal anomaly, which we do not access in $d>4$).  This statement follows naturally from the momentum constraint equation in the bulk geometry and underlies the original identification in \cite{Balasubramanian:1999re}. 

With this choice the Schwinger-Keldysh stress tensor operator has the following representation: 
\begin{equation}\label{eq:Tcftorder4}
\begin{split}
\left(\widehat{T}\indices{_v^v}\right)_{\skL/\skR}
&= 
	 -\frac{d-1}{b^d} + \int_k\, \ScS \left[\frac{2\, (d-1)^2}{b^d}\, \JoZ_{\skL/\skR} -\frac{ k^2}{d-1} \, \ (\OpZ)_{\skL/\skR} \right] , \\
\left(\widehat{T}\indices{_v^i}\right)_{\skL/\skR}
&= 
    i\, \int_k\, \frac{ k\, \omega}{d-1} \,  \ScS_i \  (\OpZ)_{\skL/\skR} \,, \\ 
\left(\widehat{T}\indices{_i^j}\right)_{\skL/\skR}
&= 
   \frac{\delta\indices{_i^j}}{b^d} + \int_k \, \ScS  \frac{2\,(d-1)^2}{b^d}\, \delta\indices{_i^j} \, \JoZ_{\skL/\skR} \\
&\qquad 
   +\int_k\left[ \frac{1}{d-1} \left(\omega^2 - \nu_s\, \Gatt \,k^2\right) \delta\indices{_i^j} \, \ScS- \frac{1}{d-2}\, \nu_s\, \Gatt \, k^2\, (\ScST)\indices{_i^j}\right] (\OpZ)_{\skL/\skR}   \,.
\end{split}
\end{equation}  
We recognize the background contribution which says that the unperturbed  planar \SAdS{d+1} black hole is a conformal plasma. The terms linear in $\JoZ$ and $\OpZ$ are the terms we should understand. 

As it is written, the stress tensor is not manifestly traceless. Nor is the conservation Ward identify obvious on the induced boundary geometry \eqref{eq:gammaLR}. The two do hold, and are, in fact, equivalent to the relation between sources and vevs \eqref{eq:JZsksol}, which picks out the sound dispersion locus:\footnote{
	The covariant derivative in the conservation equation is the one appropriate for the stress tensor density, cf., \cref{fn:tensordensity}. 
}
\begin{equation}\label{eq:sounddisp}
\expval{\nabla_\mu \widehat{T}^{\mu\nu}} = 0 =
\expval{\widehat{T}\indices{_\mu^\mu}} 
\;\; \Longrightarrow \;\; 
	 \left(\omega^2-\frac{k^2}{d-1}   - k^2\, \nu_s\,\Gatt \right) \PoZ_{\skL/\skR}  +  \frac{2\, d\, (d-1)^2}{b^d}\,  \JoZ_{\skL/\skR}  =0\,.
\end{equation}	

We should view \eqref{eq:sounddisp} as giving us the on-shell condition for the sound mode which occurs when the solution is purely normalizable, i.e., when the source contribution is set to vanish. This confirms that the operator $\KS$  gives us the dispersion relation for sound in  the plasma and identifies $\Gatt (\omega,\bk)$ as the attenuation function.
 
With this identification we can now give an  alternate presentation of the stress tensor. Let us use the dispersion relation \eqref{eq:sounddisp}  to shift the source and vev contributions in the spatial part of the stress tensor, i.e.,  use the replacement rule 
\begin{equation}\label{eq:onshellrep}
\omega^2 -  \nu_s\, \Gatt\, k^2 \;\; \mapsto \;\;  \frac{k^2}{d-1} -\frac{1}{b^2} \,\KS\,.
\end{equation}	
Re-expressing $\KS\, \PoZ$ in terms of $\JoZ$ we find that the stress tensor in \eqref{eq:Tcftorder4} can be equivalently presented as   
\begin{equation}\label{eq:TcftPefect}
\begin{split}
\expval{\widehat{T}\indices{_v^v}}_{\skL/\skR}
&= 
	-\frac{d-1}{b^d} + \int_k\, \ScS \left[\frac{2\, (d-1)^2}{b^d}\, \JoZ_{\skL/\skR} -\frac{ k^2}{d-1} \, \PoZ_{\skL/\skR} \right] , \\
\expval{\widehat{T}\indices{_v^i}}_{\skL/\skR}
&= 
    i \int_k \frac{ k\, \omega}{d-1} \,  \ScS_i \,  \PoZ_{\skL/\skR} \,, \\ 
\expval{\widehat{T}\indices{_i^j}}_{\skL/\skR}
&= 
	-\frac{1}{d-1} \, \expval{\widehat{T}\indices{_v^v}}_{\skL/\skR} \, \delta\indices{_i^j} - 
	\int_k \frac{k^2}{d-2}\, \nu_s\, \Gatt \;  (\ScST)\indices{_i^j}\, \PoZ_{\skL/\skR} \,.
\end{split}
\end{equation}  

The representation of the CFT stress tensor \eqref{eq:TcftPefect} renders the trace Ward identity manifest. Additionally, it also isolates the sound attenuation contribution captured by $\Gatt$ solely into the spatial part of the stress tensor. In fact, the contribution is governed by the longitudinal trace-free tensor structure $\ScST_{ij}$.  In this presentation, the tracelessness Ward identity is manifest, but conservation now implies the on-shell condition \eqref{eq:sounddisp}.

One could go a step further and replace the $\Gatt$ term in the spatial part of the stress tensor once again using \eqref{eq:onshellrep}, i.e., express it as
\begin{equation}\label{eq:TijOp3}
\begin{split}
\expval{\widehat{T}\indices{_i^j}}_{\skL/\skR}
&= 
	-\frac{1}{d-1} \, \expval{\widehat{T}\indices{_v^v}}_{\skL/\skR} \, \delta\indices{_i^j}  
	+ \frac{1}{d-2}\, \int_k \left(\frac{k^2}{d-1} - \omega^2\right) (\ScST)\indices{_i^j}\,  \PoZ_{\skL/\skR} \\
&\qquad	\qquad 
	- \frac{2\,d\, (d-1)^2}{(d-2)\, b^d}\,\int_k\, (\ScST)\indices{_i^j}\; \JoZ_{\skL/\skR}   \,.
\end{split}
\end{equation}	
In this manner of presentation, both the conservation and tracelessness Ward identities of the stress tensor are identically satisfied. The operator contribution encoded in $\PoZ$  has the right momentum and frequency dependent factors for the conservation to be rendered trivial, while the terms involving $\JoZ$, one can check, respect $\tensor[^\gamma]{\nabla}{_\mu}\, \widehat{T}^{\mu\nu} =0$ by themselves. 

While there appear to be three distinct parameterizations, the form given in 
\eqref{eq:TcftPefect} is the one that separates the ideal fluid contribution from the dissipative part. We will demonstrate this in \cref{sec:ideal} after writing down the stress tensor correlators. 

\subsection{Correlation functions}
\label{sec:corrfns}

From the Wilsonian influence functional \eqref{eq:SZwif} we can read off the correlation functions of the  $\OpZ$ field operator on the boundary. The retarded Green's function is given  by  
\begin{equation}\label{eq:OpZret}
\begin{split}
\expval{\OpZ(-\omega, -\bk)\, \OpZ(\omega,\bk)}^\text{Ret}
&=
 	\frac{1}{i\,c_\text{eff}\, k^2\,\Kin{\MZ}(\omega,\bk)} =-i\, \frac{2\,d\,(d-1)^2}{c_\text{eff}\,b^d}\, \frac{1}{ \bqt^2\,\KS(\omega,\bk)}\,.
\end{split}
\end{equation}
The structure of the WIF respects the KMS condition, as explained in the earlier works, implying that the Keldysh correlator satisfies the fluctuation dissipation condition, viz.,  
\begin{equation}\label{eq:OpZkel}
\begin{split}
\expval{\OpZ(-\omega, -\bk)\, \OpZ(\omega,\bk)}^\text{Kel}
&=
 	-\frac{1}{2\,c_\text{eff}}  \coth\left(\frac{\beta\omega}{2}\right) 
 		\frac{\Im\left[\Kin{\MZ}(\omega,\bk)\right]}{k^2\, \abs{\Kin{\MZ}(\omega,\bk)}^2} \,.
\end{split}
\end{equation}

We can use this information to write down the stress tensor correlators given the explicit expressions for the components in \eqref{eq:Tcftorder4}. From this expression it is clear that the result is given by the two-point functions of $\OpZ$, suitably dressed to account for the derivative operators present (the functions of $\omega,k$ in frequency/momentum domain). 

Before doing so however, we should note that we have the boundary metric determined by a single source function \eqref{eq:gammaLR}, which implies relation between sources of various components (and an absence of source for the spatial-temporal component). We will view the boundary metric as defining the source for the energy density. Thus, we can obtain the energy density two-point function naturally. Once we have this piece of data, we will use flat spacetime Ward identities, as explained in \cite{Policastro:2002tn},  to fix the remaining correlation functions. In particular, we demand that the contact terms in the retarded correlation functions are fixed so that flat spacetime momentum conservation holds -- this implies that correlation functions with at least one temporal index vanish at zero momentum.  To fix the purely spatial components we utilize full energy-momentum tensor  conservation. 

With this understanding, we now quote the result for the physical retarded correlator for the energy-momentum tensor  density operator $T_{_\text{CFT}}^{\mu\nu}$. We parameterize these as 
\begin{equation}\label{eq:TTret}
\begin{split}
\expval{T_{_\text{CFT}}^{\mu\nu}(-\omega, -\bk)\, T_{_\text{CFT}}^{\mu\nu}(\omega,\bk)}^\text{Ret} 
&=
 	\frac{2d}{i\,c_\text{eff}\,b^d}\, \frac{\mathfrak{G}^{\mu\nu,\rho\sigma}(\omega,\bk)}{\KS(\omega,\bk)} +\text{analytic} \,.
\end{split}
\end{equation}
To avoid writing involved expressions we pick a spatial direction for  sound propagation, setting $\bk = k\, \hat{x}$, decomposing $\mathbb{R}^{d-1}$ coordinates into $\{x,x_s\}$ with $s=2,\cdots,d-1$. One can check that with this choice $\ScST_{ij} = \frac{1}{d-1}\,\text{diag}\{-(d-2),1,\cdots, 1\}$. We choose a representation where $\mathfrak{G}^{\mu\nu,\rho\sigma}(\omega,\bk)$ are polynomials in $\omega, k$ and take the form:
\begin{equation}\label{eq:numpolyTT}
\begin{aligned}
\mathfrak{G}^{vv,vv} &=	\bqt^2\,,  \hspace{5.95cm} 
\mathfrak{G}^{vv,vx} =	\bqt\, \bwt\,, \\
\mathfrak{G}^{vv,xx} &=	\bqt^2\left(\frac{1}{d-1} + \nu_s\, \Gatt\right) ,\hspace{3.2cm}
\mathfrak{G}^{vv,ss} =  \bqt^2\left(\frac{1}{d-1}-   \frac{1}{d-2}\, \nu_s\,\Gatt \right) , \\
\mathfrak{G}^{vx,vx} &=	\bqt^2\left(\frac{1}{d-1} + \nu_s\, \Gatt\right) , \hspace{3.2cm}
\mathfrak{G}^{vx,xx} =	\bwt\,\bqt \left(\frac{1}{d-1} + \nu_s\, \Gatt\right) , \\
\mathfrak{G}^{vx,ss} &=	 \bwt\, \bqt \left(\frac{1}{d-1}-   \frac{1}{d-2}\, \nu_s\,\Gatt \right) ,
	\hspace{2cm}
\mathfrak{G}^{xx,xx} =    \bwt^2 \left(\frac{1}{d-1} + \nu_s\, \Gatt\right) , \\
\mathfrak{G}^{xx,ss} &=  \bwt^2\left(\frac{1}{d-1} -   \frac{1}{d-2}\, \nu_s\,\Gatt \right) , \\
\mathfrak{G}^{ss,ss} &=   \bwt^2\left(\frac{1}{d-1} -   \frac{1}{d-2}\, \nu_s\,\Gatt \right) - \frac{d-1}{d-2}\, \nu_s\,\Gatt^* \left(\bwt^2  -    \frac{d-1}{d-2}\, \bqt^2\, \nu_s\,\Gatt\right) .   
\end{aligned}
\end{equation}
Any other representation of $\mathfrak{G}^{\mu\nu,\rho\sigma}$ would differ from the above by  factors of $\KS$. The Keldysh propagator follows naturally from the fluctuation dissipation theorem.	Our choice for non-analytic part of the retarded Green's function recovers the result for the analytically continued Wightman function given in \cite{Policastro:2002tn} for $d=4$.

The physical aspect of the result which is interesting is the fact that the stress tensor correlators have a sound pole at the dispersion locus characterized by the vanishing of  $\KS$. As noted around \eqref{eq:sounddisp} this function gives us the on-shell condition for the phonon mode. At the leading order in gradients it enforces the expected equation of state condition, which fixes the speed of sound in a conformal plasma to be $\frac{1}{\sqrt{d-1}}$. At higher orders the $\Gatt $ pieces serve to attenuate the propagation and predicts its half-life.

 Solving \eqref{eq:sounddisp}, and using the expression for $\Gatt$ given in \eqref{eq:KinS}, we find\footnote{
	We used $H_k(1)  = -\frac{1}{d\,(d-2)}\, \harm{\frac{2-d}{d}}$ where $\harm{x}$ is the Harmonic number function.\label{fn:Hk1harm}
}
\begin{equation}\label{eq:genwk}
\omega = \frac{k}{\sqrt{d-1}} -i\, \frac{d-2}{d\,(d-1)}\, b\, k^2 + \frac{d-2}{2\,d^2\, (d-1)^{\frac{3}{2}}}\, \left[ d+2+ 2\, \harm{\frac{2-d}{d}} \right]\, b^2\,k^3 + \cdots\,.
\end{equation}	
We have indicated only one branch of the solution in the above expression.

As particular cases, note that with $d=4$ we recover the results for the $\mathcal{N} =4$ SYM plasmas,\footnote{ 
	We work with dimensionless frequencies and momenta defined in \eqref{eq:dimless}. This definition differs from normalizations used in earlier references. Our normalization is twice that used in   \cite{Baier:2007ix}, while \cite{Diles:2019uft} uses a normalization set by temperature and not (inverse) horizon size, which in $d=3$ differs by a factor of $\frac{3}{4\pi}$.} 

\begin{equation}\label{eq:N4disp}
\bwt 
=
	\frac{\bqt}{\sqrt{3}} - \frac{i}{6}\, \bqt^2 + \frac{3-2\, \log 2}{24\, \sqrt{3}}\, \bqt^3 + \cdots\,,
\end{equation}	
which was first obtained to cubic order in \cite{Baier:2007ix} (extending the original result of \cite{Policastro:2002tn}). Setting  $d=3$ we obtain the corresponding result  for the ABJM plasma obtained in \cite{Morgan:2009pn,Diles:2019uft}:  
\begin{equation}\label{eq:abjmdisp}
\bwt 
=
	\frac{\bqt}{\sqrt{2}} - \frac{i}{6}\, \bqt^2 + \frac{15+\sqrt{3}\, \pi - 9\, \log 3}{108\, \sqrt{2}}\, \bqt^3+ \cdots \,.
\end{equation}	
As we describe in \cref{sec:ZWasym}, from our solution we can actually extract higher order corrections to the dispersion, and quote the result accurate to quintic order in 
\eqref{eq:DispLoc}.

\subsection{A fluid dynamical perspective}
\label{sec:ideal}

We have now reproduced the physical results expected for sound propagation as encoded in the stress tensor correlators. Let us therefore try to analyze features of the solution from a hydrodynamic perspective and, in particular, attempt to understand the contributions  to the action \eqref{eq:ScontactZ}.  The gravitational calculation gives us the answer tout ensemble, but we can attempt to decompose it using hydrodynamic intuition. 

The physical question of interest is how one delineates the ideal and non-ideal parts of the effective action. Addressing this question will make clear that one should think of the action as comprising of a Schwinger-Keldysh factorized part corresponding to sound propagation, in addition to the physical influence phase, the part that governs sound attenuation, as we indicate in \eqref{eq:SosIdealpar} below.

The non-ideal part, by definition, includes all the gradient contributions in the stress tensor parameterized in hydrodynamic variables, whether or not they lead to dissipation.
We will use the stress tensor  in the form parameterized  in \eqref{eq:TcftPefect}, which judiciously isolates the  non-ideal contributions into $\Gatt$. As has been argued earlier \cite{Haehl:2015pja}, not all higher order transport is dissipative. While dissipative transport leads to entropy production, in general, there exists non-dissipative transport which is adiabatic and leads to no entropy production.  While at leading order $\Gatt(\omega,\bk)$  captures sound attenuation, which originates from the dissipative shear viscosity term,  it also includes contributions from higher order non-dissipative gradient terms.

To keep the discussion transparent, we will first identify the ideal fluid contribution, which by definition only captures zeroth order terms in the gradient expansion of the stress tensor. Having understood this part, we will then attempt to address higher order (spatial) gradient terms, which capture non-dissipative transport.

Consider an ideal  fluid with energy density and pressure related by the conformal equation of state $\epsilon = (d-1)\, p$, viz., the tensor density (nb: $\widetilde{T}$ is dimensionless for simplicity)\footnote{
	We are not keeping track of the normalization factor translating between the horizon size parameter $b$ and the physical temperature $T$, cf., \eqref{eq:ctordef}.
}
\begin{equation}\label{eq:Tideal}
T_\text{ideal}^{\mu\nu} = 
	\sqrt{-\gamma}  \left(\frac{\widetilde{T}}{b}\right)^d \left( \gamma^{\mu\nu}  + d\, u^\mu\, u^\nu\right) .	        
\end{equation}	
We claim that the part of our solution which is insensitive to sound attenuation, i.e., with $\Gatt \to 0$, describes the dynamics of such an ideal fluid on the boundary geometry (on either L or R boundary). Specifically,  we assert 
\begin{equation}
T_\text{ideal}^{\mu\nu} =  \expval{\widehat{T}^{\mu\nu}} \bigg|_{\Gatt \to \;0} \,.
\end{equation}	
We can confirm this by solving the conservation equations arising from \eqref{eq:Tideal} on the induced boundary  geometry \eqref{eq:gammaLR}. 

We can parameterize the temperature and velocity  by a field $\PoZ$, which is a-priori unrelated to the stress tensor expectation value 
\begin{equation}\label{eq:Tuideal}
\begin{split}{}
\widetilde{T}_{\skL/\skR}
&=
	1- \int_k  \left[ \frac{2\,  (d-1)}{d-2}\, \ScS\, \JoZ_{\skL/\skR} - \frac{b^{d-2} \, \bqt^2}{d\, (d-1)^2}\, \ScS\, \PoZ_{\skL/\skR} \right]\,,\\
\left(u_\mu\, dx^\mu\right)_{\skL/\skR}
&=
	\left(-1  + \int_k \, \frac{2\, (d-1)\, (d-3)}{d-2} \, \ScS\, \JoZ_{\skL/\skR}  \right) dv  -\int_k\, \frac{i\, b^{d-2}\, \bqt \,\bwt}{d\,(d-1)}\,\ScS_i\, \PoZ_{\skL/\skR} \,  dx^i\,.
\end{split}
\end{equation}	
Imposing the conservation equation one finds that $\PoZ$ and $\JoZ$ must satisfy the relation \eqref{eq:sounddisp} with $\Gatt =0$.\footnote{
	This parameterization can be motivated by considering a phonon mode for a relativistic plasma in flat spacetime.  All one needs is the statement that the dynamics is captured by conservation of the stress energy tensor. It is not hard to check that for a linearized perturbation about an equilibrium plasma in flat spacetime, 
	\[
	 T = T_0 + \int_k \frac{k^2}{d-1}\, \ScS\, \PoZ \,, \qquad 
	 u_\mu\, dx^\mu =  -dv -  \int_k\, i\,\omega\, k\, \ScS_i \, \PoZ \, dx^i \,,
	 \] 
	 satisfies the conservation law at linear order in amplitudes provided 
	$\left(-\omega^2  + \frac{k^2}{d-1}\right) \PoZ = 0$. The latter equation picks out the sound dispersion locus in the absence of a source. }  With this constraint recovered, we may identify $\PoZ$ as the physical phonon mode, i.e., as the boundary value of the non-Markovian field. We have effectively isolated the dynamical sound mode, which importantly does exist even in the absence of dissipation, on the inhomogeneous dynamical boundary spacetime \eqref{eq:gammaLR}. One can, furthermore, use the on-shell relation to write the temperature more suggestively as 
\begin{equation}\label{eq:Tempvar}
\begin{split}
\widetilde{T}_{\skL/\skR}=
	1 +\int_k \left[\frac{2\,(d-1)\,(d-3)}{d-2}\, \ScS\; \JoZ_{\skL/\skR} + \frac{b^{d-2}\, \bwt^2}{d\,(d-1)}\, \ScS\; \PoZ_{\skL/\skR} \right].
\end{split}
\end{equation}	
In this presentation, we see that the source $\JoZ$ contribution to the temperature is just the red-shift effect for a fluid propagating on \eqref{eq:gammaLR}. The contribution from $\PoZ$ parameterizes the response of the fluid.

Beyond the ideal fluid, the first correction comes from the dissipative shear viscosity term, which  physically leads to the damping of the sound in the medium. This is the leading $i\omega$ contribution to $\Gatt$. To isolate any non-dissipative contributions we must therefore switch off  time-dependence and focus on equilibrium data. In stationary equilibrium, $\Gatt(0,\bk) = \frac{d-3}{(d-1)\,(d-2)}\, b^2\, k^2 + \cdots$. The $\ScST_{ij}$ part of the spatial stress tensor can be identified as the coupling of the fluid to background curvature. For a conformal plasma it takes the form 
\begin{equation}\label{eq:kappaterm}
\TcftU^{\mu\nu}\supset \kappa \left(\tensor[^\gamma]{C}{^{\mu\alpha\nu\beta}} \, u_\alpha\, u_\beta + \sigma_\text{sh}^{<\mu\alpha}\,(\sigma_\text{sh})\indices{^{\nu>}_\alpha} + \omega_\text{vor}^{<\mu\alpha}\,(\omega_\text{vor})\indices{_\alpha^{\nu>}} \right) .
\end{equation}	
Here $\tensor[^\gamma]{C}{^{\mu\alpha\nu\beta}} $ is the Weyl tensor of the boundary geometry \eqref{eq:gammaLR},  $\omega_\text{vor}$ is the fluid vorticity,  $\sigma_\text{sh}$ is the shear tensor of the fluid, and the angle brackets indicate transverse projection.\footnote{
	We have written this term in the second order stress tensor in a form inspired by the classification of hydrodynamic transport introduced in \cite{Haehl:2015pja}. 
}
The value of the transport coefficient $\kappa = 2\,c_\text{eff}\, T^{d-2}$ is known for \SAdS{d+1} black holes \cite{Bhattacharyya:2008mz} (it was initially derived in $d=4$ in \cite{Baier:2007ix}). Using the temperature and velocity profiles identified above, one can  directly check that our  result  captures this contribution to the stress tensor.

Having understood the contributions at the level of the stress tensor we can now explain  how to interpret the contact terms in the action. Furthermore, we can also isolate terms corresponding to the Class L adiabatic action for holographic plasmas conjectured in \cite{Haehl:2015pja}, by reverse engineering a boundary action from transport data. In terms of the Weyl covariant hydrodynamic variables this action reads 
\begin{equation}\label{eq:classL}
S^\mathcal{W} = b^{-d}\,
	\int d^dx \,  \sqrt{-\gamma} \left(  \widetilde{T}^d-  b^2\, \widetilde{T}^{d-2} \left[ \frac{\WeylR }{d-2}
+ \frac{1}{2} \omega_\text{vor}^2 + \frac{1}{d}\, \mathrm{Harmonic}\left(\frac{2}{d}-1\right) \, \sigma_\text{sh}^2\right] \right).
\end{equation}  
Here $\WeylR$ is the Weyl covariant curvature scalar on the boundary.  

Let us begin with the ideal fluid part which is the leading contribution in \eqref{eq:classL}. To understand this we first note that the contribution to $S[\MZ]$ can be understood directly from the variational definition of the stress energy tensor. Specifically,  the on-shell action with $\Gatt \to 0$, which prior to our Legendre transformation is the usual generating function of correlators, is given by contracting the ideal stress tensor \eqref{eq:Tideal} with the change in the background metric from flat spacetime, viz., 
\begin{equation}\label{eq:Sideal}
S_\text{ideal}[\MZ] = \frac{1}{2}\,\int\, d^d x\,  T_\text{ideal}^{\mu\nu} \, \left(\gamma_{\mu\nu} -\eta_{\mu\nu}\right)
\;\; 
\;\; \overset{\text{\tiny{Legendre}}}{\underset{\text{\tiny{transform}}}{\longrightarrow}} \;\; 
	 S[\MZ]  \bigg|_{\Gatt \to \;0} \,.
\end{equation}	
We have dropped the background constant free energy part and focused on the pieces arising from the solution to the linearized equations of motion. In particular, the contact term contribution in its entirety originates  from the propagation of an ideal fluid on 
\eqref{eq:gammaLR}. It should now be clear that the curious factor of $(d-6)$ is just a numerical accident; it arises due  to the relation between the metric components in $\gamma_{\mu\nu}$. There is nothing special about relativistic conformal fluids in six spacetime dimensions, nor are \SAdS{7} black holes (and the dual $(0,2)$ SCFT plasma) in any way singled out.

Having understood the connection between the gravitational on-shell action and the stress tensor, we can connect to the adiabatic  effective action \eqref{eq:classL}.  Prior to Legendre transformation the ideal part is simply the free energy evaluated on the sound mode solution. On the grsK solution we can represent it using the rescaled thermal vector 
$\vb{b}^\mu_{\skL/\skR} = \left(\frac{u^\mu}{T}\right)_{\skL/\skR}$, which is
\begin{equation}\label{eq:b2def}
\begin{split}
\vb{b}^\mu_{\skL/\skR} = 
	b\left[
	\left(1- \int_k\, \ScS \, \frac{b^{d-2}\, \bwt^2}{d\,(d-1)}\, \PoZ_{\skL/\skR}\right)\pdv{v} - 
	\int_k\, i\, \frac{b^{d-2}\, \bqt\, \bwt}{d\,(d-1)}\, \ScS_i\, \PoZ_{\skL/\skR}\, \pdv{x^i}\right] .
\end{split}
\end{equation}
In order to  compute the quadratic part of the action it will suffice to know rescaled thermal vector accurate to linear order in amplitude, consistent with our identification using the stress tensor.  The reason is that the amplitude expansion of the ideal fluid action 
\begin{equation}\label{eq:SidMZ}
S_\text{ideal}[\MZ] =
 	\int \, d^dx\, \sqrt{-\gamma_\skR} \, \left[ -(\gamma_{\skR})_{\mu\nu}\, \vb{b}_\skR^\mu\, \vb{b}_\skR^\nu\right]^{-\frac{d}{2}} - 	\int \, d^dx\, \sqrt{-\gamma_\skR} \, \left[ -(\gamma_{\skR})_{\mu\nu}\, \vb{b}_\skR^\mu\, \vb{b}_\skR^\nu\right]^{-\frac{d}{2}} \,,
\end{equation}	
results in two terms. One is the contribution which leads to the ideal fluid stress tensor in \eqref{eq:Sideal}, and the other the `heat current' term, which originates from the change of the  rescaled thermal vector (the $\delta \vb{b}^\mu = \vb{b}^\mu - \frac{1}{b}\partial_v$ variation), cf., \cite{Haehl:2015pja}. Using the on-shell relation for the ideal fluid one can check that these two contributions nicely sum up (up to the aforementioned Legendre transform) into the terms arising from \eqref{eq:SidMZ}.

This is structurally similar to the Wilsonian influence functional in the vector sector, which captures the shear  modes driving momentum diffusion in \cite{Ghosh:2020lel}.  The main difference in that case was that since there was no propagating mode;  the ideal piece was purely expressible in terms of a contact term, and moreover could be isolated directly from the boundary terms of the Einstein-Hilbert action.  

For the sound mode we can re-express the on-shell action as an ideal piece and a term that captures sound attenuation. To wit, 
\begin{equation}\label{eq:SosIdealpar}
\frac{1}{c_\text{eff}} \,S[\MZ] = S_\text{ideal,LT}[\MZ]  -  \frac{d-2}{ d^2 \, (d-1)^3}\, 
\int_k \,  b^d\, k^4  \left(\PoZ^\dag_d \, \Gatt(\omega,\bk)  \left[\PoZ_a + \left(\nB+\frac{1}{2}\right) \PoZ_d\right] + \text{cc} \right)   .
\end{equation}	
Here the `LT' term in the subscript is present to remind us that one should  Legendre transform the ideal fluid part to account for the fact that we have a sound pole. From the $\Gatt$ term we can also reproduce the Weyl curvature contribution in \eqref{eq:classL} above, as promised. One finds 
\begin{equation}\label{eq:SWeylJZ}
S[\MZ] \supset 
	- \int_k\, \frac{8\,(d-1)^2\,(d-3)}{b^d}   \bqt^2  \left(\JoZ^\dag_\skR\, \JoZ_\skR -\JoZ^\dag_\skL\, \JoZ_\skL \right) .
\end{equation}	
We have used the on-shell relation \eqref{eq:JZsksol} between $\PoZ$ and $\JoZ$  to make clear that this term arises from the spatial curvature of the boundary geometry.

\section{Discussion}
\label{sec:discuss}

In this paper, we have extended the analysis of open quantum systems with memory to include Goldstone modes with a decay width. The earlier analysis of \cite{Ghosh:2020lel,He:2021jna} focused on diffusive modes in thermal plasmas, which does not incorporate such propagating modes. Specifically, we analyzed the dynamics of energy transport and the physics of associated sound modes in a relativistic thermal plasma. While there is a physical difference in the nature of the long-lived modes, our analysis confirms the general paradigm articulated in these earlier works. Namely, the long-lived modes can be captured into a Wilsonian Schwinger-Keldysh  effective action, which we derived for conformal relativistic  plasma with a holographic dual. The key result is the Wilsonian influence functional parameterized directly in terms of the boundary expectation value of the energy flux operator $\expval{(\TcftU)\indices{_v^i}}$.

While the field theory result shares many characteristics with the corresponding effective action for diffusive modes, there are interesting technical peculiarities in the gravitational description. For the diffusive modes  one was able to distill the bulk dynamics into non-minimally coupled designer scalar fields (one per polarization), where the non-minimal coupling was captured by an auxiliary dilaton, whose primary characterization was its asymptotic fall-off rate (the Markovianity index). This auxiliary dilaton however modulated only the interactions in the radial direction, i.e., as a function of energy scale in the field theory, but was spatially homogeneous. This no longer holds for the bulk dual of the sound mode; the auxiliary dilaton has a non-trivial modulation along the spatial directions of the boundary. It nevertheless remains true that the dual field has a non-Markovianity index $3-d$ for spatially inhomogeneous modes.

Furthermore, the gravitational analysis gives a beautiful picture for the dynamics of energy transport. The physical phonon degree of freedom is part of the ideal fluid, and thus should be captured by the hydrodynamic sigma model (Class L) actions of \cite{Haehl:2015pja}.  Owing to the presence of a gapless mode, one should not construct the sigma model action directly, but rather write the Wilsonian analog, which effectively captures the two derivative kinetic term of the Goldstone mode. Since this part is conservative, the resulting Schwinger-Keldysh effective action is factorized into L and R pieces, as indicated in \eqref{eq:SosIdealpar}. The dynamical information, viz., the dispersion relation \eqref{eq:JZsksol}, is obtained from this action by the constrained variational principle outlined in the aforementioned reference. The Class L action also captures higher order non-dissipative contributions, like the background curvature coupling \eqref{eq:SWeylJZ}.\footnote{ There is a specific prediction for fourth order (in gradients) transport data contained in  the $\bqt^4$ terms of \eqref{eq:Gatt4}. We have not attempted to classify the terms in the Class L action that are responsible for it.}

Once we have separated out the propagating mode, what is left is the physics of sound attenuation. Since this is driven by the leading order dissipative terms, the shear viscosity of a conformal plasma, the structure is isomorphic to that found for momentum diffusion in \cite{Ghosh:2020lel}. In other words, the physical influence phase of the sound mode is the non-factorized part of the Wilsonian influence functional, with a physical kernel $\Gatt(\omega,\bk)$. The dissipative part of this kernel is not captured by the Class L sigma model actions, as it should be; it is these frictional effects which drive the plasma to behave as an open quantum system. So in a sense,  $\Gatt(\omega,\bk)$ is the physical influence functional for phonons, though their complete dynamics also requires the kinetic operator arising from the conservative part of the action.

Owing to the spatial modulation of the designer field dynamics in the gravitational description, one finds there to be an interesting discontinuity in the dynamics between vanishing and non-vanishing spatial momentum. As explained at the outset, in order to isolate the physics of sound, it suffices to imagine there being an infra-red cut-off in momenta and to study modes which are long-wavelength above this cut-off scale. To understand the physics of the soft zero modes however, needs a bit more work.  These modes can be understood as large diffeomorphisms of the background, but we have not attempted to quantize this system. It would be interesting to do so.  Alternately, one could work with a physical cut-off, say by  placing the plasma on a compact spatial volume, e.g., on 
$\mathbf{S}^{d-1}$.  

While our analysis was focused on sound modes in a neutral plasma, it can be readily extended to include additional conserved charges. For example, in a  charged plasma we have fluctuations of both energy and charge density; while the former leads to sound, the latter leads to charge diffusion in the scalar sector. In this case, one  has intricate dynamics where two long-lived modes are coupled to each other. It is nevertheless possible, as in the case of vector perturbations discussed in \cite{He:2021jna}, to decouple these two modes and construct the desired effective action. Our preliminary investigations suggest that the general paradigm explained here continues to hold; we hope to report on this in the near future \cite{He:2021ab}.

Finally, it is worth contrasting the analysis of real-time fluctuating hydrodynamics with the earlier work on the fluid/gravity correspondence \cite{Bhattacharyya:2008jc,Hubeny:2011hd}. The focus in that work and extensions thereof was to construct the gravity dual of a fluid flow of the boundary CFT. In particular, given a holographic system whose stress tensor one-point function can be parameterized in terms of hydrodynamic variables, viz., temperatures and velocities, obeying the conservation equations,  the fluid/gravity paradigm constructs a spacetime geometry characterized by this hydrodynamic data. By virtue of focusing on thermal one-point functions, that analysis had a technical advantage of being able to work with $SO(d-1) $ tensor decomposition, but more importantly was fully non-linear in amplitudes of departures from thermal equilibrium. 

The open effective field theory paradigm however addresses a slightly different question: ``What is the gravitational dual of a fluctuating plasma''? More precisely, realizing that the plasma consists of  both short-lived and long-lived modes, we seek to  parameterize its dynamics in terms of the sources for the former and the operators (or fields) corresponding to the latter. This was the philosophy outlined \cite{Ghosh:2020lel} for the study of the Wilsonian influence functional of the plasma.  One has to not only keep track of the dissipative pieces which relate to infalling quanta in the dual gravity, but also the Hawking quanta that correspond to stochastic fluctuations. But this is precisely what has been achieved in terms of the designer fields, which now parameterize the fluctuating bulk metric. While they are not manifestly $SO(d-1)$ covariant, and our analysis thus far has been restricted to linear order in amplitudes, the close resemblance of the ingoing part of our solutions to those obtained in the fluid/gravity literature makes it highly suggestive that it should be possible to bootstrap onto a non-linear solution. It would be fascinating if this goal can be realized.

\section*{Acknowledgements}
It is a pleasure to thank Christian Ferko,  Veronika Hubeny, Shivam Sharma, Omkar Shetye, and Spenta Wadia for helpful discussions. We would also to thank Jewel Ghosh, Siddharth Prabhu, and V.~Vishal for collaboration during the initial stages of the project.

TH was supported by  U.S. Department of Energy grant DE-SC0020360 under the HEP-QIS QuantISED program. RL and AS acknowledge support of the Department of Atomic Energy, Government of India, under project no. RTI4001. They would also like to acknowledge their debt to the people of India for their steady and generous support to research in the basic sciences. MR and JV were supported  by U.S.\ Department of Energy grant DE-SC0009999. TH, MR, and JV also acknowledge support from the University of California. 

\appendix

\section{Dynamics of scalar gravitons}
\label{sec:dynamicsderive}

Our starting point for analyzing the action is simply the Einstein-Hilbert action with its Gibbons-Hawking variational boundary term and appropriate counterterms. We are going to be working to quartic order in gradients. A-priori we expect that we would need counterterms accurate to that order. However, as we shall see there are some additional subtleties in this system which will allow us to obtain certain finite results from the quadratic counterterms alone.  Irrespective of this we will quote here the full counterterm action accurate to fourth order in boundary derivatives.\footnote{
	The fourth order counterterms will turn out to be the leading regularization for $\MZ$ which receives no corrections at lower orders.}

The gravitational dynamics  we consider is  governed by\footnote{
	We eschew the overall normalization by $\frac{1}{16\pi G_N}$ to keep the expressions simple. Boundary quantities will be obtained by multiplying by $c_\text{eff}$ at the end.}
\begin{equation}\label{eq:SEH}
\begin{split}
S_\text{grav} 
&= 
	\int d^{d+1} x\,\sqrt{-g} \,\left( R + d(d-1) \right)  + 2\,\int\, d^dx\, \sqrt{-\gamma} \, K  + S_\text{ct}\\
S_\text{ct} 
&=  
    \sqrt{-\gamma} \left( -2 (d-1) - \frac{1}{d-2} \, \tensor[^\gamma]{R}{} 
    - \frac{1}{(d-4)\,(d-2)^2} \left(
        \tensor[^\gamma]{R}{_{\mu\nu}} \, \tensor[^\gamma]{R}{^{\mu\nu}}  - \frac{d}{4(d-1)}\, \tensor[^\gamma]{R}{}^2
    \right) \right)     .
\end{split}
\end{equation}
We will first examine the equations of motion which we write as 
\begin{equation}\label{eq:AdSEomL}
\EEq_{AB} = R_{AB} - \frac{1}{2}\, R\, g_{AB}-  \frac{1}{2}\, d(d-1)\, g_{AB} = 0 \,, 
\end{equation}  
and then proceed to analyze the variational principle. 

\subsection{Gauge invariant data and time-reversal}
\label{sec:ginvtr}

To understand the dynamics of the scalar gravitational perturbations and deduce that the dynamics can be captured by a single field $\MZ$ we will proceed in a series of steps.  Our first task will be to identify the diffeomorphism invariant combinations of the metric perturbations for the ansatz \eqref{eq:hscalarpert}.  A natural way to capture this information is to look at the curvature tensors which we  write in terms of  orbit space tensors.  It will be convenient to  define a connection on this part of the geometry:\footnote{We will use lowercase early alphabet Latin characters to indicate orbit space tensors in addition to the conventions specified in \cref{fn:conventions}.}
\begin{equation}\label{eq:OrbitC}
\begin{split}
\gOrb &\equiv \dv{r}(r^2 f)=d\, r - (d-2) \,rf\  .
\end{split}
\end{equation}
Some useful identities which we have used to simplify the expressions are:
\begin{equation}\label{eq:Dup}
\left(\Dz_+ - \frac{1}{2} \Upsilon\right) \mathfrak{F} 
    =r\sqrt{f}\ \Dz_+  \left(\frac{\mathfrak{F}}{r\sqrt{f}}\right) \,, \qquad 
\left(\Dz_+ - \frac{1}{2}\,\gOrb\right)(r\, \mathfrak{F}) = r\left(\Dz_+-\frac{1}{2}\,r^2f' \right) \mathfrak{F}
 .
\end{equation}  

We start with the metric parameterized in terms of $\HH_{AB}$ as presented in \eqref{eq:hscalarpert}. For this geometry the gauge invariant  combinations organized into 
 the orbit space tensors are  \cite{Kodama:2003jz}:
\begin{itemize}[wide,left=0pt]
\item An orbit space vector  $\FX^a$ whose dual one-form has components
\begin{equation}\label{eq:dinvX}
\begin{split}
\FX_v  &\equiv k\,  r \,\HH_{vx}-i\omega\, r^2 \,\HHT\ ,\quad
\FX_r \equiv  k \,r \,\HH_{rx}+ r^2\, \dv{\HHT}{r}  \,.
\end{split}
\end{equation}
\item An orbit space symmetric traceless rank $2$ tensor,  $\FY_{ab}$,  with components
\begin{equation}\label{eq:dinvY}
\begin{split}
\FY_{vv} 
&\equiv  
    k^2 \, (\HH_{vv} -2\,r^2f\, \HS) -2\ i\omega\, \FX_v -\gOrb\,  (\FX_v +r^2 f\,\FX_r) \ ,\\
\FY_{vr} 
&\equiv 
    \FY_{rv} \equiv  k^2 \,(\HH_{vr}+2\,\HS)  +(\gOrb -i\omega)\ \FX_r + \dv{\FX_v}{r}   \ ,\\
\FY_{rr} 
&\equiv  
    k^2\,  \HH_{rr} + 2\, \dv{\FX_r}{r} \,.
\end{split}
\end{equation}
\item And finally, we have orbit space scalars
\begin{equation}\label{eq:dinvHscalar}
\begin{split}
\FYS &\equiv 
    k^2 \left(\HS +\frac{\HHT}{d-1}\right) +\frac{1}{r} \,  (\FX_v +r^2 f\, \FX_r)\ ,\\
\frac{1}{2}\, \FY^a_a
&=  
\FY_{vr} +\frac{1}{2} r^2 f\, \ \FY_{rr} 
= k^2 \left(\HH_{vr} +2\,\HS +\frac{1}{2} \,r^2 f\ \HH_{rr}\right)+\frac{d \FX_v}{dr}+(\mathbb{D}_++\gOrb)\FX_r \,.
\end{split}
\end{equation}
\end{itemize}
 
To understand the time-reversal properties of these combinations we use the observation that on the orbit space time-reversal is just a diffeomorphism. Hence we conclude that $\FY^a_a$ and $\FYS$ have even time-reversal parity. For the reminder we use the fact that the orbit space vectors can be decomposed into the basis adapted to time-reversal introduced above \eqref{eq:Dz},  
\begin{equation}\label{eq:XTR}
\FX_a\, dx^a  = \FX_v \left(dv-\frac{dr}{r^2f}\right) + (\FX_v +r^2 f\, \FX_r)\frac{dr}{r^2f} \,, 
\end{equation}
and use the fact that  $dv-\frac{dr}{r^2f}$ is odd under time-reversal and $\frac{dr}{r^2f}$ is even. Similar decomposition for the tensors leads to
\begin{equation}\label{eq:HTR}
\begin{split}
\FY_{ab}\, dx^a\, dx^b  
&= 
    \FY_{vv}\left\{\left(dv-\frac{dr}{r^2f}\right)^2+\left(\frac{dr}{r^2f}\right)^2\right\}  + 
    2\, r^2 f \left( \FY_{vr} +\frac{1}{2} r^2 f\ \FY_{rr}  \right) \left(\frac{dr}{r^2f}\right)^2 \\
& \qquad 
    +\; 2\left(r^2f \FY_{vr} +\FY_{vv}  \right)\frac{dr}{r^2f}\left(dv-\frac{dr}{r^2f}\right) .
\end{split}
\end{equation}

For the purposes of analyzing the equations of motion it is helpful to define some rescaled combinations of fields which have definite time-reversal parity. We introduce:
\begin{equation}\label{eq:rEOSdef}
\begin{split}
\PHE = r^{d-3}\, \HH_{vv}\,, \quad \PHO = r^{d-3} \left(\HH_{vv} + r^2f\, \HH_{vr}\right) \,,
\quad 
\PHW = 2\, r^{d-2}\, \HS \,.
\end{split}
\end{equation}  
We summarize the essential information from this analysis  in  \cref{tab:trparity}.

\begin{table}
\centering
\begin{tabular}{ ||c|c|c|| c||}
\hline\hline
\shadeB{TR Parity} & \shadeR{Gauge invariants} & \shadeR{Metric components}  & 
\shadeR{Debye gauge data }\\
\hline\hline
\multirow{3}{4em}{Even}
    & $\FYS,\ \FY_{vv}$
        & $\HS, \ \HH_{vv}, \ \HHT $  &  $\PHE, \ \PHW$ \\
    & $\FY_{vr} +\frac{1}{2}\, r^2 f\, \FY_{rr} $
        & $ \HH_{vr} + \frac{1}{2}\, r^2f\, \HH_{rr}$ & $\MW, \ \MZ$\\
   & $\FX_v +r^2 f\, \FX_r$ 
        &  $\HH_{vx} + r^2f\, \HH_{rx}$ & 
 \\ \hline  
\multirow{2}{4em}{Odd}
    & $ r^2f\, \FY_{vr} +\FY_{vv}$
        & $\HH_{vv} + r^2f\, \HH_{vr}$ & $\PHO$ \\
    & $\FX_v$ 
        & $\HH_{vx}$ & \\
\hline  \hline 
\end{tabular}
\caption{Time-reversal parity of the scalar perturbation of \SAdS{d+1}. }
\label{tab:trparity}
\end{table}
%

\subsection{Dynamics in the Debye gauge}
\label{sec:Debye}

In \cref{sec:ginvtr} we introduced the gauge invariant combinations of metric perturbations. One can however fix some of these metric functions by using the diffeomorphism freedom. A-priori we can gauge fix three functions, leaving behind four of the seven perturbation functions appearing in \eqref{eq:hscalarpert}.  We will implement  this by working with  a set of $4$ functions   $\{\HS,\HH_{vv}, \HH_{vr},  \HH_{rr}\}$ by first rescaling out a factor of $k^2$ from the gauge invariant scalar and tensor data, $\FYS$ and $\FY_{ab}$, i.e., setting $\FX_{a} = \HHT=0$.
Equivalently, we have the gauge conditions
\begin{equation}
\text{Debye Gauge :}\quad \HH_{vx}=\HH_{rx}=\HHT=0\ .
\end{equation}
We can interpret $\FYS$ and $\FY_{ab}$ in terms of  metric components in the scalar sector in a Debye gauge.\footnote{ 
	This statement is true for spatially inhomgeneous modes. For spatially homogeneous modes all the invariants $\FY_{ab}$ and $\FYS$  are determined in terms of the vector invariant $\FX_a$, which has been set to zero here by our gauge choice. Most of our analysis will be for  $k\neq 0$ where this is not an issue. We will highlight this when we study the  homogeneous modes in \cref{sec:zeromodes}. \label{fn:k0invars}
}    
This is a coordinate chart such that metric has no derivatives of scalar plane waves under plane wave decomposition, viz., the perturbation can be recast into the form 
\begin{equation}\label{eq:Debyemetric}
ds^2_{(1)}
=
\int_k \left\{(\HH_{vv} -2r^2f\, \HS) \, dv^2+2 \,(\HH_{vr} +2\,\HS)  \,dv dr
	+\HH_{rr}\,  dr^2+ 2\, r^2 \, \HS \, 
d\bx^2\right\}\ScS\ .
\end{equation}
This was the gauge choice adopted in \cite{Kodama:2003jz}.

We will now present the linearized Einstein equations in terms of these scalars by decomposing \eqref{eq:AdSEomL} into plane waves. Employing the definitions in \eqref{eq:rEOSdef} and further introducing the combination: 
\begin{equation}\label{eq:rBTdef}
\HH_{rr}	= - \frac{1}{r^{d+1} f^2} \left[2(\PHO-\PHE) + r f \,(d-1)\, \PHW + \PHB\right] ,
\end{equation}	
we end up the metric which at linear order takes the form:
\begin{equation}\label{eq:EOWDeb}
\begin{split}
 ds_{(1)}^2
 &=  
 	\frac{\PHE-r f\,\PHW}{r^{d-3}}\, dv^2 
	 +\frac{2}{r^{d-1}f} \left(\PHO-\PHE +rf\,\PHW\right) dv\,dr
	 + r^2\, \frac{\PHW}{r^{d-2}}\, d\vb{x}^2 \\
& \qquad \qquad 
	-\frac{1}{r^{d+1}f^2}\left[2(\PHO-\PHE) + r f \,(d-1)\, \PHW + \PHB\right] dr^2  \,.
\end{split}
\end{equation}	

With this choice of gauge the vector gauge invariants vanish $\FX_a =0$, while the remaining  scalar and tensor invariants  simplify in the parameterization \eqref{eq:EOWDeb} to
\begin{equation}\label{eq:STgaugeinvarsPGF}
\begin{split}
\FYS 
&\equiv 
	 \frac{k^2}{2\,r^{d-2}} \,\PHW \,, \\
\FYE
&\equiv
	\FY_{vv}   
		= \frac{k^2}{r^{d-3}} (\PHE-rf\, \PHW)  \,,\\
\FYO 
&\equiv 
	\FY_{vv} + r^2f\, \FY_{vr} 
	=
	 \frac{k^2}{r^{d-3}} \PHO   \,, \\
\FYB
&\equiv  
	\FY_{vr} + \frac{1}{2}\, r^2f\, \FY_{rr} =	
	-\frac{k^2}{2\,r^{d-1}f}\left(\PHB + (d-3)\,rf \, \PHW\right)   .
\end{split}
\end{equation}

It will be helpful to assemble the equations of motion \eqref{eq:AdSEomL} into time-reversal invariant orbit space tensor combinations  as above. We first have the scalar equation, which involves only $\PHB$ and takes a simple form:\footnote{
	With $\HHT \neq0$,  this equation gets modified to  
	\[
		\EEqT
	 =- \frac{k^2}{2\,r^{d-1} }  \PHB + \left[\frac{1}{r^{d-1}}\, \Dz_+\left(r^{d-1}\, \Dz_+\right) + \omega^2 + \frac{d-3}{d-1}\, k^2 f\right]  \HHT\,.
	\]
	Now $\HHT$ is a Markovian field of index $\ann = d-1$, albeit one with an analytically continued momentum $k^2 \to -\frac{d-3}{d-1} \, k^2$ and sourced by $\PHB$. The operator acting on $\HHT$  is the one acting on  $\MZ$ in \eqref{eq:ZMeqn} with the specification $\Lk = \frac{d-1}{2}\, r^3f'$.  The Markovian part of the $\MZ$ solution in \eqref{eq:Zsolution} is homogeneous solution of this operator.
}
\begin{equation}\label{eq:eeqT}
\EEqT
 =- \frac{k^2}{2\,r^{d-1}}  \PHB  \,.
\end{equation}	
This equation is actually a simple algebraic constraint on the invariants: $\EEqT = \FYB + (d-3)\, \FYS$.

The orbit space tensor equations assembled again into time-reversal invariant combinations take the form:
\begin{equation}\label{eq:eeq12B}
\begin{split}
\EEq_1 
&\equiv 
	\frac{2\,r^{d-1}}{d-1}\, \EEq_{vv} \\ 
&=
	\Dz_+ \left(\Tone - r\, \PHB\right) + \frac{k^2}{d-1}\, \left(\PHE - \PHB  \right) , \\ 
\EEq_2
&\equiv
	\frac{2\,r^{d-1}}{d-1}\,  (\EEq_{vv} + r^2f\, \EEq_{vr}) \\
&=
	-i\omega 	\left(\Tone - r\, \PHB\right) + \frac{k^2}{d-1}\PHO\,, \\  
\EEqB
&\equiv
	\frac{2\,r^{d+1}f}{d-1}\, \left(\EEq_{vr} + \frac{1}{2}\, r^2f\, \EEq_{rr} \right)\\
&= 
	-\Dz_+ \left(\Tone - \frac{r}{2}\, \PHB \right)
	 - i\omega\, r\, \PHO 
	 - \frac{r}{2} \left(\Dz_+ - \gOrb + r f \right) \left[\Dz_+ \PHW - (d-2) \, rf\, \PHW \right] \\ 
&\qquad \quad
	+ \frac{r}{2}\,(\omega^2-k^2f)\, \PHW + \frac{k^2+ d(d-1)\, r^2}{2\,(d-1)}  \, \PHB   \,.
\end{split}
\end{equation}	
We introduced here the quantity $\Tone$, which is defined to be
\begin{equation}\label{eq:T1def}
\Tone
\equiv 
	r \, \PHE - \left(\Dz_+ - \frac{\gOrb}{2}\right) (r\,\PHW) 
=	
	r\left[\PHE - \Dz_+ \PHW + \frac{r^2f'}{2}\PHW\right] .
\end{equation}	
This leaves us with the vector equations which being coefficients of $\ScS_i$ have an explicit momentum factor, keeping track of which will be important for understanding spatial zero modes. We find:
\begin{equation}\label{eq:eeq45}
\begin{split}
\EEq_4\
&\equiv
	2\,r^{d-1} f\, \EEq_{vi}  = i k_i\, \widetilde{\EEq}_4 \\ 
&=
	ik_i\left[\Dz_+ \PHO + i\omega \left(\PHE - \PHB \right)\right] , \\
\EEq_5
&\equiv
	2\,r^{d-1} f\, \left(\EEq_{vi}  + r^2f\, \EEq_{ri} \right) =i k_i\, \widetilde{\EEq}_5\\ 
&=
	ik_i\left[\Dz_+ \PHE+ i\omega\, \PHO - (d-1) \left(\Dz_+ - \frac{1}{2}\, \gOrb\right)(rf\PHW) 
		-\frac{r}{2}\left(d+(d-2)f\right) \PHB\right] .
\end{split}
\end{equation}	
The tilded equations strip out the momentum factor which is convenient to do. The remaining equations which are orbit space scalars picking out the trace and the $\ScS_{ij}$ part of the spatial harmonics can be naturally expressed in terms of them as 
\begin{equation}\label{eq:eeq67}
\begin{split}
\EEq_6
&\equiv 
	- \frac{2\,r^{d-1} f}{d-1}\, \sum_{i=1}^{d-1}\, \EEq_{ii} \\	
&=
	\Dz_+ \left(\frac{\widetilde{\EEq}_5}{f}\right) 	 + i\omega \, \frac{\widetilde{\EEq}_4}{f} +2 \,\frac{d-2}{d-1} \, r^{d-1}\, \EEqT \,, \\
\EEq_7
&\equiv 
	f\,\EEq_{ij} = \frac{k_i\,k_j}{k^2}\, \EEqT \,.
\end{split}
\end{equation}
Finally, a natural way to combine the equations  involves taking a particular combination of $\EEqB$ and $\EEq_5$:  
\begin{equation}\label{eq:eeq3}
\begin{split}
\EEq_3 
&=
	 \frac{2}{r} \,\EEqB+ \widetilde{\EEq}_5  \\
&=
	(\Dz_+ + 2 rf )\left[\Dz_+ \PHW - \PHE +\PHB\right] - i\omega\, \PHO 	+\frac{\Lk}{(d-1)\,  r} \PHB \\
& \qquad 
	+ \left(\omega^2 - k^2 f + \frac{d-3}{2}\, r^3 ff'\right) \PHW \,.
\end{split}
\end{equation}	
%

\subsection{Parameterizing the solution space: \texorpdfstring{$k\neq 0$}{inhomogeneous}}
\label{sec:psolspace}

Since there are only four physical functions, we should only have to use four of the equations of motion. One can check that not all the equations given above are independent (explicitly visible for example in $\EEq_6$ and $\EEq_7$), which suggests that a judicious choice of four equations should suffice to distill the dynamics into a manageable form. For $k\neq 0$ an efficient choice turns out to be the set $\{\EEqT, \EEq_1, \EEq_2, \EEq_3\}$, satisfying which will ensure that the reminder are also upheld. We will now analyze the equations introducing $\MW$ and $\MZ$ to simplify the dynamics in the process. 

Let us begin with $\EEqT =0$ which says that $\PHB(r,\omega,k) = 0$, as long as we focus on non-zero $k$, spatially inhomogeneous modes. We will use this to set $ \PHB =0$  this subsection and return to the case where we have a spatially homogeneous function in \cref{sec:zeromodes}.

In the rest of the section we will give a brief discussion of how one simplifies the dynamics to that of a single scalar field.   We have three independent linearized Einstein's equations in the set \eqref{eq:eeq12B}, \eqref{eq:eeq45} for the fields $\{\PHE,\PHO,\PHW\}$.  It will be convenient to pick the following combinations as our independent Einstein's equations, setting $\PHB =0$ in the process to simplify our expressions:
\begin{equation}\label{eq:ssindL}
\begin{split}
\mathbb{E}_1 
&= 
	\Dz_+ \Tone+ \frac{k^2}{d-1}\, \PHE  \,, \\
\mathbb{E}_2
&=
    -i \omega\, \Tone+ \frac{k^2}{d-1}\, \PHO \,,\\
\mathbb{E}_3
&= 
       \left( \Dz_+  + 2\,rf \right)\left[\Dz_+ \PHW-\PHE \right] - i\omega\,\PHO   + \left(\omega^2-k^2f +\frac{(d-3)}{2}\, r^3f f'\right) \PHW\,.
\end{split}
\end{equation}
We will see shortly that $\EEq_4$ in \eqref{eq:eeq45} will be accounted for (actually it can be eliminated algebraically using an algebraic identity). The combination $\EEq_3$ above will simultaneously take care of $\EEq_5$ and $\EEqB$ by definition.

\paragraph{The Weyl factor and momentum flux fields:} To solve these equations we adopt a strategy similar to the one employed in the analysis of gauge field equations in \cite{Ghosh:2020lel,He:2021jna}. One notes that $\EEq_2$ is the energy conservation equation; in fact $\PHO$ is the only time-reversal odd field which is related to the momentum flux. This suggests we should algebraically solve this equation by letting $\PHO \propto \omega$. We express $\mathscr{T}$ in terms of the same variable and then fix $\PHE$ using the first equation. To wit, 
\begin{equation}\label{eq:T123Xpar}
\Tone= -\frac{k^2}{d-1}\, \MW \,, \qquad \PHO= -i\omega\, \MW\,, \qquad \PHE =   \Dz_+ \MW \,.
\end{equation}  
This choice ensures that the first two equations in \eqref{eq:ssindL} are satisfied. We are then left with third equation $\mathbb{E}_3$, which can be viewed as a relation between $\PHW$ and $\MW$. This gives a constraint on the parameterization, isolating the true dynamical equation.

\paragraph{The designer field for sound:} At this point, based on the experience with vector polarizations and diffusive mode, one would expect that $\MW$ is the physical variable that should parameterize the designer field dual to the sound mode in the plasma. While this is physically correct (as we will justify) there is however a technical obstacle. The parameterization \eqref{eq:T123Xpar} does not immediately give an autonomous equation for $\MW$ but rather leads to a coupled system between $\MW$ and $\PHW$ from $\mathbb{E}_3$.

One can however isolate a new field $\MZ$ by realizing that $\PHW$ and $\MW$ are not independent but related to each other through the relation  
\begin{equation}
\PHE- \frac{1}{r} \, \Tone= \left(\Dz_ + - \frac{1}{2}\, r^2\, f'\right) \PHW = \left(\Dz_+ +  \frac{k^2}{d-1}\, \frac{1}{r}\right)\, \MW\,.
\end{equation}  
The first equality follows from \eqref{eq:T1def} and the second from \eqref{eq:T123Xpar}. This can be solved by introducing an auxiliary field $\MZ$ and solving for $\PHW$ and $\MW$ in terms of it.\footnote{To do so we use the observation that equations of the form $(\partial + A ) X = (\partial+B) Y$ can be solved by setting $X =\frac{1}{A-B} (\partial +B) Z $  and $Y = \frac{1}{A-B} (\partial +A) Z$.} This results in the expression \eqref{eq:XiHSZ} quoted above.

This explains the origin of the designer field $\MZ$ and the momentum dependent factor $\Lk$, which originates during decoupling the $\MW$ and $\PHW$ dynamics.  Once we arrive here, it is  straightforward to check that the final constraint equation on the system is the equation of motion for $\MZ$ given earlier in \eqref{eq:ZMeqn}. It is easy to see that the resulting equation is second order once one appreciates that $\MW$, $\MZ$, and $\PHW$ satisfy a linear relation from \eqref{eq:XiHSZ}
\begin{equation}\label{eq:linearXSZ}
\MW = \PHW- \frac{1}{(d-1)} \, \MZ \,.
\end{equation}  
 This allows us to write $\Dz_+ \PHW-\PHE =\Dz_+\PHW - \Dz_+\MW = \frac{1}{d-1}\,\Dz_+\MZ$
 which then reduces $\EEq_3$ to 
\begin{equation}\label{eq:masterL}
\frac{r^{d-3}\,\Lk^2}{f} \, \left(\Dz_+-\frac{\gOrb}{2}\right)
        \left[ \frac{f}{r^{d-3}\,\Lk^2} \, \left(\Dz_+-\frac{\gOrb}{2}\right)   \right]  (r\, \MZ)
            + \left(\omega^2 - k^2 f + \frac{d^2}{4}\, r^2 (1-f)^2\right)  r \,\MZ  =0 \,.
\end{equation}  
A slight simplification of \eqref{eq:masterL} using the explicit expression for $\Lk$ and \eqref{eq:Dup} leads to the equation of motion \eqref{eq:ZMeqn} quoted in the main text. We emphasize that the dynamics is governed by a time-reversal invariant equation, which as explained in \cite{Jana:2020vyx}, allows one to construct smooth solutions on the grSK geometry.

The parameterization of the metric functions in terms of $\MZ$ is easily obtained to be 
\begin{equation}\label{eq:EOSMZ}
\begin{split}
\PHE 
&=
     \Dz_+ \left(\frac{r}{\Lk} \left[\Dz_+ -\frac{r^2\, f'}{2}\right]  \MZ \right)\,,  \\
\PHO 
&= 
    - \frac{i\omega\, r}{\Lk} \left[\Dz_+  -\frac{r^2\, f'}{2}\right]   \MZ \,,\\
\PHW 
&=
    \frac{1}{ \Lk}\left[r\,\Dz_+ + \frac{k^2}{d-1} \right]\, \MZ\,.
\end{split}
\end{equation}
This suffices to determine the linearized geometry \eqref{eq:MZscalarpert} once we know the solution for $\MZ$. The change of variables involves $\MZ$, $\Dz_+ \MZ$, and $\Dz_+^2 \MZ$ and appears to be necessary to ensure that the classical phase space is only two-dimensional, parameterized by an effective source and a corresponding dual CFT plasma operator. We will focus on parameterizing the phase space for the present by boundary values of $\MZ$ and subsequently argue that the physical solution space is best parameterized by the stress tensor component $(\TcftU)_v^i$ or equivalently by $\MW$.

\section{Variational principle in the scalar sector}
\label{sec:actionderive}

We have indicated in \eqref{eq:SEH} that $L$ will refer to the Lagrangian including the measure factor. Unless explicitly indicated, we will write the terms in the action in a series of steps below, quoting at each stage this Lagrangian in momentum space. Integrations over momenta and over the bulk radial coordinate can thus be avoided in the expressions, which themselves tend to be pretty long. We also use $\dag$ to indicate the frequency and momentum reversed field, viz.,  $\Phi^\dag(\omega,\bk) = \Phi(-\omega,-\bk)$, and thus use $+\;\text{cc}$ to account for symmetrization. This analysis is restricted to $\bk \neq 0$ as we seek to establish the variational principle for $\MZ$ at the end of the day.

\subsection{Action for time-reversal invariant fields}
\label{sec:SEOS}

Since the background \SAdS{d+1} solution has a non-vanishing on-shell action, when we expand the Einstein-Hilbert action with the perturbation ansatz, we will have terms starting at the zeroth order in the amplitudes of the perturbation. We will separate out the zeroth and first order contributions out ab-initio -- they do not contribute to the dynamics of the linearized modes. Rather, these terms correspond to the background free energy and represent the ideal fluid contribution of the boundary action.\footnote{
	In the analysis of the tensor and vector modes in \cite{Ghosh:2020lel,He:2021jna} we in fact even extracted a part of the quadratic terms which assembled nicely to give ideal fluid contribution at the outset. In the present case, given the relative complexity of the dynamics, we find it useful to keep the quadratic pieces together and only isolate the part which involves terms at most linear in the fluctuation fields. \label{fn:quadSid}} We therefore will write:
\begin{equation}\label{eq:Sgrav012}
S_\text{grav} = S_\text{grav,lin} + S_\text{grav,quad}\,.
\end{equation}	
We work with the fields  $\{\PHE, \PHO, \PHW\}$ having chosen to eliminate $\HH_{rr}$ using \eqref{eq:rBTdef} (and use \eqref{eq:eeqT} to set $\PHB =0$). 

To begin, let us look at the  contribution from the background and the linear terms in the fluctuations takes the form
\begin{equation}\label{eq:Sgrav01Phi}
\begin{split}
S_\text{grav,lin} 
&= 
	\int\, d^d x \,\Bigg\{
	r^d\left(d+(d-2)\,f -2\,(d-1)\,\sqrt{f}\right)\\
&\qquad \qquad 
	+ (d-1)\left[ d\left(\frac{1+f}{2} -\sqrt{f} \right)r^2 \, \PHW
	- r\,\PHE \left(1-\frac{1}{\sqrt{f}}\right)  \right]
	\Bigg\} \,.
\end{split}
\end{equation}	
Up to this order, we can express this result as an ideal fluid action on the induced boundary geometry. This is similar to the earlier discussion in the vector and tensor cases \cite{Ghosh:2020lel}. One can equivalently write 
\begin{equation}\label{eq:Sgrav01ideal}
S_\text{grav,lin} = \int \, d^dx\, \sqrt{-\gamma} \, \left[ -\gamma_{\mu\nu}\, \vb{b}^\mu\, \vb{b}^\nu\right]^{-\frac{d}{2}}\,,  \qquad \vb{b}^\mu \, \partial_\mu = b\,\partial_v.
\end{equation}	
Here $\vb{b}^\mu$ is the (rescaled) thermal vector that picks out the inertial frame. We will later see that on our solution there are corrections to the thermal vector, which we will need to account for, to get the correct ideal fluid action at quadratic order as noted in \cref{fn:quadSid}. 

Turning to the quadratic part, we will proceed in a series of steps, outlining independently the contributions of the bulk  Einstein-Hilbert and Gibbons-Hawking terms.  This will suffice to demonstrate that the dynamics is governed by the familiar Dirichlet boundary conditions for the aforementioned fields (and hence for  $\HH_{AB}$). Finally, we will outline the contribution from the counterterms that render the on-shell action finite.

The bulk Einstein-Hilbert term can be shown to decompose into a bulk piece, a boundary term, and a total temporal derivative term, viz.,
\begin{equation}\label{eq:SEHsplit}
\begin{split}
\sqrt{-g}\, \left(R + d(d-1)\right) =  L_{_\text{EOW}}^\text{\tiny{bulk}} + \pdv{r}L_{_\text{EOW}}^\text{\tiny{bdy}}  + \pdv{v} L_{_\text{EOW}}^\text{\tiny{dot}}\,.
\end{split}
\end{equation}	
 We start with the bulk term which can be simplified to the form:
\begin{equation}\label{eq:LEHPhi}
\begin{split}
& L_{_\text{EOW}}^\text{\tiny{bulk}}  [\PHE,\PHO,\PHW] \\
&= 
	\frac{d-1}{4\,f\,r^d} \bigg\{ 
	d\,r\,  \Dz_+ \PHW \, \Dz_+ \PHW^\dag - \frac{\Dz_+ \PHW^\dag \,\Dz_+ \PHE + \text{cc} }{f} 
	+ (d-3)\, r\left( \PHE^\dag \Dz_+ \PHW +\text{cc} \right)\\
&\qquad
	+ (d-2)\, r\left( \PHW^\dag \Dz_+ \PHE +\text{cc} \right)+
	r^2 \big(d -(d-2) \,(d+1)\, f\big)\, \left( \PHW^\dag \Dz_+ \PHW +\text{cc} \right) \\
&\qquad 
	+ \frac{2 i\omega}{(d-1) f} \left[ \left( (d-2)\, \PHW^\dag \,\Dz_+ \PHO  - \PHO\, \Dz_+ \PHW^\dag
	- \text{cc}\right)
	-\frac{1}{r f}\left(\PHE^\dag \,\Dz_+ \PHO+\PHO\, \Dz_+ \PHE^\dag - \text{cc}\right)
	\right]	\\
&\qquad 
	+ \frac{i\omega\,r}{(d-1)\,f} \left(d\,(d-3) + (d^2-5\,d+8)\,f\right) \left(\PHW\PHO^\dag -\text{cc}\right)
	+ \frac{4i\omega\,(d-f)}{(d-1)\, r^2} \left(\PHE^\dag\,\PHO -\text{cc})\right)\\
&\qquad 
	+\frac{2k^2}{(d-1)\,r f} \left(\PHO^\dag\, \PHO - \PHE^\dag \, \PHE\right)-
	\frac{\omega^2-k^2f + (d-2)(d-3)\, r^2f^2}{f} (\PHE^\dag \PHW + \text{cc}) \\
&\qquad 
+ r \left[(d-2) (\omega^2-k^2 f) +r^2f \left(-2\,d (2d-5) + (d+2)\, (d-2)^2\,f\right) \right]	\, \PHW^\dag \, \PHW
	\bigg\}\,.
\end{split}
\end{equation}
This part of the action is obtained by direct  evaluation and integrating by parts to isolate the boundary terms. It is interesting to observe that only $\PHW$ has a quadratic kinetic term and $\PHE$ appears in the kinetic part only coupled to $\PHW$. Since $\PHO$ is the only time-reversal odd field its appearance in the action is highly constrained (and it only shows up with explicit time-derivatives).   Note also that the field $\PHW$ has a wrong sign kinetic term (from the last line) reflecting the familiar issue with the conformal mode in gravity. This already suggests that despite appearances $\PHW$ is not the physical field. 

The temporal boundary term $L_{_\text{EOW}}^\text{\tiny{dot}} $ does not enter the analysis and can be dropped ab-initio. The radial boundary term  cancels against a similar contribution from the Gibbons-Hawking term. The precise form of these terms will therefore not be necessary for us. 
\begin{equation}\label{eq:SGH}
2\,  \sqrt{-\gamma}\, K  
=  
	L_{_\text{EOW}}^\text{\tiny{GH}}   -L_{_\text{EOW}}^{\text{\tiny{bdy}}}  \,.
\end{equation}
This is a good consistency check for our computation ensuring that the Einstein-Hilbert action together with the Gibbons-Hawking term has a good variational principle. The remaining part of the Gibbons-Hawking term turns out to be:\footnote{In \cite{Ghosh:2020lel,He:2021jna} this contribution was referred to as the ideal piece, since in those cases, it corresponds to the bare ideal fluid action. We will refrain from employing that notation here; the ideal fluid contribution to the on-shell action is a bit more involved in the scalar sector, owing to the presence of a propagating mode.}
\begin{equation}\label{eq:LGHPhi}
\begin{split}
L_{_\text{EOW}}^\text{\tiny{GH}}[\PHE,\PHO,\PHW] 
&= 
	\frac{1}{2\,r^d\,f} \bigg\{
	 i\omega\, (d-2) \, r^2 \left(\PHO^\dag \PHW -\text{cc}\right)
	 -	 i \omega\, \frac{r}{f}  \left(\PHO^\dag \PHE -\text{cc}\right) \\
&\qquad 
	 - (d-1)\, (d-2)\, r^3f\, \left(\PHE^\dag\, \PHW +\text{cc}\right)\\
&\qquad 
	+\frac{d-1}{2} \, r^4f \left[ d \, (d-3) +(d+1)\, (d-2)\,  f\right] \PHW^\dag\, \PHW 
	\bigg\}\,.
	\end{split}
\end{equation}

The variation of \eqref{eq:LEHPhi} gives us three independent equations. One of these is the momentum conservation equation $\EEq_1$ from \eqref{eq:ssindL} which comes from varying 
$\PHO$. The other two equations are linear combinations of the ones we have given above. It is interesting to note that the variation does not directly produce the $\EEq_3$ equation which was crucial to derive the autonomous second order equation for $\MZ$.   One aspect that is clear from the variational analysis is that the fields $\{\PHE, \PHO, \PHW\}$ obey Dirichlet boundary conditions. This is manifest from the structure of the Gibbons-Hawking term which is a quadratic form in these three fields.  

Finally, the counterterm action is given as
\begin{equation}\label{eq:LctPhi}
\begin{split}
L_{_\text{EOW}}^\text{ct}[\PHE,\PHO,\PHW] 
&= 
	\frac{1}{4\,\,r^d\,f^\frac{3}{2}} \Bigg\{
	\left[(d-1)\, r^2 - \frac{1}{(d-1)\,(d-2)\,(d-4)}\, \frac{k^4}{r^2} \right]  \PHE^\dag\, \PHE	\\
&\qquad
	+r f\, \left[k^2 + (d-1)\, (d-2)\, r^2\right] \left(\PHE^\dag\, \PHW + \text{cc}\right)\\
&\qquad
	+(d-1) \, r^2f \left[ \left(\omega^2 - k^2 f\right) - d\, (d-2)\, r^2 f\right] \PHW^\dag\, \PHW 
	\Bigg\}\,.
\end{split}
\end{equation}
We will quote results accurate to quartic order in the gradient expansion for which it suffices to include the boundary counterterm that is quadratic in derivatives (i.e., it only includes the boundary Einstein-Hilbert term in \eqref{eq:SEH}). We have included here the contribution from the quartic counterterm for completeness.

\subsection{The designer scalar action}
\label{sec:Zaction}

We would now like to distill the action in terms of the designer scalar field $\MZ$. To do so we can directly compute the terms the bulk Lagrangian and boundary terms from the Einstein-Hilbert dynamics defined in \eqref{eq:SEH} with the metric ansatz \eqref{eq:MZscalarpert}  and expand to quadratic order. We could equivalently begin with the action given in terms of the $\{\PHE, \PHO, \PHW\}$ fields and use the substitutions given in \eqref{eq:EOSMZ}. This is a bit more useful, since we have already removed in the process redundant boundary terms. We can therefore focus on just the three terms computed earlier:
  $L_{_\text{EOW}}^\text{\tiny{bulk}} $, $L_{_\text{EOW}}^\text{\tiny{GH}} $, and $L_{_\text{EOW}}^\text{ct} $.

Direct substitution of \eqref{eq:EOSMZ} into the bulk term $L_{_\text{EOW}}^\text{\tiny{bulk}}$ leads to formidable expression, denoted as $L_{_\text{EH}}[\MZ]$. There is however a nice structure beneath this mess. Lets first see why $L_{_\text{EH}}[\MZ]$ reduces into a two-derivative action in the bulk, with the complications relegated to the boundary terms. Since $\PHE$ appears with a single radial derivative in \eqref{eq:LEHPhi} and the change of variables to $\MZ$ involves a double radial derivative \eqref{eq:EOSMZ}, we end up with a action with higher derivative terms. The highest derivative term is  $\dv[3]{r}\MZ^\dag\, \dv[2]{\MZ}{r}$. If we naively vary this action with respect to the field $\MZ$ we expect to get a quintic order equation of motion which should be implied by \eqref{eq:masterL}, i.e., the resulting equation must be some combination of $\mathbb{E}_{\MZ}$ and derivatives thereof. Carrying out the exercise we find however 
\begin{equation}\label{eq:SEHOKIvar}
\left(-\dv[3]{r}\fdv{\MZ'''} + \dv[2]{r}\fdv{\MZ''} - \dv{r}\fdv{\MZ'} +\fdv{\MZ}\right) L_{_\text{EH}}[\MZ] 
=\frac{1}{4}\, \frac{d-2}{d-1}\, \frac{k^4}{r^{d-1}\, f\, \Lk^2} \, \mathbb{E}_{\MZ}^\dag \,.
\end{equation}	

The fact that the higher order action leads to a second order equation of motion is a sign of the hidden simplicity. Once one knows this it is a matter of corralling the higher derivative terms and showing they are total derivatives. With some effort one can show that 
\begin{equation}\label{eq:MZactionsplit}
\begin{split}
L_{_\text{EOW}}^\text{\tiny{bulk}}\left[\PHE,\PHO,\PHW\right] \longrightarrow
L_{_\text{EH}}[\MZ]  
&= 
	L[\MZ]  +  \dv{r} L_{\partial}[\MZ] \,. 
\end{split}
\end{equation}
The bulk action in momentum space is given in \eqref{eq:Zaction} which we reproduce here 
\begin{equation}\label{eq:MZactA}
\begin{split}
L[\MZ]
&= 
	-  \frac{\sqrt{-g}}{4}\left(\frac{d-2}{d-1}\right) \,   \frac{k^4}{r^{2(d-2)}\, \Lk^2} \\
&\qquad \qquad 
\times 	\left[
		\frac{ \Dz_+ \MZ^\dag\, \Dz_+ \MZ }{r^2f}
	-\left( \frac{\omega^2}{r^2f} -\frac{k^2}{r^2} \left(1-\frac{(d-2)\, r^3\,f'}{\Lk}\right) \right) \MZ^\dag\, \MZ \right] .  
\end{split}
\end{equation}	

The complicated boundary terms can be understood as follows. Firstly, the leading  $\dv[3]{r}\MZ^\dag\, \dv[2]{\MZ}{r}$ term being absent in \eqref{eq:MZactA} suggests that $L_{_\text{EH}}^{\partial}[\MZ]$ begins with  $\dv[2]{r}\MZ^\dag\, \dv[2]{\MZ}{r}$.   The Gibbons-Hawking term $L_{_\text{EOW}}^\text{\tiny{GH}}$  does not have a corresponding term with this high derivative order, but the counterterm does (from $\PHE\, \PHE^\dag$).
The cleanest presentation of the boundary terms turns out to be to combine the contributions from $L_{\partial}[\MZ]$ and $L_{_\text{EOW}}^\text{\tiny{GH}}$ and express the result as a general quadratic form in the variables $\Dz_+ \MW \sim \Dz_+^2\MZ$,  $\PHW \sim \Dz_+ \MZ$, and $\MZ$ itself. We will refer to this total collection of boundary terms as the variational boundary terms of $\MZ$ and write:
\begin{equation}\label{eq:LZbdy0}
L_\text{var}[\MZ] = L_{_\text{EOW}}^\text{\tiny{GH}} \left[\PHE,\PHO, \PHW\right] + L_{\partial}[\MZ]\,.
\end{equation}	
We find  
\begin{equation}\label{eq:LZbdy}
\begin{split}
L_\text{var}[\MZ] 
&= 
	-\frac{d-1}{4\,r^{d-2} f} 
	\bigg\{ 
		\Dz_+ \MW\, \Dz_+\MW^\dag + c_1 \left( \PHW^\dag\, \Dz_+ \MW +\text{cc}\right) +   c_2 \left( \MZ^\dag\, \Dz_+ \MW +\text{cc}\right)   \\
&	\qquad \qquad \qquad 
		 + c_3\, \PHW^\dag \, \PHW + c_4\, \left(\MZ^\dag \, \MW+ \text{cc}\right) + c_5\, \MZ^\dag \, \MZ 
	\bigg\}\,,
\end{split}
\end{equation}
with coefficient functions  
\begin{equation}\label{eq:LZbdycfs}
\begin{split}
c_1 &= 
	\frac{k^2}{(d-1)\,r} +\frac{1}{2}\, r\left(d + (d-4)\, f\right) \,,\\
c_2 &= 
	-\frac{k^2}{(d-1)^2\, r}\,, \\
c_3 &=
	-\frac{d(d-2)}{2} \, r^2f \, \left(1+f\right) + \frac{d}{2\,(d-1)}\,(1-f) \, \Lk  + (\omega^2-k^2\,f )\,,  \\
c_4 &= 
	-\frac{d\, k^2}{2\,(d-1)^2}\,(1-f) \,, \\
c_5 &=
	- \frac{d^3\,(d-2)\,r^6\, f\,(1-f)^3}{8 \, \Lk^2} 	
	+ \frac{d^2\, r^4 \,(1-f)^2\left(d + (3\,d-8) \,f\right)}{8\,(d-1)\,  \Lk}
	-\frac{k^2\, (\omega^2-k^2\,f)}{(d-1)^2\, \Lk}   \\
&\qquad 
	 -\frac{d\,(d-2)}{2\,(d-1)^2}\,r^2f\,(1-f)  - \frac{d}{2\, (d-1)^3}\, (1-f) \, \Lk\,.
\end{split}
\end{equation}
One can check directly that 
\begin{equation}\label{eq:LZEHcheck}
L_{_\text{EOW}}^\text{\tiny{bulk}}+ \dv{r} L_{_\text{EOW}}^\text{\tiny{GH}} 
= L[\MZ] + \dv{r} L_\text{var}[\MZ] \,.
\end{equation}	

Having dealt with the bulk and boundary terms let us turn to the counterterms. These can be evaluated by direct substitution, though we note that the presence of the $\PHE\, \PHE^\dag$ does mean that the counterterms are functionals of $\Dz_+^2 \MZ$, which is somewhat unusual.
The counterterm action can be corralled into:
\begin{equation}\label{eq:LZct}
\begin{split}
L_\text{ct}[\MZ] 
&= 
	\frac{d-1}{4\, r^{d-2}\, f^{3/2} }
	\bigg\{
		b_0\, \Dz_+\MW^\dag \, \Dz_+ \MW + b_1 \, \left( \PHW^\dag \Dz_+ \MW + \text{cc} \right)
		+ b_2 \,  \PHW^\dag \, \PHW  \bigg\} \,,
\end{split}
\end{equation}
with coefficient functions
\begin{equation}\label{eq:LZctcfs}
\begin{split}
b_0 &=
	1 -\frac{k^4}{(d-1)^2\, (d-2)\, (d-4)\, r^4} \,, \\
b_1 &= 
	\frac{1}{2}\, \left(d-4 + d \,f \right) r f  + \frac{f }{(d-1)\, r} \, \Lk\,,\\
b_2 &=
	-d\, (d-2)\, r^2\, f^2 + (\omega^2-k^2\, f) \,f \,.
\end{split}
\end{equation}
%

\subsection{The variational principle for \texorpdfstring{$\MZ$}{Z}}
\label{sec:ZvarP}

We have all the pieces at our disposal to deduce the variational principle for the field $\MZ$. Our first task will be to work out the variational principle that leads to the generating function of boundary correlators, viz., the usual   boundary conditions in the AdS/CFT parlance. We will then work out the appropriate Legendre transform that computes the WIF of the boundary theory from the grSK contour.

For the variational analysis we will treat the factor $\frac{k^4}{\Lk^2}$ as an overall pre-factor that we will account for at the end of the day. Equivalently, we work with an auxiliary system for $\MZ$ where the action has this factor scaled out. 

With this understanding let us first record the momentum conjugate the field $\MZ$. From \eqref{eq:MZactA} we find:
\begin{equation}\label{eq:Zmom}
\PiZ = - \frac{d\,\nu_s}{8} \, \frac{1}{r^{d-3}} \, \Dz_+ \MZ  \,.
\end{equation}	
Further, using the Green's function \eqref{eq:Zsolution} we can check that the asymptotically the conjugate momentum is constant and parameterized by the inverse Green's function $\KS$. With our conventions for $\Gin{\MZ}$ we have 
\begin{equation}\label{eq:PiZasyn}
\PiZ = - \frac{k^2\, \KS(\omega,\bk)}{2\,d\,(d-1)^2}  \, b^{d-2}\,  \text{coeff}_{\br^0} \left[\Gin{\MZ}\right]+ \cdots  \,.
\end{equation}	
We see that the conjugate momentum $\PiZ$ is finite; the ellipses is \eqref{eq:PiZasyn} denote the subleading terms of $\order{\br^{2-d}}$. 

This behaviour of the field $\MZ$ and its conjugate $\PiZ$ is indeed what one expects from a non-Markovian field based on the analysis of \cite{Ghosh:2020lel}. At this point we can even guess that the boundary conditions for $\MZ$ are Neumann for the purposes of computing correlation functions. We will however want to compute the Wilsonian Influence Functional (WIF) parameterized by the boundary moduli fields for $\MZ$, which will turn out to be computed by quantizing the field with renormalized Dirichlet boundary conditions. 

We will now argue for this  directly by analyzing the variational principle for the action $S[\MZ]$. Recall that we have organized the classical action for the designer field as 
\begin{equation}\label{eq:Zactionfull}
S_\text{grav}[\MZ] = \int_k\, \int dr\, L[\MZ] + \int_k \, \left(L_\text{var}[\MZ] + L_\text{ct}[\MZ]\right) .
\end{equation}	
Let us begin with the bulk term whose variation is simply:
\begin{equation}\label{eq:LEHZvar}
\delta L[\MZ] = \frac{k^4}{\Lk^2} \, \PiZ\, \delta \MZ^\dag\,.
\end{equation}	
The variation of the boundary term $L_\text{var}[\MZ]$ produces various terms which can be expressed a combinations of $\delta Z$, $\delta \Dz_+ \MZ \propto \delta \PiZ$ and $\delta^2\Dz_+ \MZ$. To deduce this we note that $\MW$ and $\PHW$ contain $\Dz_+ \MZ$ while $\Dz_+ \MW$ has a piece that behaves as $\Dz_+^2 \MZ$. Putting this together we expect that the stationarity of the action demands
\begin{equation}\label{eq:bcZgenr}
\frac{k^4}{\Lk^2}\left[ \PiZ^\dag\, \delta \MZ  + A_1\, \delta \MZ + A_2\, \delta\, \Dz_+ \MZ + A_3\, \delta\, \Dz_+^2 \MZ\right] =0\,,
\end{equation}	
where $A_i$ depends on $(\MZ, \Dz_+\MZ, \Dz_+^2\MZ)$ and background metric data through the coefficient functions defined in \eqref{eq:LZbdycfs}. We factored out the $\frac{k^4}{\Lk^2}$ piece as advertised which will be helpful when we use our knowledge of the solution in the gradient expansion (at this stage it was not strictly necessary). This complicated second order boundary condition is what is necessary for ensuring the stationarity of the action at a generic radial hypersurface. We are however interested in an asymptotically locally AdS geometry, so we should understand what the behaviour of the boundary condition is in the $r\to \infty$ limit (see similar discussion in \cite{He:2021jna}). For this we need not only the asymptotics of the  coefficient functions $c_i$ in \eqref{eq:LZbdycfs} (which is clear from their definition), but also the large $r$ behaviour of the functions $\MW$, $\PHW$, $\Dz_+\MW$. These are of course easy to extract given the solution \eqref{eq:Zsolution} for $\MZ$.

Generally, for the purposes of the variational principle it suffices to focus on the bulk and the boundary terms as we have done above. One ignores the counterterms -- they are  important to ensure that we have a finite norm on the phase space. Crucially, they should be expressed in terms of the intrinsic data on the boundary that are being held fixed by the boundary conditions. Since we inherit the counterterm action from those for the Einstein-Hilbert dynamics \eqref{eq:SEH} they are naturally expressed in terms of variables that are held fixed. This is easy to see from our discussion of the dynamics in terms of the triple $\{ \PHE,\PHO,\PHW\}$ in \cref{sec:SEOS}. We have a complicated boundary action as evidenced above, and if we wish to discern the asymptotic behaviour of \eqref{eq:bcZgenr} it will be helpful to work with a regularized phase space. Consequently, we will include the counterterms in our variation and discern what they tell us about the variational principle asymptotically.

Before we evaluate the variation of the  boundary terms and counterterms it will helpful to corral them into a nicer form. Using \eqref{eq:LZbdy} and \eqref{eq:LZct} we find:
\begin{equation}\label{eq:LZbdytotal}
\begin{split}
L_\text{var+ct}[\MZ] 
&\equiv 
	L_\text{var}[\MZ] + L_\text{ct}[\MZ] \\
&= 
	-\frac{d-1}{4\,r^{d-2}f} \bigg[
	\left(1-\frac{b_0}{\sqrt{f}}\right) \Dz_+ \MW^\dag\, \Dz_+ \MW 
	+ \left(c_1 - \frac{b_1}{\sqrt{f}}\right)  \left[\PHW^\dag\, \Dz_+ \MW +\text{cc} \right] \\
&\qquad\qquad\quad  	
	+ \left(c_3 - \frac{b_2}{\sqrt{f}}\right) \PHW^\dag\, \PHW + \left[\MZ^\dag\left(c_2\, \Dz_+\MW +c_4\, \MW\right) +\text{cc} \right] + c_5\, \MZ^\dag \, \MZ
	\bigg] \,.
\end{split}
\end{equation}	
We now simplify this in the following steps:
\begin{itemize}[wide,left=0pt]
\item The coefficient functions are explicitly known, so we can use the background data to estimate the leading large $r$ behaviour. To do so we must carefully extract the pieces that scale as $r^{-2(d-1)}$ since the functions are highly divergent. Carrying out this exercise we notice that some terms drop out (eg.,  the $\PHW^\dag\, \PHW$ term).
\item Next using the asymptotics of the solution  given in \eqref{eq:PHWasym}, \eqref{eq:PHEasym}, and \eqref{eq:PiZasyn} we learn that at large $r$  
\begin{equation}\label{eq:asymrelns}
\frac{\PHW}{r^{d-2}} = \frac{1}{d-2}\, \frac{\Dz_+ \MW}{r^{d-1}} + \cdots \,, \qquad 
k^2\, \frac{\Dz_+ \MW}{r^{d-1}} = -4\,(d-1)\, \PiZ + \cdots \,.
\end{equation}	
This implies  that we can for purposes of the large $r$ behaviour carry out the replacements above. We will not directly eliminate all occurrences of $\Dz_+ \MW$ since the relation to $\PiZ$ involves a factor of $k^2$. 
\end{itemize}
Implementing this we find the boundary action  \eqref{eq:LZbdytotal} reduces to
\begin{equation}\label{eq:LZbdytotalSimp}
\begin{split}
L_\text{var+ct}[\MZ] 
& 
= 
	 \frac{(d-1)\,(d-6)}{8\,(d-2)\,b^d} \left(\frac{\Dz_+ \MW^\dag}{r^{d-1}}\, \frac{\Dz_+ \MW}{r^{d-1}} \right)- \left(\PiZ^\dag \, \MZ +\text{cc}\right) 	+\text{subleading counterterms}\,.
\end{split}
\end{equation}
The two terms that are indicated above are the leading contribution near the boundary while the terms we have dropped have subleading pieces that serve as counterterms. For the purposes of ascertaining the variation principle the terms we have retained suffice. 

With this simplified boundary term in hand, we can now understand the full variational principle. Taking a variation of \eqref{eq:LZbdytotalSimp} and including the bulk contribution \eqref{eq:LEHZvar} we find:
\begin{equation}
\delta S[\MZ] = \int_k\, \frac{k^4}{\Lk^2}\left[\frac{(d-1)\,(d-6)}{8\,(d-2)} \,\frac{\Dz_+ \MW}{r^{2(d-1)}}\, \delta\Dz_+\MW^\dag  -\MZ\, \delta\PiZ^\dag \right] .
\end{equation}	
Finally, using \eqref{eq:asymrelns} we conclude that the first term is subdominant and the total variation of the action is proportional to $\MZ\, \delta \PiZ^\dag$.\footnote{
	Strictly speaking this is not necessary; from \eqref{eq:asymrelns} we note that $\delta \Dz_+ \MW \sim \delta \PiZ$ so the total variation is indeed proportional to $\delta \PiZ^\dag$ implying the Neumann boundary condition deduced above.} 
Thus, the conventional boundary conditions used to compute the generating function of stress tensor correlators is a Neumann boundary condition for $\MZ$. 

We emphasize that this is a highly non-trivial statement relying on the nature of the asymptotic fall-offs for the fields involved. At finite radial position we would have a complicated mixed boundary condition fixing some relation of the schematic form given in \eqref{eq:bcZgenr}. In fact, this is what would have been suggested if we examined the field redefinitions \eqref{eq:EOSMZ} and used the fact that $\{\PHE, \PHO, \PHW\}$ have Dirichlet boundary conditions imposed on them. 
It is also not something we have imposed by hand, but rather it is completely inherited from the original gravitational dynamics \eqref{eq:SEH}.\footnote{
	This is analogous to what was seen earlier in the analysis of vector modes in \RNAdS{d+1} background, see \cite{He:2021jna} for details.}

\section{Boundary observables}
\label{sec:bdyobs}

We now have understood the dynamics of the gravitational system encoded in $\MZ$. This information can be used to decipher the boundary observables directly.  We describe the computation of the on-shell action, the boundary stress tensor, and then turn to a brief discussion of the relative merits of the field $\MZ$ versus $\MW$. We will continue to drop an overall factor of $c_\text{eff}$ which we have restored in the main text.

\subsection{The boundary sources and operators}
\label{sec:bsourcevev}

We have determined that the $\MZ$ should be quantized with Neumann boundary conditions for purposes of computing correlation functions. This suffices for us to see that $\MZ$ does behave like a regular non-Markovian field introduced in \cite{Ghosh:2020lel}.  The field $\MZ$ should, by virtue of this boundary condition, limit to the dual boundary operator $\OpZ$. Taking the limit on the grsK geometry leads to the statements asserted in \cref{sec:grsKZ}, in particular, \eqref{eq:ZOpDef} and \eqref{eq:PoZdef}. 

The field $\MZ$ is only divergent starting at $\order{r^{d-4}}$, which is lower than what one would expect for a field that is supposed to encode the physics of the boundary stress tensor. A consequence of this is that the renormalized field $\MZ$  is not modified by the boundary counterterms up to the quartic order (i.e., it is uncorrected by the boundary cosmological constant and Einstein-Hilbert counterterm). It only gets renormalized by the quartic $R^2$ counterterm in \eqref{eq:SEH} which is contained in the coefficient $b_0$ at $\order{k^4}$ in \eqref{eq:LZctcfs}. Taking this into account we learn that the correction comes from the $k^4$ contribution to $b_0$ in \eqref{eq:LZctcfs}. The renormalized field $\MZ_{_\text{ren}}$ can be determined to be
\begin{equation}\label{eq:Zren}
\begin{split}
\MZ_{_\text{ren}} 
&= 
	\MZ - \frac{k^2}{(d-2)\, (d-4)}\, \frac{\Dz_+ \MW}{r^{d-1}}\, r^{d-4} + \cdots \\ 
&= 
	\MZ +\frac{4\,(d-1)}{(d-2)\, (d-4)}\,  r^{d-4} \, \PiZ+ \cdots \,.
\end{split} 
\end{equation}	

Having identified the boundary conditions for $\MZ$ and its renormalized counterpart $\MZ_{_\text{ren}}$, let us turn to identifying the source.  Since $\MZ$ is quantized with Neumann boundary condition, the source should be defined in terms of the conjugate momentum $\PiZ$. However, the relations in \eqref{eq:asymrelns} suggests that the conjugate momentum can be traded for the fields  $\frac{\Dz_+ \MW}{r^{d-1}}$ and $\frac{\PHW}{r^{d-2}}$.  This is consistent with the fact that these fields determined the induced boundary geometry. Indeed, including the background piece the induced metric on the boundary evaluates (on either boundary) to 
\begin{equation}\label{eq:indbdymetA}
\gamma_{\mu\nu}\, dx^\mu\, dx^\nu = \left(1+\frac{\PHW}{r^{d-2}}\right) \, \eta_{\mu\nu}\, dx^\mu\, dx^\nu  + \frac{\Dz_+\MW}{r^{d-1}}\, dv^2 \,.
\end{equation}	
It therefore makes sense to identify the temporal and spatial components of the boundary metric as the sources. These are however not independent from each other or from that conjugate momentum $\PiZ$ owing to   \eqref{eq:asymrelns}. We define therefore the boundary source as in \eqref{eq:JZdef} and note that 
\begin{equation}\label{eq:JZWWB}
\JoZ_{\skL/\skR} = \lim_{r\to\infty\pm i0}\, \frac{1}{4\,(d-1)} \, \frac{\Dz_+\MW}{r^{d-1}} 
=  \lim_{r\to\infty\pm i0}\, \frac{d-2}{4\,(d-1)} \, \frac{\PHW}{r^{d-2}} \,.
\end{equation}	
This definition of the boundary source for $\MZ$ in terms of $\Dz_+\MW$ has a proper gradient expansion, unlike $\PiZ$ which has an additional $k^2$. As noted above, it is also the physically correct variable; the temporal component of the boundary metric $\HH_{vv}$ that couples to the energy density is indeed  $\PHE = \Dz_+ \MW $ up to a  factor  of $r^{d-1}$. Fixing all the normalization factors we find that the induced boundary metric can be expressed as in \eqref{eq:gammaLR}.

\subsection{The boundary stress tensor}
\label{sec:BYT}

The boundary stress tensor density is given by varying the boundary Gibbons-Hawking term and the counterterms given in  \eqref{eq:SEH}. This leads to the following expression accurate to quartic order in gradients
\begin{equation}\label{eq:TBYcft}
\begin{split}
 \TcftD_{\mu\nu} 
& =
 \lim_{r\to\infty}
 	\frac{2\sqrt{-\gamma}}{r^2} \left[ 
 	 K\, \gamma_{\mu\nu}  -K_{\mu\nu} 
 	 - (d-1)\, \gamma_{\mu\nu} + \frac{1}{d-2}  \, \tensor[^\gamma]{G}{_{\mu\nu}}
 	\right. 	\\
&\left.
	+\frac{1}{(d-2)^2\,(d-4)} \Bigg(\tensor[^\gamma]{\nabla}{^2}\, \tensor[^\gamma]{R}{_{\mu\nu}}
	+ 2\, \tensor[^\gamma]{R}{_{\mu\rho\nu\sigma}}\, \tensor[^\gamma]{R}{^{\rho\sigma}} 
	\right.\\
& \left. \quad	
	+ \frac{1}{2\,(d-1)}\, \left[-
		(d-2)\,\tensor[^\gamma]{\nabla}{_\mu}\, \tensor[^\gamma]{\nabla}{_\nu} \tensor[^\gamma]{R}{}  - 
		 d\, \tensor[^\gamma]{R}{} \, \tensor[^\gamma]{R}{_{\mu\nu}}
		  \right]
	\right.  \\
& \left. \quad
	-\frac{1}{2} \,\gamma_{\mu\nu} \left(
		 \tensor[^\gamma]{R}{_{\rho\sigma}}\,  \tensor[^\gamma]{R}{^{\rho\sigma}} 
		  -\frac{d}{4(d-1)}\,  \tensor[^\gamma]{R}{^2} 
		  +\frac{1}{d-1}\, \tensor[^\gamma]{\nabla}{^2}\, \tensor[^\gamma]{R}{}
		\right) \Bigg) \right] .
\end{split}
\end{equation}	

We will now present the result for $(\TcftD)_\mu^{\ \nu}$ which makes it easier to see the traceless condition by inspection.  At the first order in amplitudes one evaluates the components of the stress tensor from the Brown-York analysis supplemented with counterterms \eqref{eq:TBYcft}. We  quote the results for the individual components in turn. 

First up, the spatio-temporal pieces are
\begin{equation}\label{eq:Tupdnvi}
\begin{split}
\left(\TcftD\right)\indices{_v^i} 
 &=
 	-f \left(\TcftD\right)\indices{_i^v} 
 		=  -\lim_{r\to\infty\pm i0} \,  ik\,\ScS_i\, T_1 \,, \\
T_1 &= 
		  i\, \PHO - \omega \,\sqrt{f}\, \PHW
		-\frac{\omega\,k^2}{(d-1)\,(d-2)\, (d-4)} \frac{1}{r^3\, \sqrt{f}}\, \PHE   ,  \\
&=
 	- \frac{\omega}{d-1}  \, \MZ_{_\text{ren}}  + \omega\, \left(1-\sqrt{f}\right) \PHW  \,,
\end{split}
\end{equation}	
where $\MZ_{_\text{ren}} $ was defined in \eqref{eq:Zren}. Notice that the $\PHW$ term limits to zero as we approach the asymptopia. As such, from the original expression it appears that we should regard $\MW$ as the field dual to the energy flux operator, since $\sqrt{f} \, \PHW$ is a counterterm contribution. However, as assembled in the last line, it is somewhat transparent that the boundary operator is $\MZ$ which is cleanly renormalized by the quartic counterterm. We will return to this in \cref{sec:Eredef}.

The temporal component is a bit more complicated  but can be evaluated straightforwardly. We find:
\begin{equation}\label{eq:Tupdnvv}
\begin{split}
\left(\TcftD\right)\indices{_v^v}
&=  
		-\frac{d-1}{b^d} + \lim_{r\to\infty\pm i0}\, \ScS \, T_2\,, \\
T_2 
&=
	(d-1)\, r \left[ 
		\Dz_+ \PHW - \left(2 -\frac{1}{\sqrt{f}}\right)\, \PHE 
		- d\,  (1-\sqrt{f})\, r \sqrt{f}\,\PHW + \PHB\right]  \\
&\qquad \quad 
  - k^2\, \sqrt{f}\, \PHW - \frac{k^4}{(d-1)\,(d-2)\,(d-4)}\, \frac{1}{r^3\,\sqrt{f}}\, \PHE \,,  \\
&=
	-(d-1) \left[
		 \left(1 -\frac{1}{\sqrt{f}}\right)\, r\,\PHE +  \Tone-r\,\PHB + \frac{k^2}{d-1}\, \MW \right] -
		 \frac{k^2}{d-1}\, \MZ_{_\text{ren}}  \\  
&  \qquad \quad 
 +  (1-\sqrt{f}) \left[\frac{d\,(d-1)}{2}\, r^2\, (1-\sqrt{f})  + k^2 \right]\PHW  \,.
\end{split}
\end{equation}
Once again we have combined terms suitably; in the last line the $\PHW$ terms are vanishing in the limit $r \to \infty \pm i0$. If we also  use \eqref{eq:T123Xpar} and set $\PHB =0$ as we are allowed by the $\EEqT$ equation, the first parenthesis also simplifies to $\frac{\PHE}{r^{d-1}}$  which we recognize as a source contribution.

Finally, the spatial stress tensor has contributions from two tensor structures, $\delta_{ij}$ and $\ScST_{ij}$. We will write these as the pressure and shear-stress contribution as follows:
\begin{equation}\label{eq:Tupdnij}
\begin{split}
\left(\TcftD\right)\indices{_i^ j}  
&= 
	 \frac{1}{b^d}\, \delta\indices{_i^j} 
	 + \lim_{r\to\infty\pm i0} \, \left[\frac{1}{f}\, T_P\, \delta\indices{_i^j} \,\ScS+ T_Y\, (\ScST)\indices{_i^j} \right]  \\
T_P
&=
	 (d-1) r \, \sqrt{f} (1-\sqrt{f}) \, \PHE 	-i\omega\, \PHO + \omega^2\, \sqrt{f}\, \PHW  + \frac{k^2}{d-1}\, \frac{\sqrt{f}}{r} \left(\PHE  -  (d-2) \, r f  \, \PHW  \right)	   \\
&\qquad
		+ \frac{d\,(d-1)}{2}   \left(1-\sqrt{f}\right)^2 r^2 f\,\PHW - \widetilde{\EEq}_5
	\\	
T_Y 
&=
	- \frac{k^2}{r\sqrt{f}} \left[rf \,\PHW - \frac{\PHE}{d-2}\right]  .
\end{split}
\end{equation}
We  refrained from writing the quartic order in gradient term which renormalizes $\MZ$  and also have exploited the fact that there is an explicit contribution proportional to the $\EEq_5$ equation to simplify the answer.  The first summand in $T_P$ simplifies to $\frac{\PHE}{2\,r^{d-1}}$, which is a source term, while the designer field $\MZ$ assembles from the subsequent pieces involving $\PHO$ and $\PHW$ which constitute part of the operator contribution. One gets a term proportional to  $\omega^2 \MZ$ from them. The contribution from $\PHE-(d-2)\, r\, \PHW$ is however also an operator contribution, the leading order terms in this difference cancels, as the reader can verify from \eqref{eq:PHWasym} and \eqref{eq:PHEasym}.

With this information we can evaluate the stress tensor on our solution parameterized by the designer field. The result of this exercise  is what is reported in \eqref{eq:Tcftorder4} where we have adhered to the identification of source and vev terms. Specifically, contributions of the form 
$\frac{\PHE}{r^{d-1}}$ and $\frac{\PHW}{r^{d-2}}$ are written in terms of $\JoZ$ using \eqref{eq:JZWWB}.

\subsection{The on-shell action}
\label{sec:Zosbdy}

Once we have identified the boundary conditions for $\MZ$, we can evaluate the on-shell action on the grSK solution. At quadratic order this is just a boundary term and can be easily evaluated on the grSK geometry. One will obtain from this the generating function of boundary correlation functions with sources on the two boundaries of the grSK geometry, viz., 
\begin{equation}
S_\text{grav}[\MZ] = S_\skR^S - S_\skL^S \,.
\end{equation}	
We would however  like to  evaluate the Wilsonian influence function, for which we need to perform a Legendre transformation of the generating functional. This requires one to evaluate the on-shell action after including a suitable boundary term to carry out the Legendre transform: 
\begin{equation}\label{eq:wifZA}
S[\MZ] =
 \left[S_\text{grav}[\MZ] + \int_k \left(\PiZ \, \MZ^\dag + \text{cc}\right)\right]_\text{on-shell} .
\end{equation}	
As noted in \cite{Ghosh:2020lel} this amounts to quantizing $\MZ$ without the additional variational boundary term, i.e., we quantize $\MZ$ with (renormalized) Dirichlet boundary conditions as have noted in \eqref{eq:Zren}.

We will find two distinct contributions to the on-shell action: one arises from terms of the form $\PiZ\, \MZ^\dag$ and originates from a combination of  bulk action, various boundary terms, and  the Legendre transform. The other contribution will turn out to be purely a functional of the source originating from the $\Dz_+ \MW^\dag\, \Dz_+ \MW$ in \eqref{eq:LZbdytotalSimp}. Using \eqref{eq:JZdef} this piece can be written as a factorized source contribution on the two boundaries of the grSK geometry. We end up with the result quoted in \eqref{eq:SoSZsplit} in the form of a contact term (the source contribution) and the genuine influence functional. 

The explicit evaluation of the on-shell action is straightforward since we have already deduced the asymptotic behaviour of the field and the conjugate momentum. Let us start with the non-contact term and record the influence functional. The grsK solution for $\MZ$ is given in 
\eqref{eq:Zsksol}. Accounting for the contribution from the Legendre transform we find this evaluates to
\begin{equation}\label{eq:wifZB}
\begin{split}
 S_\text{WIF}[\MZ] 
&=
	 \int_k\, \frac{1}{2} \,\left(\MZ^\dag_\text{ren} \, \PiZ + \PiZ^\dag\,\MZ_\text{ren}  \right) \bigg|_{r = r_c +i0}^{r=r_c  -i0} \\
&=
	-\int_k\, k^2 \left(\PoZ_d^\dag\,  \frac{b^{d-2}}{2\,d\,(d-1)^2}\, \KS(\omega,\bk)\,  \left[\PoZ_a + \left(\nB+\frac{1}{2}\right) \PoZ_d\right] +\text{cc}\right) ,
\end{split}
\end{equation}	
where we introduced a large radius regulator at $r=r_c$. This is the result quoted in \eqref{eq:SZwif}. 

The contact term may likewise be evaluated directly. From the definition of the sources \eqref{eq:JZdef} and the form of the boundary terms in \eqref{eq:LZbdytotalSimp} we see that this is simply given in terms of the boundary source. The contributions furthermore factorize  leading us to the following expression at quadratic order: 
\begin{equation}\label{eq:Scontact}
S_\text{contact} [\MZ] 
= \int_k\, \frac{2\,(d-1)^3\,(d-6)}{ (d-2)\, b^d} \, \left[\JoZ_\skR^\dag\, \JoZ_\skR - \JoZ_\skL^\dag\, \JoZ_\skL\right]  .
\end{equation}	
This contact term contribution is quite peculiar. However, as we describe in \cref{sec:ideal}, it is nothing but the contact term part of an ideal fluid on the boundary geometry \eqref{eq:indbdymetA}.  We also explain there how one can isolate various hydrodynamic transport data from our answer and connect to the discussion of Class L adiabatic fluid lagrangians of \cite{Haehl:2015pja}.

\subsection{On field redefinitions and boundary operators}
\label{sec:Eredef}

The solution for the designer sound field $\MZ$ on the grSK contour can be repackaged directly in terms of field theory data. Recall that we expect a single mode of the boundary stress tensor that captures the effective dynamics in the low energy limit. The conservation equation of the stress tensor turns out to be a constraint on the field $\MW$, cf., the discussion around \eqref{eq:ssindL} and \eqref{eq:T123Xpar}. Equivalently, examination of the induced boundary stress reveals that the leading contribution comes directly from $\MW$,
\begin{equation}\label{eq:TviMW}
(\TcftU)\indices{_v^i} =   \int_k \ScS\, \omega\, k_i\, \left(\MW + \text{counterterms}\right) .
\end{equation}  
The full expression for the stress tensor including the counterterms can be found in 
\eqref{eq:Tupdnvi},  where the $\sqrt{f} \PHW$ term can be seen to arise from counterterm contributions.

These arguments suggest to us that the holographic dual of the sound mode in the plasma should be identified with $\MW$. However, $\MW$ does not by itself have simple dynamics.\footnote{We were able to derive a third order radial differential equation for $\MW$ directly. At each order in the gradient expansion this equation turned out to be a second order inhomogeneous equation for $\dv{r}\MW$ which suggests again that there is further simplification possible by passing onto $\MZ$. }  Consequently, we rely on $\MZ$ as an intermediate auxiliary field to analyze the problem and translate the physical data back onto $\MW$ therefrom.

The relative choice between $\MZ$ and $\MW$ is effectively a field redefinition in the boundary.  
To appreciate this, let us obtain the grSK solution for $\MW$ by first constructing the inverse Green's function $\Gin{\MW}$ for $\MW$ which is reported in \eqref{eq:MWasym}. We find 
\begin{equation}\label{eq:MWsksol}
\MW^\text{SK}(\ctor, \omega, \bk) = \Gin{\MW} \, \PoZ_a +\left[ \left(\nB+\frac{1}{2}\right) \, \Gin{\MW } -\nB\, e^{\beta\omega(1-\ctor)} \, \Grev{\MW} \right] \PoZ_d \,.
\end{equation}	
The key point to note is that the coefficient of radially homogeneous mode $\Gin{\MW}$ from \eqref{eq:MWasym} is  
\begin{equation}\label{eq:GinMWfin}
\text{coeff}_{\br^0}\left[\Gin{\MW}\right] = -\frac{1}{d-1}\, \left(1+ \frac{2}{d}\,\Gatt\right) ,
\end{equation}	
 which suggests that 
\begin{equation}
\begin{split}
\EOp_{\skL,\skR} 
&= \lim_{r\to \infty \pm i0} \left[ \MW + \text{counterterms}\right] \\
&= 
 -\frac{1}{d-1}\, \left(1+ \frac{2}{d}\,\Gatt\right)\, \PoZ_{\skL,\skR} \,.
 \end{split}
\end{equation}	
One could, if one wished to do so, convert the expressions in the main text directly to expressions involving $\EOp$, but we have refrained from doing so to avoid complicating the already involved discussion.

\section{Further details of the gradient expansion solutions}
\label{sec:gradexpfns}

The solution for the designer field $\MZ$ which satisfies \eqref{eq:ZMeqn} with ingoing boundary conditions was given in \cref{sec:MZgradexp}. This form was chosen to make direct contact with the functions appearing in the fluid/gravity literature \cite{Bhattacharyya:2008mz} and the earlier analysis in \cite{Ghosh:2020lel}. We compile some useful results about the functions appearing in the expansion in this appendix.

The functions $\{F,H_\omega,H_k,I_\omega,I_k,J_\omega,J_{\omega k},J_k\}$ that parameterize the solution for $\MZ$ in \eqref{eq:Zsolution} have compact integral expressions tabulated in \cref{tab:gradfns}. In \cref{sec:gradfnsasym} we collect several useful facts about them  and determine their asymptotic behaviour. Using this data we record  in \cref{sec:ZWasym} the  asymptotic expansions for the fields $\MZ$, $\MW$, and the metric functions $\PHE, \PHW$ which will prove useful for the phase space and boundary condition analysis.   Subsequently, in \cref{sec:exppargrad} we present the solution in the alternate form parameterized in \cite{He:2021jna} for ease of comparison. 

\subsection{Asympotics of the solution}
\label{sec:gradfnsasym}

As noted in \cref{sec:MZgradexp} the functions parameterizing the solution for $\MZ$ in gradient expansion are cleanly written in terms of a double-integral transform of a source function $\mathfrak{J}$, cf., \eqref{eq:gensol}. The sources for the various functions are collated in \cref{tab:Mgradsol}. Examining this data we immediately see that there are some useful relations:
\begin{itemize}[wide,left=0pt]
\item First, the sources for the functions $H_\omega$ and $H_k$ determine an useful identity for $I_k$:
\begin{equation}\label{eq:IkHreln}
H_\omega(\br)+H_k(\br)= \frac{d-2}{2} \,I_k(\br) \,.
\end{equation}
\item In some cases, the inner integral in \eqref{eq:gensol} can be performed, resulting in a representation involving only one integral, e.g.,
\begin{equation}\label{eq:FHIintrep}
\begin{split}
F(\br)&\equiv 
    \int_\br^\infty\frac{y^{d-1}-1}{y(y^d-1)}dy\,, \\ 
H_k(\br)
&\equiv 
    \frac{1}{d-2}\int_\br^\infty\frac{y^{d-2}-1}{y(y^d-1)}dy \,,\\
H_\omega(\br) 
&\equiv  
    -H_k(\br) +(d-2)\int_\br^\infty\frac{H_k(y)-H_k(1)}{y(y^d-1)}dy\,,\\
I_\omega(\br) 
&\equiv  
    2\,\int_\br^\infty\frac{H{}_\omega(y)-H_\omega(1)}{y(y^d-1)}dy= -I_k(\br) +(d-2)\int_\br^\infty\frac{I_k(y)-I_k(1)}{y(y^d-1)}dy\ .
\end{split}
\end{equation}
\end{itemize}

These integral expressions allow us to write down the asymptotic solution for the functions quite efficiently at low orders in gradient expansion. As we proceed to higher orders this structure is lost, and we have to use the nested integral representation \eqref{eq:gensol} to deduce the asymptotics. 
We now record the behaviour of the functions to a sufficiently large order to ensure that we can recover the part of the metric functions  which contribute to finite boundary data. 

Up to the third order in the gradient expansion, we have the functions $\{F,H_k,H_\omega,I_k, I_\omega\}$, whose asymptotics can be determined from \eqref{eq:FHIintrep} to be
\begin{equation}\label{eq:FHIasym}
\begin{split}
F(\br)
&= 
    \br^{-1}- \frac{\br^{-d}}{d}+\frac{\br^{-d-1}}{d+1}- \frac{\br^{-2d}}{2d}+\frac{\br^{-2d-1}}{2d+1}+\cdots\,,\\
H_k(\br)
&= 
    \frac{\br^{-2}}{2(d-2)}-\frac{\br^{-d}}{d(d-2)}+\frac{\br^{-d-2}}{d^2-4}-\frac{\br^{-2d}}{2d(d-2)}+\frac{\br^{-2d-2}}{2(d+1)(d-2)}+\cdots\,,\\
H_\omega(\br) &=-\frac{\br^{-2}}{2(d-2)}-\frac{(d-2)^2 H_k(1)-1}{d(d-2)}\br^{-d}+\frac{d-4}{2(d^2-4)}\br^{-d-2}\\
&\qquad-
    \frac{d(d-2)^2 H_k(1)-2}{2 d^2(d-2)}\br^{-2d} +\frac{d^2-12}{4(d+1)(d^2-4)}\br^{-2d-2}+\cdots\,, \\
I_k(\br)
&=
    -\frac{2H_k(1)}{d}\br^{-d}+\frac{\br^{-d-2}}{d^2-4} -\frac{d(d-2)H_k(1)+1}{d^2(d-2)}\br^{-2d}+\frac{(d+4)\,\br^{-2d-2}}{2(d+1)(d^2-4)}+\cdots\,,\\
I_\omega(\br)
&=
    -\frac{2H_\omega(1)}{d}\br^{-d}-\frac{\br^{-d-2}}{d^2-4} \\
&\qquad\quad
    -\frac{d(d-2)H_\omega(1)+(d-2)^2H_k(1)-1}{d^2(d-2)}\br^{-2d}-\frac{3\,\br^{-2d-2}}{(d+1)(d^2-4)}+\cdots\ .
\end{split}
\end{equation}
In addition at the quartic order we have four functions. Three of them  $\{ J_\omega,J_k, J_{\omega k}\}$ are finite and have the following asymptotic behaviour:
\begin{equation}\label{eq:Jasym}
\begin{split}
J_\omega(\br)
&=
    \frac{\br^{-4}}{8(d-2)(d-4)}+ \frac{4H_\omega(1)+\lambda_\omega d}{d^2}\, \br^{-d}
    +\frac{1-(d-2)^2H_k(1)}{2d(d^2-4)}\,\br^{-d-2} \\
&\qquad 
    -\frac{d-16}{4(d^2-4)(d^2-16)}\, \br^{-d-4} + \frac{\lambda_\omega}{2d}\,\br^{-2d}
    -\frac{2(d^2-2d-1)+d(d+5)(d-2)^2H_k(1)}{4d^2(d^2-4)(d+1)}\br^{-2d-2}\\
&\qquad
    -\frac{5(d^2-20d-32)}{8(d^2-4)(d^2-16)(d+1)(d+2)}\br^{-2d-4}+\cdots\,,\\
J_{\omega k}(\br)
&=
    -\frac{\br^{-4}}{4(d-2)(d-4)}+ \frac{4H_k(1)+\lambda_{\omega k} d}{d^2}\br^{-d}
    -\frac{2-(d-2)^2H_k(1)}{2d(d^2-4)}\br^{-d-2} \\
&\qquad 
        -\frac{6}{(d^2-4)(d^2-16)}\br^{-d-4} + \frac{\lambda_{\omega k}}{2d}\br^{-2d}
        +\frac{d^2-7d-2+d(d+3)(d-2)^2H_k(1)}{4d^2(d^2-4)(d+1)}\br^{-2d-2}\\
&\qquad
    +\frac{d^3-16d^2-144d-192}{8(d^2-4)(d^2-16)(d+1)(d+2)}\br^{-2d-4}+\cdots\,,\\
J_k(\br)
&=
    \frac{\br^{-4}}{8(d-2)(d-4)}+ \frac{\lambda_k}{d}\br^{-d} +\frac{1}{2d(d^2-4)}\br^{-d-2}+\frac{d^2+6d-16}{4(d-2)(d^2-4)(d^2-16)}\br^{-d-4}\\
&\qquad
     + \frac{\lambda_k}{2d}\br^{-2d} +\frac{d+3}{4d(d+1)(d^2-4)}\br^{-2d-2} +\frac{d^3+11d^2+48 d+48}{8(d^2-4)(d^2-16)(d+1)(d+2)}\br^{-2d-4}+\cdots\,,
\end{split}
\end{equation}
while the fourth function is the one that captures all the divergences in $\MZ$ and asymptotes to 
\begin{equation}\label{eq:Vasym}
V_k(\br)=
     - \frac{\br^{d-4}}{d-4}+\frac{\br^{-4}}{4}- \frac{\br^{-d}}{d}+\frac{\br^{-d-4}}{d+4}- \frac{\br^{-2d}}{2d}+\frac{\br^{-2d-4}}{2d+4}+\cdots\ .
\end{equation}  
In writing these expressions we have introduced three numerical constants $\{\lambda_\omega,\lambda_{\omega k},\lambda_k\}$, which are defined via the large $r$ limits:
\begin{equation}\label{eq:lamdef}
\begin{split}
\lambda_\omega +\frac{4}{d}H_\omega(1) 
&\equiv
    \lim_{y\to\infty}\Bigl\{
    -\frac{y^{d-4}}{2(d-2)(d-4)}+\int_1^y\frac{\mathfrak{J}_\omega (z)\ dz}{z(z^d-1)} 
    \Bigr\}\,,\\
\lambda_{\omega k} +\frac{4}{d}H_k(1) 
&\equiv
    \lim_{y\to\infty}\Bigl\{ 
    \frac{y^{d-4}}{(d-2)(d-4)}+\int_1^y\frac{\mathfrak{J}_{\omega k} (z)\ dz}{z(z^d-1)} 
    \Bigr\}\,,\\
\lambda_k  
&\equiv
    \lim_{y\to\infty}\Bigl\{
    -\frac{y^{d-4}}{2(d-2)(d-4)}+\int_1^y\frac{\mathfrak{J}_k (z)\ dz}{z(z^d-1)} 
    \Bigr\}\,,
\end{split}
\end{equation}
where $\{\mathfrak{J}_\omega,\mathfrak{J}_{\omega k},\mathfrak{J}_k\}$ are the source functions for $\{J_\omega,J_{\omega k},J_k\}$ given in \cref{tab:Mgradsol}.
 
\subsection{The metric functions in gradient expansion}
\label{sec:ZWasym}

Armed with the expressions for the asymptotics of the functions appearing in the gradient expansion, we can estimate the near-boundary behaviour of $\MZ$.
Firstly, for $\MZ$ we find the asymptotic expansion:
\begin{equation}\label{eq:Zasym}
\begin{split}
\Gin{\MZ}(\br,\omega,\bk) 
&=
    \frac{2\,\bqt^2\, \KS (\omega,\bk) \, \br^{d-4}}{d\,(d-1)\,(d-2)(d-4)}
            + 1-\frac{i\bwt}{\br} + \sum_{i=2}^4\frac{\mathfrak{a}_i}{\br^i}    - \frac{1}{d} \frac{\Ost}{\br^d}    +\order{\br^{-d-1}}\,,
\end{split}
\end{equation}
where
\begin{equation}\label{eq:Zasymcfs}
\begin{split}
\mathfrak{a}_2
&=
    \frac{2\,\bqt^2\,\Ost}{d(d-1)} +\frac{d-3}{2\,(d-2)}\left(\frac{\bqt^2}{d-1}-\bwt^2\right)\\
\mathfrak{a}_3
&=
       i\bwt\left[  \frac{\Ost}{d^2\,(d^2-1)}  
    \left( \frac{(d^2-1)\,(d+3)}{3}\, \bwt^2-(d^2+2\,d+5)\,\bqt^2\right)\right. \\
&\left.\qquad  \quad
      - \frac{d-3}{2(d-2)}
        \left(\frac{\bqt^2}{d-1}-\frac{d-5}{d-3}\,\bwt^2\right)
+\frac{i\bwt}{d^2}
        \left(\frac{d^2-5}{d^2-1}\,\bqt^2+(d+3)\frac{\bwt^2}{3}\right)\right]\\
\mathfrak{a}_4
&= 
    \frac{1}{8\,(d-2)(d-4)}
    \left[\frac{(d-5)\,(d-7)\,\bwt^4}{3}+\left(\frac{16}{d\,(d-1)}+d-5\right)\bqt^2
    \left(\frac{\bqt^2}{d-1}-2\,\bwt^2\right) 
    \right]  .
\end{split}
\end{equation}
Furthermore, we have introduced a new function $\Ost $, which is determined up to the quartic order in gradients to be
\begin{equation}\label{eq:Gatt4}
\begin{split}
\Ost
&\equiv 
  	\Gatt 
    +  2i\bwt \left(H_k(1)\,\kaps^2 +H_\omega(1)\, \bwt^2\right)  \\
&\qquad 
  - \frac{2\,\bqt^2\, \KS}{d\,(d-1)\,(d-2)}
    -\left(\lambda_\omega+\frac{4}{d} H_\omega(1)\right) \bwt^4 \\
&\qquad 
   - \left(\lambda_{\omega k}+\frac{4}{d}H_k(1)\right) \bwt^2\, \kaps^2
    - \lambda_k\left(\kaps^2+\frac{4(d-2)^2}{d(d-3)}\bwt^2\right)  +\cdots\,.
\end{split}
\end{equation}
We will have more to say about this function below, but note that it agrees with $\Gatt$ up to quadratic order in gradients.

The rescaled metric functions $\{\PHE,\PHO, \PHW\}$ and the field $\MW$ can be recovered from the above solution for $\MZ$ using \eqref{eq:EOSMZ}. We can express these functions in terms of the ingoing  boundary to bulk Green's function. We  normalize this inverse propagator using the solution for $\MZ$ so the asymptotic values are obtained in terms of the modulus field $\PoZ_{\skL,\skR}$ on the grSK geometry. 

Denoting the  ingoing Green's function of the Weyl factor $\PHW$  as $\Gin{\text{W}}$ we  can evaluate directly (we report all the divergent non-normalizable terms, but only  the leading  normalizable ones)
\begin{equation}\label{eq:PHWasym}
\begin{split}
\Gin{\text{W}} (\br,\omega,\bk)
&= 
	    \frac{1}{ \Lk}\left[r\,\Dz_+ + \frac{k^2}{d-1} \right]\, \Gin{\MZ} (\br,\omega,\bk)\\
&= 	  
    \frac{2\,\KS }{d\,(d-1)\,(d-2)}\, \br^{d-2}
    \left[1-\frac{i\bwt}{\br}-\frac{(d-5) (d-1)\,\bwt^2+(d-3)\, \bqt^2}{2\,(d-1)\,(d-4)} \, \frac{1}{\br^2}\right] \\
&\qquad 
    +\frac{2\,\Ost}{d\,(d-1)} \,
    \left[ 1-\frac{i\bwt}{\br} \right]  +\order{\br^{-2}}\,.
\end{split}
\end{equation}

The data above is sufficient to obtain $\MW$,  which after all is just a linear combination of $\MZ$ and $\PHW$ from \eqref{eq:linearXSZ}. Scaling $\MW$ with a factor of $i\bwt$ gives us the function $\PHO$. For completeness let us record the leading terms in $\MW$:
\begin{equation}\label{eq:MWasym}
\begin{split}
\Gin{\MW}(\br,\omega,\bk)
&= 
	\frac{r}{\Lk}\, \left(\Dz_+ -\frac{1}{2}\, r^2\, f'\right)  \Gin{\MZ} (\br,\omega,\bk) \\
&=
    \frac{2\,\KS }{d\,(d-1)\,(d-2)}\, \br^{d-2}
    \left[1-\frac{i\bwt}{\br}
    -\frac{(d-5) \,\bwt^2+ \bqt^2}{2\,(d-4)} \, \frac{1}{\br^2}\right] \\
&\qquad 
	- \frac{1}{d-1} \left(1 - \frac{2}{d}\,\Ost \right)    \left[ 1-\frac{i\bwt}{\br}\right] + \order{\br^{-2}}\,.
 \end{split}
\end{equation}	

The final piece of data is the metric function $\PHE$ (we don't need to evaluate $\PHO$ since it is just $\MW$ up to a factor of $i \omega$).  The easiest way to obtain its inverse Green's function, denoted $\Gin{\text{E}}$, is by  using $\PHE = \Dz_+ \MW$. We find that it has the following asymptotic expansion:
\begin{equation}\label{eq:PHEasym}
\begin{split}
\Gin{\text{E}} (\br,\omega,\bk)
&= 
	\Dz_+ \left(\frac{r}{\Lk} \left[\Dz_+ -\frac{r^2\, f'}{2}\right]  \right)\Gin{\MZ} (\br,\omega,\bk)
	= \Dz_+ \Gin{\MW}(\br,\omega,\bk)\\
&= 
	\frac{2\,\KS}{d\,(d-1)\,b} \,  \br^{d-1}
	\left[ 1 - \frac{i\bwt}{\br} -\frac{(d-3)\, \bwt^2 + \bqt^2}{2\,\br^2}
	 +i\,\bwt\, \frac{(d-5) \,\bwt^2+ \bqt^2}{2\,(d-4) \,\br^3}
	\right] +\order{\br^0} \,.
\end{split}
\end{equation}	

As noted in the main text below \eqref{eq:kappas}, by examining the leading non-normalizable mode of $\PHW$, or equivalently $\MW$, we deduce the coefficient $K_s$ accurate to quartic order in gradients, enabling us to get the sound attenuation function $\Gatt(\omega,\bk)$ 
to quadratic order.
However, if one  parameterizes the solution by a function $\Gatt(\omega,\bk)$, which is at least first order in derivatives, then we find that the constant mode in $\PHW$ and $\MW$, given in terms of $\Ost$ above, can equivalently be expressed in terms of this attenuation function. Our explicit expression determines this coefficient to be $\Ost$, suggesting that $\Gatt$ and $\Ost$ agree not just up to quadratic order, but rather that $\Ost$ determines $\Gatt(\omega,\bk)$ all the way to  quartic order, with the specific relation $\Ost = \Gatt(\omega,\bk)$. While we have not checked this statement explicitly, we conjecture this to be true to all orders in the gradient expansion. As evidence we offer that the expressions for the on-shell action and the stress tensor can be entirely parameterized in terms of $\Gatt$, suggesting that the solution must likewise be given in terms of it.

\paragraph{Sound dispersion to quintic order:} If, as conjectured above, the function $\Ost$ is indeed  the attenuation function $\Gatt$, one can deduce the dispersion locus 
$\omega(k)$ to quintic order in momenta. This is because  with the knowledge of $\Ost$ to quartic order, we actually have $\KS$ accurate to sextic order in gradients. Assuming our conjecture $\Ost = \Gatt(\omega,\bk)$, we find
\begin{equation}\label{eq:DispLoc}
\begin{split}
\bwt(\bqt) 
&=
	 \frac{\bqt}{\sqrt{d-1}} 
	 -i\, \frac{\nu_s}{2}\bqt^2 
	 + \frac{\nu_s}{2\,\sqrt{d-1}} \left(1-\frac{d-1}{4}\, \nu_s-(d-2)\, H_k(1)\right)\bqt^3 \\
&\qquad \qquad
	 + i \nu_s \, \mathfrak{h}_4 \, \bqt^4 +\frac{\nu_s}{\sqrt{d-1}} \, \mathfrak{h}_5 \, \bqt^5\,,
\end{split}
\end{equation}
with
\begin{equation}\label{eq:DispLoch45}
\begin{split}
\mathfrak{h}_4
 &= 
 	\frac{\nu_s}{2\, (d-2)}+ \frac{1}{d-1} \, H_\omega(1) -\frac{d-4}{d(d-1)}\, H_k(1),\\
\mathfrak{h}_5
&=
	 -\frac{(d+2)}{8\,(d-2)}\, \nu_s + \frac{(d-1)\left(4\,d\,(\,d+4)-4\right)}{64\,d\,(d-2)}\,\nu_s^2
	  - \frac{3\,d^3-38\,d^2+60\,d+24}{4\,d^2\,(d-1)}\, H_k(1)\\
 &\qquad 
 	+ \frac{3}{8} \, (d-2)^2\, \nu_s \,H_k(1)^2 + \frac{3\,d-8}{d\,(d-1)}\, H_\omega(1) - \frac{1}{2(d-1)} \, \lambda_{\omega}
 	+\frac{d-2}{2\,(d-1)} \, \lambda_{\omega k}\\
&\qquad\quad
	-\frac{d^4-9\, d^3+31\,d^2-43\,d+16}{2\,d\,(d-1)\,(d-3)} \, \lambda_k\,.
\end{split}
\end{equation}
The constants $H_k(1)$ is known in terms of the Harmonic number function as noted in  \cref{fn:Hk1harm}, while $H_\omega(1)$ has an expression in terms of an infinite sum (cf., Eq.~(A.28) of \cite{Ghosh:2020lel}). We have not attempted to derive similar expressions for the constants $\lambda_\omega$. $\lambda_k$, and $\lambda_{\omega k}$ defined in \eqref{eq:lamdef}.

\subsection{The designer field solution repackaged}
\label{sec:exppargrad}

To facilitate comparisons with the analysis of \cite{He:2021jna} we present first the solution of $\MZ$
in a slightly different form, using the exponentiated form of the gradient expansion ansatz. We introduce
\begin{equation}\label{eq:Zexpgrad}
\MZ(r) =  \frac{1}{b^{d-2}}\, \exp\left(\sum_{n,m=1}^\infty  (-i)^m \, \bwt^m\, \bqt^n \, \Mser{\MZ}{m,n}(\br)\right).
\end{equation}
The functions $ \Mser{\MZ}{m,n}(\br)$ can be determined almost entirely in terms of the solution for the $\ann=d-1$ Markovian scalar, $\Mser{d-1}{m,n}(\br)$.  The deviations from Markovian behaviour only occurs for the momentum dependent pieces, as explained in \cref{sec:MZgradexp}. Therefore, 
\begin{equation}\label{eq: expsolm0}
\Mser{\MZ}{m,0}(\br) = \Mser{d-1}{m,0}(\br) \,.
\end{equation}

These functions $\Mser{d-1}{m,n}(\br)$ are compiled in Table 1 of \cite{He:2021jna} for general Markovianity index $\ann$ and can be specialized to $\ann = d-1$. To write compact expressions we introduce an integral transform:\footnote{The data given in \cite{He:2021jna} is written in terms of the inverse radial variable $\ri = \frac{1}{\br}$ which we have translated here to the dimensionless radial variable $\br$. } 
\begin{equation}\label{eq:MItransform}
\mathfrak{T}\big[\mathfrak{g}\big](\br) \equiv \int^\infty_\br\, \frac{dy}{y^2\, f}\, \mathfrak{g}(y) , \ \hspace{0.5cm} 
\hat{\mathfrak{T}}\big[\mathfrak{g}\big](\br) \equiv \int^1_\br\, \frac{dy}{y^2\,f}\, \mathfrak{g}(y) \,.
\end{equation}  
In terms of these,  the auxiliary functions $\Dfn{d-1}{m,n}(\br)$  are defined at low orders in the gradient expansion as
\begin{equation}\label{eq:DeltaFns}
\begin{split}
\Dfnh{\ann}{2,0} (\br) 
& = 
     \hat{\mathfrak{T}}\left[\br^{1-d} - \br^{d-1}\right] \,,\\
\Dfnh{\ann}{1,2} (\br) 
& = 
    -\hat{\mathfrak{T}}\left[\br^{-1}\, \Dfnh{\ann}{2,0}(\br) \right]  \,,\\
\Dfnh{\ann}{3,0} (\ri) 
& = 
    - \hat{\mathfrak{T}}\left[\br^{1-d}\, \Dfnh{\ann}{2,0}(\br)^2 \right] \,.
\end{split}
\end{equation}
The data entering the solution can then be presented compactly in \cref{tab:Mgradsol}.

\renewcommand{\arraystretch}{1.8}
\begin{table}[th!]
\centering
\begin{tabular}{|c|c|c|}
\hline
\shadeR{$\mathfrak{T}\big[\mathfrak{g}\big]$}   &   \shadeB{$\mathfrak{g}$ }        &       \shadeB{Asymptotics} \\     \hline

$\Mser{d-1}{1,0}$  & 
    $1 - \br^{1-d}$    &  
        $\frac{1}{\br} - \frac{\br^{-d}}{d} $ \\ \hline

$\Mser{d-1}{0,2}$  & 
    $\frac{1}{d-2}\frac{1}{\br}\left(1-\br^{2-d}\right)$     &  
            $\frac{1}{d-2}\left(\frac{1}{2\br^2} -\frac{\br^{-d}}{d}\right)$ \\ \hline

$\Mser{d-1}{2,0}$  & $-\br^{1-d}\Dfnh{d-1}{2,0}(\br)$                            & $\frac{1}{2(d-2)}\frac{1}{\br^2} - \frac{\Dfn{d-1}{2,0}(1)}{d}                                                                                                                                   \br^{-d}$ \\ \hline

$\Mser{d-1}{3,0}$  & $2\br^{1-d}\Mserh{d-1}{2,0}(\br)$                           & $-\frac{2\Mser{d-1}{2,0}(1)}{d}\br^{-d}-                                                                                                                            \frac{\br^{-2-d}}{(d-2)(d+2)}$ \\ \hline

$\Mser{d-1}{1,2}$  & $2\br^{1-d}\Mserh{d-1}{0,2}(\br)$                       &  $\frac{2\Mser{d-1}{0,2}(1)}{d}\br^{-d}-                                                                                                                                                   \frac{\br^{-2-d}}{(d-2)(d+2)}$ \\ \hline

$\Mser{d-1}{4,0}$  &   $2\br^{1-d}\left(\Mserh{d-1}{3,0}(\br)+\frac{1}{2}\Dfnh{d-1}{3,0}(\br)\right)$ 
    & 
        \scriptsize $-\frac{1}{4(d-4)(d-2)^2}\frac{1}{\br^4}+\frac{\Dfn{d-1}{3,0}(1)+
        2\Mser{d-1}{3,0}(1)}{d}\br^{-d}$\normalsize 
                                            \scriptsize $+\frac{\Dfn{d-1}{2,0}(1)}{(d-2)(d+2)}                                                                                                                                           \br^{-2-d}$\normalsize    \\ \hline

$\Mser{d-1}{2,2}$  & \scriptsize$2\br^{1-d}\left(\Mserh{d-1}{1,2}(\br)-\frac{1}{d-2}\left(\Dfnh{d-1}{1,2}(\br)-\Mserh{d-1}{2,0}(\br)\right)\right)$\normalsize                                                                              &   \scriptsize$-\frac{1}{2(d-4)(d-2)^2}\frac{1}{\br^{4}}-\frac{1+(d-2)\Dfn{d-1}{2,0}(1)}{(d+2)(d-2)^2}\br^{-2-d}$                                                                                                                                           \normalsize \\ 
                                        &                                                                                   & \scriptsize $-\frac{2\left(\Mser{d-1}{2,0}(1)-\Dfn{d-1}{1,2}                                                                                                                                        (1)+(d-2)\Mser{d-1}{1,2}(1)\right)}{d(d-2)}                                                                                                                                           \br^{-d}$ \normalsize\\ \hline

$\Mser{d-1}{0,4}$  & $-\frac{1}{d-2}\left(\Mserh{d-1}{0,2}(\br)-\Mserh{3-d}{0,2}(\br)\right)$ 
    & 
        \scriptsize $-\frac{1}{4(d-4)(d-2)^2}\frac{1}{\br^4}+\frac{\Mser{d-1}{0,2}(1)
            -\Mser{2-d-1}{0,2}(1)}{d(d-2)}           \br^{-d}+\frac{\br^{-2-d}}{(d+2)(d-2)^2}$                                                                                                                                           \normalsize \\ \hline
\end{tabular}
\caption{The functions appearing in the gradient expansion of the Markovian $\sen{d-1}$ up to the fourth order in gradients, given in the form of an integral transform defined in Eq.~(\ref{eq:MItransform}). We  also present the leading asymptotic behaviour of the functions which is used for computing boundary observables.}
\label{tab:Mgradsol}
\end{table}

The solutions for the remaining functions with $n\neq0$ up to quartic order can be determined to be
\begin{equation}
\begin{split}
\Mser{\MZ}{0,2}(\br) 
&=  
    \Mser{d-1}{0,2}(\br) - \frac{2\,(d-2)}{d-1}\, \Mser{d-1}{0,2}(\br) \,, \\
\Mser{\MZ}{1,2}(\br) 
&=  
     \Mser{d-1}{1,2}(\br) -\frac{2\,(d-2)}{d-1} \, \Mser{d-1}{1,2}(\br) - \frac{4\, (d-2)}{d\,(d-1)} \, \Mser{d-1}{0,2}(\br)\,, \\
\Mser{\MZ}{2,2}(\br) 
&= 
      \Mser{d-1}{2,2}(\br) - \frac{2\,(d-2)}{d-1} \, \Mser{d-1}{2,2}(\br) - \frac{4}{d\,(d-1)} \, \Mser{d-1}{2,0}(\br)+ \frac{4}{d\,(d-1)} \, \Mser{d-1}{0,2}(\br) \\
&
    + \frac{2}{d\, (d-1)} \,\Dfn{\MZ}{2,2}(\br) + \frac{4}{d\,(d-1)}\,\Dfn{d-1}{1,2}(\br)- \frac{2}{d\,(d-1)\,(d-2)\,(d-4)} \, \br^{d-4} \,,\\
\Mser{\MZ}{0,4}(\br) 
&=  
     \Mser{d-1}{0,4}(\br) - \frac{4\,(d-2)}{(d-1)^2} \, \Mser{d-1}{0,4}(\br)  + \frac{4\,(d-3)}{d(d-1)^2} \, \Mser{d-1}{0,2}(\br) + \frac{2}{d\,(d-1)^2(d-2)}\, \Dfn{\MZ}{0,4}(\br)\,.
\end{split}
\end{equation}
As indicated we could express the solution almost completely in terms of the Markovian data computed earlier, but had to introduce two additional functions:
\begin{equation}
\begin{split}
\Dfn{\MZ}{0,4}(\br) 
&= 
    \int_\infty^\br\, \frac{dy}{y^2f} \left(\frac{1}{y^{d-1}} - y^{d-3}\right)  , \\
\Dfn{\MZ}{2,2}(\br)
&=
    \int_\infty^\br\, \frac{dy}{y^2f} \, \left[\frac{1}{y^{d-1}}\left(2\,\Dfnh{\MZ}{0,4}(y) -\frac{1}{d-2}\right) +\frac{1}{d-2}\, \frac{1}{y^3}\right] .
\end{split}
\end{equation}
Of these, only $\Dfn{\MZ}{0,4}(\br) $ has a divergent behaviour at large $\br$. It is defined using the earlier functions in \eqref{eq:Dmark} by analytic continuation. In this parameterization,  
both $ \Mser{\MZ}{2,2}(\br)$ and $\Mser{\MZ}{0,4}(\br)$  have divergent corrections (to quartic order), while that used in \cref{sec:MZgradexp} the divergence was isolated into $V_k(\br)$.

\section{Spatially homogeneous modes}
\label{sec:zeromodes}

This appendix is somewhat outside the main line of development of the paper and is included for completeness. As we saw above in \cref{sec:psolspace} the dynamics of spatially inhomogeneous modes can be distilled into that of a single non-Markovian scalar $\MZ$ satisfying a second order non-Markovian differential equation. We wish now to analyze  what the equations of motion imply for spatially homogeneous modes.

We will carry out the analysis in two steps. First we examine the large diffeomorphisms of the \SAdS{d+1} solution that respect the Debye gauge choice. Subsequently, we look at the solution space for spatially homogeneous modes looks like and parameterize it in terms of the most general allowed data compatible with asymptotically locally AdS asymptotics.  

We will find that the two sets of analyses lead to the same set of zero modes. The surprise will be that there are more zero modes than those that can be lifted to physical moduli captured by $\MZ$.

\subsection{Large diffeomorphisms of the background}
\label{sec:largediffeos}

The background \SAdS{d+1} geometry is parameterized by $b$ which is a measure of the black hole temperature or mass. One has additionally chosen a particular Weyl frame, by making a suitable choice of the radial coordinate $r$. Consider the now the following diffeomorphism and parameter shift on the orbit space, leaving spatial homogeneity intact: 
\begin{equation}\label{eq:alphadiffeo}
\begin{aligned}
r  &\mapsto r  + \chi_r(v,r ) \,, \qquad &
\vb{x} & \mapsto (1+C_x) \, \vb{x} \,, \\
v & \mapsto v + \chi_v(v,r) \,, \qquad & 
b & \mapsto (1+C_b)\,  b\,.
\end{aligned}
\end{equation}
We have two functions on the orbit space and two constants $C_x$ and $C_b$ which the reader will recognize is precisely the freedom to rescale spatial length scales and the boundary temperature homogeneously.

To check this explicitly, we simply implement this change on the background solution, 
\begin{equation}
ds_{(0)}^2  = -r^2f\, dv^2 + 2 \, dv \,dr + r^2\, d\vb{x}^2\,.
\end{equation}	
Retaining terms to linear order in the $\chi$'s  and $C$'s to be consistent with our linearized analysis, we find that the metric remains in the Debye gauge and can be cast in the form of our linearized ansatz. To wit, 
\begin{equation}
\begin{split}
ds_{(0)}^2 
&\mapsto ds_{(0)}^2 + 
 	\frac{\PHE-r f\,\PHW}{r^{d-3}}\, dv^2 
	 +\frac{2}{r^{d-1}f} \left(\PHO-\PHE +rf\,\PHW\right) dv\,dr
	 + r^2\, \frac{\PHW}{r^{d-2}}\, d\vb{x}^2 \\
& 
\qquad\qquad
	-\frac{1}{r^{d+1}f^2}\left[2(\PHO-\PHE) + r f \,(d-1)\, \PHW + \PHB\right] dr^2 
\end{split}
\end{equation}
with the metric functions taking the form:
\begin{equation}\label{eq:metricchi}
\begin{split}
\PHW 
&= 
	2\, C_x\, r^{d-2} + 2 \, r^{d-3}\, \chi_r  \,, \\
\PHO 
&= 
	- r^{d-1} f\, \Dz_+ \left(\chi_v  - \frac{ \chi_r }{r^2f}\right) + r^{d-3} \pdv{v} \chi_r - r^d f'\, C_b \,, \\
\PHE \ 
&= 
	 -2 \,r^{d-1} f\, \pdv{v}(\chi _v - \frac{\chi_r}{r^2f}) - r^{d-1} f' \, \chi_r + 2 \, r^{d-1} f\, C_x - r^d f'\, C_b \\
\PHB 
&=
	-2  \,r^{d-1}  f\, \pdv{v}(\chi_v-\frac{\chi_r}{r^2f} ) -2\, \Dz_+\left(r^{d-3}\, \chi_r\right) -2\, (d-3)\, r^{d-1} f\, C_x \,.
\end{split}
\end{equation}	

To summarize, we have two functions on the orbit space, which along with two constant parameters characterize the space of large diffeomorphisms. As such, demanding that the spacetime be asymptotically locally \AdS{d+1} constrains the two functions. It is not hard to see that 
$\chi_r(v,r) \to r\, \chi_r^\infty(v)$, while $\chi_v(v,r) \to \chi_v^\infty(v)$. The former corresponds to a choice of (time-dependent) Weyl frame, while the latter is the boundary time reparameterization mode. 

Before we turn to the dynamical equations it is useful to examine the quantity $\Tone$. We find 
\begin{equation}\label{eq:TphiBzero}
\Tone - r \, \PHB 
= 
	\frac{d \, (C_x - C_b)}{b^d}\,. 
\end{equation}	
We recall that in our solutions this parameter is vanishing which suggests that in our solution space for $k\neq 0$ we only have access to the locus $C_x = C_b$. This says that we are only allowed to change the background temperature (which is rescaled by $C_b$) provided that we concertedly change the spatial length scales/volume (set by $C_x$). As presaged early on in our discussion, the overall rescaling of temperature in non-compact space requires injecting an infinite amount of energy, which is unphysical.
 
It is interesting to evaluate the boundary stress tensor for this family of large diffeormorphisms. One finds using the results of \cref{sec:BYT}
\begin{equation}\label{eq:Tchi}
\begin{split}
T\indices{_v^v}
&=
	\lim_{r\to \infty} 
	 \frac{d-1}{2}\, \frac{\PHE}{r^{d-1}}    - \frac{d\,(d-1)}{b^d}\, (C_x-C_b) \,, \\
T\indices{_i^j}
&=
	 \lim_{r\to \infty} \frac{d-1}{2}\, \frac{\PHE}{r^{d-1}}  \, \delta\indices{_i^j} \,.
\end{split}
\end{equation}
Notice that the contribution from the asymptotic value of the source $\PHE$ is exactly what the spatially inhomogeneous modes pick up.  Our large diffeomorphisms however have an additional contribution in the energy density which comes from $\Tone - r\, \PHB$ using \eqref{eq:TphiBzero}; it is this contribution that is vanishing in our designer solution.

\subsection{Parameterizing the solution space: \texorpdfstring{$k= 0$}{homogeneous}} 
\label{sec:parsolhom}

We give now a short complementary perspective on the spatially homogeneous modes from the dynamical equations of motion. As one might anticipate the physical solution space is already fully characterized by the large diffeomorphism modes, so we expect to see the same degree of freedom in the solution space, as we shall verify below. The manner in which this happens is that the dynamics of the system is modified at $k=0$ resulting in additional moduli in the problem. Technically, at $k=0$ some of the equations degenerate. For one, the scalar equation $\EEq_T$, the spatial vector equations $\EEq_4$ and $\EEq_5$, and the spatial tensor equation $\EEq_7$ are trivially satisfied, each being explicitly proportional $k_i$.

We are then left with a simpler  set of equations, which we write as\footnote{
	We use the superscript $0_k$ to remind us that we are looking at spatially homogeneous modes of our fields.}
\begin{equation}\label{eq:zero12B6}
\begin{split}
\EEq_1 
&=
	\Dz_+ \left(\Tone^{0_k} - r\, \PHB^{0_k}\right)   , \\ 
\EEq_2
&=
	  \pdv{v}(\Tone^{0_k} - r\, \PHB^{0_k}) \,, \\  
\EEqB
&= 
	-\Dz_+ \left(\Tone^{0_k} - \frac{r}{2}\, \PHB^{0_k} \right)
	 + r\, \pdv{v}\PHO^{0_k} 
	 - \frac{r}{2} \left(\Dz_+ - \gOrb + r f \right) \left[\Dz_+ \PHW^{0_k} - (d-2) \, rf\, \PHW^{0_k} \right] \\ 
&\qquad \quad
	+ \frac{r}{2}\,\left(-\pdv[2]{v}\, \PHW^{0_k} + d  \,r\, \PHB^{0_k} \right)  \,, \\
\EEq_6
&=
	\Dz_+ \left(\frac{\widetilde{\EEq}_5}{f}\right) 	 + i\omega \, \frac{\widetilde{\EEq}_4}{f}  \,, \\
\end{split}
\end{equation}
where we have written the last equation succinctly using the parameterization defined in \eqref{eq:eeq45} for convenience. 

The first two equations in \eqref{eq:zero12B6} imply that the combination $\Tone^{0_k} - r\, \PHB^{0_k}$ must be a constant, 
\begin{equation}\label{eq:CTzero}
\Tone^{0_k} - r\, \PHB^{0_k} = \mathfrak{C} _T \;\; \Longrightarrow \;\; \Dz_+ \PHW^{0_k}  = \PHE^{0_k} - \PHB^{0_k}  + \frac{r^2f'}{2}\, \PHW^{0_k}  - \frac{\mathfrak{C} _T}{r} \,.
\end{equation}	
We are then left with two equations for effectively three variables. However, the remaining two equations are not independent, for
\begin{equation}\label{eq:eeq60dep}
\EEq_6 + \Dz_+ \left(\frac{2\,\EEqB}{rf}\right) + \frac{rf'}{f} \,  \EEqB =0\,.
\end{equation}	
The one remaining equation can be written as
\begin{equation}\label{eq:eeqB0}
r^{d-1} f\, \Dz_+ \left(\frac{1}{r^{d-1}\, f}\, \left(\Dz_+\PHW^{0_k} + \PHB^{0_k}\right)  \right)	
	+  \left( \pdv[2]{v}\,\PHW^{0_k}  - 2\,  \pdv{v} \PHO^{0_k} \right) =0 \,.
\end{equation}	

The solution space is parameterized by two functions $\PHW^{0_k}$ and $\PHO^{0_k}$. The other two functions $\{\PHE^{0_k}, \PHB^{0_k}\}$ can be solved  in terms of them using 
 \eqref{eq:CTzero}  and \eqref{eq:eeqB0}. Inspired  by \eqref{eq:metricchi}, we can w.l.o.g parameterize the functions $\PHW^{0_k}$ and $\PHO^{0_k}$ 
\begin{equation}\label{eq:WOparzero}
\begin{split}
\PHW^{0_k} 
&=
	2\, \mathfrak{F}_1(v,r) +   2\, C_x\, r^{d-2}  \,, \\ 
\PHO^{0_k} 
&= 
	\pdv{v}\mathfrak{F}_1+r^{d-1}f\, \Dz_+ \mathfrak{F}_2(v,r)  - \frac{d\, C_b}{b^d\, r} \,,
\end{split}
\end{equation}	
where we demanded that the solution be asymptotically AdS.  Here $C_x$ is a coefficient that will related to integration constants momentarily. We recover the other two functions as given in \eqref{eq:metricchi} as\footnote{
		The operator $\Dz_+$ annihilates $e^{-\frac{1}{2}\, \bwt\, \ctor}$ but this solution does not satisfy ingoing boundary conditions and hence we restrict to allowing an integration constant $\mathfrak{C}_B$.
}
\begin{equation}\label{eq:EBsolzero}
\begin{split}
\PHB^{0_k}
&=
	-2\,\Dz_+\left(\mathfrak{F}_1 +C_x\, r^{d-2} \right) + 2 \,r^{d-1} f\, \pdv{v}\mathfrak{F}_2  + r^{d-1} f\, \mathfrak{C}_B \,, \\
\PHE^{0_k} 
&= 
	2 \,r^{d-1} f\, \pdv{v}\mathfrak{F}_2  -  r^2f'   \bigg(\mathfrak{F}_1  +  C_x\, r^{d-2}  \bigg) +  \frac{\mathfrak{C}_T}{r}  + r^{d-1} f\, \mathfrak{C}_B \,.\\
\end{split}
\end{equation}	

One can readily see that we may identify $\mathfrak{F}_1 = r^{d-3} \,\chi_r $ and $\mathfrak{F}_2 = -\chi_v + \frac{\chi_r}{r^2f} $ recovering the background diffeomorphisms. This implies that the time-independent pieces in
$\PHB^{0_k}$ and  $\PHE^{0_k}$ are
\begin{equation}\label{eq:EBconstants}
\PHB^{0_k}(r) \sim \left(\mathfrak{C}_B - 2 \, (d-2)\, C_x \right)  r^{d-1} f \,, \qquad  
\PHE^{0_k}(r) 	\sim   \frac{\mathfrak{C}_T -  b^{-d}\, d\,C_x}{r} +  r^{d-1} f\, \mathfrak{C}_B  \,.
\end{equation}
Comparing with \eqref{eq:metricchi} we conclude that we can parameterize the integration constants by defining them in terms of $C_x$ and $C_b$:
\begin{equation}\label{eq:constmatch}
\mathfrak{C}_B =  2 \,   C_x \,, \qquad \mathfrak{C}_T = \frac{d(C_x - C_b)}{b^d}   \,.
\end{equation}	
We thus recover the full set of large diffeomorphisms from the analysis of the equations of motion.

\paragraph{Zero modes and designer field:} One can ask if this solution is related to the zero momentum solution of the designer field   $\MZ$.  The ingoing Green's function given  in \eqref{eq:Zsolution} requires $\mathfrak{C}_T = \mathfrak{C}_B =0$, which can be inferred by noting that the source terms vanish at $\omega =0$. In addition, we also note that $\PHB$ vanishes identically in our parameterization by $\MZ$, which demands a relation between $ \mathfrak{F}_1$ and $\mathfrak{F}_2$, viz.,
\begin{equation}\label{eq:relBZ}
\begin{split}
\Dz_+ \mathfrak{F}_1 = r^{d-1}f\, \pdv{v} \mathfrak{F}_2\,.
\end{split}
\end{equation}
This is consistent with the relations \eqref{eq:EOSMZ} between $\PHE, \PHO, \PHW$ and $\MZ$ at $k=0$,  which in turn require that $\MZ$ is given in terms of the diffeomorphism functions 
$\mathfrak{F}_1 $ and $\mathfrak{F}_2$ as
\begin{equation}\label{eq:relZ12}
\frac{1}{d-1}\, \Dz_+ \MZ^{0_k} = r^2f'\, \mathfrak{F}_1 \,, \qquad \frac{1}{d-1}\, \pdv{v}\MZ^{0_k} = \pdv{v}\mathfrak{F}_1 - r^{d-1}f\, \Dz_+ \mathfrak{F}_2\,.
\end{equation}	
Solving for $\mathfrak{F}_1$ and $\mathfrak{F}_2$ using  \eqref{eq:relBZ} and the first equation of \eqref{eq:relZ12}, we can then write an autonomous equation for $\MZ^{0_k}$. The resulting equations turns out to be implied by the  Markovian wave equation (with $k=0$)
\begin{equation}\label{eq:ZmarkE}
\frac{1}{r^{d-1}}\, \Dz_+\left(r^{d-1} \, \Dz_+ \MZ^{0_k}\right)  -\pdv[2]{v} \MZ^{0_k} =0 \,,
\end{equation}	
which is of course the zero momentum limit of our designer field equation \eqref{eq:ZMeqn}. This is also the zero momentum limit of a minimally coupled massless scalar field in \SAdS{d+1}.

We thus see that a part of the large diffeomorphisms is indeed the homogeneous solution for the designer field. In particular, given a solution for $\MZ(r,\omega,\bk)$ satisfying \eqref{eq:ZmarkE}, we can determine large diffeomorphism functions $\mathfrak{F}_1$ and $\mathfrak{F}_2$. However, one does not recover the full set of large diffeomorphisms. Specifically, the part parameterized by the constants $\{\mathfrak{C}_T, \mathfrak{C}_B\}$ (or equivalently $\{C_x,C_b\}$) is not recovered from the designer field dynamics.


\providecommand{\href}[2]{#2}\begingroup\raggedright\endgroup

\end{document}